\documentclass[aps,superscriptaddress,showpacs,showkeys,preprintnumbers,amsmath,amssymb,amsfonts,nofootinbib,10pt]{revtex4}

\usepackage{graphicx}

\newcommand{\be}{\begin{eqnarray}}
\newcommand{\ee}{\end{eqnarray}}
\newcommand{\Tr}{\mathrm{Tr}}

\begin{document}

\title{Scaling functions for the O(4)-model in d=3 dimensions}
                                         
\author{Jens~Braun}
\affiliation{TRIUMF, 4004 Wesbrook Mall, Vancouver, BC V6T 2A3, Canada}
\author{Bertram~Klein}
\affiliation{Physik Department, Technische Universit\"at M\"unchen, James-Franck-Strasse 1, 85748 Garching, Germany}

\date{\today}

\begin{abstract}
A non-perturbative Renormalization Group approach is used to calculate scaling functions for an O(4) model in $d=3$ dimensions in the presence of an external symmetry-breaking field. These scaling functions are important for the analysis of critical behavior in the O(4) universality class. For example, the finite-temperature phase transition in QCD with two flavors is expected to fall into this class.
Critical exponents are calculated in local potential approximation.
Parameterizations of the scaling functions for the order parameter and for the longitudinal susceptibility are given. Relations from universal scaling arguments between these scaling functions are investigated and confirmed. The expected asymptotic behavior of the scaling functions predicted by Griffiths is observed. Corrections to the scaling behavior at large values of the external field are studied qualitatively. These scaling corrections can become large, which might have implications for the scaling analysis of lattice QCD results.
\end{abstract}

\preprint{TUM/T39-07-24}

\maketitle

\section{Introduction}

The investigation of phase transitions and critical phenomena has been an important motivation in the development of the Renormalization Group (RG) \cite{Wilson:1971bg,Wilson:1971dh,Wegner:1972ih,Wilson:1973jj,Wetterich:1992yh,Polchinski:1983gv,Liao:1994cm,Liao:1994fp,Liao:1995nm}, and it remains one of the core areas of the application of RG methods.
Due to the universal behavior of different systems in the vicinity of a critical point, the results of relatively simple systems, such as spin models with an O(N)-symmetry, play an important role in the analysis of phase transitions in much more complicated systems.

Among the systems of interest is also QCD, where scaling behavior appears in various contexts.
An example is the scaling behavior observed for the chiral phase boundary in the plane of temperature and number of massless quark flavors, near the critical number of flavors above which QCD
is still asymptotic free but chirally symmetric in the infrared regime \cite{Braun:2005uj,Braun:2006jd}.
For two light quark flavors, the phase transition at finite temperature is expected to fall into the O(4) universality class \cite{Pisarski:1983ms}. 
While the dynamics of the non-Abelian gauge fields and quarks in QCD are difficult to describe in detail and require a non-pertubative treatment, universality arguments can be brought to bear on the analysis of the behavior near the phase transition.
QCD lattice simulations remain indispensable as non-perturbative calculations in which all fermionic and gauge degrees of freedom are taken into account. However, since quark masses are large and explicitly break the symmetry, and since the simulation volumes are finite, the observation of critical behavior on the lattice is difficult. Strictly speaking, a phase transition cannot even occur in a finite volume. 

For this reason, the finite-volume behavior and the scaling behavior with an external symmetry-breaking field of a theory in the same universality class are of great interest for the analysis of the critical behavior in lattice simulations. 
Different methods can be used to study the scaling behavior in simpler model systems. However, since long-range fluctuations are essential for the critical behavior, even here a non-perturbative approach is needed. Renormalization Group methods are a natural choice for this problem. The RG approach used in this paper allows to study quantum field theories in an effective-potential framework in infinite as well as finite volume. It also allows the inclusion of an external symmetry-breaking field so that the dependence on the strength of such a field can be studied. 
A natural upper bound for the possible size of the field in such a calculation is set by the ultraviolet cutoff of the theory.
For a spin model, such as the Ising model, such an external field corresponds to an external magnetic field $H$, whereas in QCD the bare quark masses fill the role of the symmetry-breaking field. In the case of a spontaneously broken symmetry, the external field controls the mass of the pseudo-Goldstone modes; in QCD, this is the mass of the pions.   
The renormalization group allows us to vary the external parameters of a given theory, 
such as volume size or external symmetry breaking, over a wide range in parameter space, from a rather
deformed theory to the theory of physical interest.
This makes it possible to close a gap between scaling behavior in a finite volume and with a given external field and the scaling behavior in the limit of interest. 

Even though current lattice calculations are performed in large volumes and close to the physical value of the pion mass, it remains difficult to extract the order of the phase transition, and this question is still not conclusively settled \cite{Aoki:2006we,D'Elia:2005bv,Cossu:2007mn}.
Usually, only the known critical exponents are used as input for a finite-size scaling analysis of lattice data. Including information about the scaling function as well would increase the power of the analysis \cite{Toussaint:1996qr,Schulze:2001zg, Engels:2001bq, Mendes:2006zf}.
The question of the size of the scaling region also remains open, at least for staggered fermions current volumes might still be to small to observe finite-size scaling \cite{Kogut:2006gt}.

The main results of the present paper are accurate determinations of the scaling functions of
an O(4) symmetric model in three dimensions over a wide parameter range for the external symmetry breaking field.
Scaling functions for the order parameter and the longitudinal susceptibility are calculated, and the relations between these functions are investigated. We also investigate qualitatively scaling corrections for large values of the external field.
The discussion of finite-size scaling functions from a finite-volume calculation  \cite{Klein:2007gu,Klein:2007qh} is the subject of a forthcoming paper.
The discussion of other O(N)-symmetric models will also be postponed to a future publication.

The paper is organized as follows: In Sec.~\ref{sec:scaling}, we provide an overview over the scaling behavior in the three-dimensional O(4) model, known results for the scaling function and relations between the different forms of the scaling function.
The RG method is introduced in Sec.~\ref{sec:RG} and technical details and the setup of our calculation are discussed. 
We determine the critical exponents for our calculation in Sec.~\ref{sec:critex} to ensure a consistent analysis of our results. A comparison to values from other methods allows to assess the systematic errors inherent in our calculation.
Our main results for the scaling functions are presented in Sec.~\ref{sec:orderparameter} for the order parameter, and Sec.~\ref{sec:susceptibility} for the susceptibility.
Sec.~\ref{sec:masses} contains a brief discussion of the masses of the longitudinal and transverse fluctuations, $m_\sigma$ an $m_\pi$, and we present our concluding remarks in Sec.~\ref{sec:conclusions}.

\section{Scaling in the O(4) model in $d=3$} 
\label{sec:scaling}

Scaling behavior can be observed in the vicinity of a critical point.
Critical behavior is characterized by a diverging correlation length $\xi \to \infty$. Close to a critical point,  there is no relevant length scale due to the critical long-range fluctuations, and such a system is invariant under a change of length scale. As a consequence, the behavior of thermodynamic observables in the vicinity of the critical point is characterized by critical exponents which are universal for systems in the same universality class.

The critical behavior is governed by the singular part of the free energy, and the behavior of thermodynamic variables can be derived from the free energy.
We consider an O(4) model with an external symmetry-breaking field $H$ and the temperature $T$ as the two relevant coupling constants. The invariance of the singular free energy density under a rescaling of the length with a factor $\ell$ close to the critical point can be expressed as 
\be
f_s(t, h) &=&  \ell^{-d} f_s(\ell^{y_t} t, \ell^{y_h} h). 
\label{eq:scaling-free-energy}
\ee
Possible corrections to the scaling behavior due to irrelevant couplings are neglected. The dimensionless couplings $t$ and $h$ are defined as
\be
t=\frac{T-T_c}{T_0} \quad \mathrm{and} \quad h=\frac{H}{H_0}, 
\ee
and the critical point is located at $(T, H)=(T_c, 0)$ or $(t, h)=(0, 0)$. 
The critical  exponents can be expressed in terms of $y_t$ and $y_h$ and are connected by a number of scaling laws, so that there are only two independent ones:
\be
y_t=\frac{1}{\nu}, \quad y_h=\frac{\nu}{\beta \delta},\nonumber
\ee
\be
\gamma = (2-\eta) \nu, \quad \gamma = \beta(\delta -1), \quad \beta = \frac{1}{2}(d-2+\eta)\nu, \quad \nu d = \beta(1+\delta).
\ee
 
The critical exponent $\nu$ governs the behavior of the correlation length $\xi$ and the critical exponent $\gamma$ governs the behavior of the (longitudinal) susceptibility $\chi$ as the critical temperature is approached, 
\be
\xi \sim |t|^{-\nu}, \quad \quad
\chi \sim |t|^{-\gamma}.
\ee

The critical exponents $\beta$ and $\delta$ describe the behavior of the order parameter $M$ as the critical point is approached from either the $h$- or the $t$-direction:
\be
M(t=0, h) = h^{1/\delta}, \quad \quad M(t, h=0) = (-t)^\beta \;\;\; \mathrm{for} \;\;\; t<0.
\label{eq:normalization}
\ee
In the usual convention, which we will use in our analysis, the normalization constants $H_0$ and $T_0$ are determined by these two relations.  

Starting from the scaling form of the free energy \eqref{eq:scaling-free-energy}, the scaling behavior of the order parameter $M$ and the longitudinal susceptibility $\chi$ can be derived using their thermodynamic definitions
\be
M &= &-\frac{\partial f_s}{\partial H} \label {eq:Mdef} \\ 
 \chi&=& \frac{\partial M}{\partial H} \label{eq:chiLdef}
\ee
by choosing the scaling factor $\ell$ in an appropriate way. The most intuitive and most commonly used form of the scaling functions is obtained by choosing $\ell^{y_h} h = 1$. The free energy density then becomes a function of $h$ and the scaling variable $z = t/(h^{1/(\beta \delta)})$ only, and the order parameter can be expressed as
\be
M = h^{1/\delta} f(z) \Leftrightarrow  \frac{M}{h^{1/\delta}} = f(z)
, \quad \quad z\equiv \frac{t}{h^{1/(\beta \delta)}}, 
\label{eq:Mzscaling}
\ee
where $f(z)$ is a universal scaling function. With the normalization \eqref{eq:normalization},  $f(z)$ has the properties $f(0) =1$ and $f(z) \to (-z)^\beta$ for $z \to -\infty$. 

From the definition of the longitudinal susceptibility $\chi$, it follows that its behavior can also be described in terms of a scaling function $f_\chi(z)$, which in turn is determined from the scaling function of the order parameter and its derivative:
\be
\chi = \frac{1}{H_0} h^{1/\delta -1} \frac{1}{\delta} \left[f(z) -\frac{z}{\beta} f^\prime(z)\right] \Leftrightarrow H_0\, h^{1-1/\delta} \chi = f_\chi(z) = \frac{1}{\delta} \left[f(z) -\frac{z}{\beta} f^\prime(z)\right].
\ee 
The scaling function for the susceptibility is therefore completely determined if one knows the critical exponents and the scaling function for the order parameter. This relation presents an additional, very non-trivial test for the scaling behavior.

While the scaling function $f(z)$ expressed in terms of the scaling variable $z$ is convenient for visualizing the behavior of the order parameter, it is not the most stringent test of the scaling behavior as it is relatively insensitive to scaling violations. In addition, we find that it is not very convenient for parameterizing the scaling behavior. Better for this purpose is the Widom-Griffiths parameterization \cite{Widom:1965xx,Griffiths:1967xx} of the equation of state as $y=y(x)$ in terms of the scaling variables
\be
x \equiv \frac{t}{M^{1/\beta}}, \quad\quad y \equiv \frac{h}{M^\delta}.
\ee
(For definiteness, we will take $H \ge 0$ and $M \ge 0$ throughout.) This parameterization can be obtained from the scaling form of the free energy \eqref{eq:scaling-free-energy} by taking $\ell^{y_t} t =1$. It is equivalent to the first one, and the two parameterizations are related by
\be
f(z) = \frac{1}{y^{1/\delta}}, \quad 
z = \frac{x}{y^{1/(\beta \delta)}}.
\label{eq:translation}
\ee
With the normalization \eqref{eq:normalization}, the function $y=y(x)$ satisfies the normalization conditions $y(0)=1$ and $y(-1)=0$.

Assuming that the equation of state in the form $H = H(M, T)$ is analytic everywhere away from the first-order line $H=0, T< T_c$ and the critical point $H=0, T=T_c$, Griffiths showed \cite{Griffiths:1967xx} that for $H>0$, $T> T_c$ and $M>M_0>0$ for some value $M_0$, i.e. $x>0$, the equation of state can be expressed in a convergent expansion as
\be
y(x) &=& \sum_{n=1}^{\infty} c_n\, x^{\gamma - 2 \beta(n-1)} = x^\gamma \left(c_1 + c_2 x^{-2 \beta} + c_3 x^{- 4\beta} + \ldots \right). 
\label{eq:Griffithsasymptotic}
\ee 
For asymptotically large values of the scaling variable $x$, this expression allows us to determine the critical exponent $\gamma$ from the leading term and thus provides an additional check for the consistency of the scaling behavior. It also suggests a parameterization of the equation of state $y=y(x)$ for large $x$. 

As for the first parameterization, the scaling behavior of the susceptibility can be expressed in terms of the scaling function $y=y(x)$, its derivative and the critical exponents. One finds
\be
\chi =  \frac{\partial M}{\partial H} = \frac{1}{M^{\delta-1}} \frac{1}{H_0}   \left(\delta \, y(x) -\frac{1}{\beta} x\, y^\prime(x)\right)^{-1}   \Leftrightarrow \left[ H_0 M^{\delta -1} \chi \right]^{-1} = \delta \, y(x) -\frac{1}{\beta} x\, y^\prime(x).
\ee
We consider later the inverse rescaled expression $\left[ H_0 M^{\delta -1} \chi \right]^{-1}$ to avoid the divergence of the universal result which appears at $x=0$ in this parameterization.

With Griffiths' expression \eqref{eq:Griffithsasymptotic}, and using the scaling law $\gamma=\beta(\delta-1)$, one finds for the rescaled susceptibility for large $x$ the expression
\be
 \left[ H_0 M^{\delta -1} \chi \right]^{-1} &=& \delta \left( y(x) -\frac{1}{\beta\delta } x\, y^\prime(x) \right) = x^\gamma \left(c_1 + 3 c_2 x^{-2 \beta} + 5 c_3 x^{-4 \beta} + \ldots \right)\nonumber\\
 &=& \sum_{n=1}^\infty c_n \, (2n-1) x^{\gamma-2\beta(n-1)}, 
 \label{eq:Griffithsasymptoticsusceptibility}
\ee
with the known coefficients $c_n$ from \eqref{eq:Griffithsasymptotic}. Remarkably, this means that for large $x$ the leading-order terms from the scaling function of the order parameter $y(x) = c_1 x^\gamma + \ldots$ and the scaling function of the inverse rescaled susceptibility, $\delta y(x) -\frac{1}{\beta} x y^\prime(x) = c_1 x^\gamma+ \ldots$ should be exactly the same, provided the scaling laws for the critical exponents hold.

The transverse susceptibility
\be
\chi_T &=& \frac{M}{H} 
\label{eq:chiTdef}
\ee
also satisfies a universal scaling relation. With eq.~\eqref{eq:Mzscaling}, one finds that the scaling function is given by the one for the order parameter, after $\chi_T$ has been rescaled in the same way as the longitudinal susceptibility:
\be
\chi_T &=& \frac{1}{H_0} h^{1/\delta-1} \left(\frac{M}{h^{1/\delta}}\right) =  \frac{1}{H_0} h^{1/\delta-1}  f(z) \Leftrightarrow H_0 h^{1-1/\delta} \chi_T = f(z).
\ee
Using the Widom-Griffiths scaling form, one finds likewise that the rescaled susceptibility is given by the scaling function $y(x)$:
\be
\chi_T &=& \frac{1}{H_0 M^{\delta-1}}\frac{1}{y}
\Leftrightarrow
\left[H_0 M^{\delta -1} \chi_T \right]^{-1}= y.
\ee
Since we only recover the scaling function of the order parameter from these relations, the transverse susceptibility does not provide any additional test of the scaling behavior. In our implementation of the model, the relation \eqref{eq:chiTdef} follows from the determination of the order parameter from the minimum of the potential and is explicitly satisfied by construction. For this reason we do not analyze  the scaling relation for $\chi_T$ separately.

There are few results for the scaling functions in the literature. The equation of state in the form $y=y(x)$ was calculated in the $\epsilon$-expansion to ${\mathcal O}(\epsilon^2)$ by Br\'ezin, Wallace and Wilson \cite{Brezin:1972fb}, and a parameterization of the result close to the coexistence point $x=-1$ (corresponding to the first-order line $H=0, T<T_c$) is provided by Wallace and Zia \cite{Wallace:1975vi}.
Monte-Carlo simulations of the O(4)-symmetric spin model on the lattice were used by Toussaint \cite{Toussaint:1996qr} to obtain the scaling function $f(z)$, and by Engels and Mendes \cite{Engels:1999wf}  to obtain a parameterization for the equation of state $y(x)$ (see also \cite{Engels:2001bq}), and by Cucchieri and Mendes for a comparison in a parametric form \cite{Cucchieri:2004xg}.

The result obtained from the $\epsilon$-expansion \cite{Brezin:1972fb} for $N=4$ in $d=3$ becomes after expansion around $x=-1$   \cite{Wallace:1975vi}
\be
y(x) &=& (1+x) \left\{1+\epsilon [A \log(1+x) + B]   +\epsilon^2[C \log^2(1+x)+ D \log(1+x) + E
 ] +{\mathcal O}(\epsilon^3)\right\}\\
\mathrm{with}&& A=0.125, \;\; B=-0.0270494, \;\;C=-0.015625,\;\; D=0.148902, \;\; E=0.138468.
 \nonumber
\ee
This expression is valid for $(1+x) \ll 1$.
In \cite{Wallace:1975vi} it is argued that the expansion around $x=-1$ can be inverted to give an equation of state $x=x(y)$ in the form
\be
x(y)+1 &=& \tilde{c}_1 y + \tilde{c}_2 y^{1-\epsilon/2} + \tilde{d}_1 y^2+ \tilde{d}_2 y^{2-\epsilon/2} + \tilde{d}_3 y^{2-\epsilon} + \ldots, 
\label{eq:generalepsiloninverted}
\ee
where the coefficients are determined only to ${\mathcal O}(\epsilon)$ while the original expansion is known to ${\mathcal O}(\epsilon^2)$. For $N=4$ and $d=3$ ($\epsilon =1$) the perturbative result from \cite{Wallace:1975vi} is
\be
&x(y)+1 = \tilde{c}_2 y^{1/2}  + ( \tilde{c}_1 + \tilde{d}_3) y & \mathrm{with}\\
 &\tilde{c}_2 = 0.250+0.280 \epsilon = 0.530, \;\; \tilde{c}_1=0.750-0.253\epsilon = 0.497, \;\; \tilde{d}_3=0.03125, \;\; \tilde{c}_1 + \tilde{d}_3 =0.528.& \nonumber
 \label{eq:epsiloninverted}
\ee
The normalization condition $y(0)=1$ is only satisfied to about $10 \%$ ($y(0)=0.92$ for the inverted expansion), and the expansion is very accurate only close to the expansion point $x=-1$. In general, it is known that the $\epsilon$-expansion satisfies the large-$x$ behavior from Griffiths' expansion \eqref{eq:Griffithsasymptotic} only order by order in $\epsilon$, but not explicitely \cite{Brezin:1972fb}, and thus even the full expression is not expected to be accurate for $x\gg 1$.

While critical exponents from the $\epsilon$-expansion are available to higher order in $\epsilon$ \cite{Guida:1998bx}, for consistency we use in the comparisons involving the scaling function the critical exponents calculated to ${\mathcal O}(\epsilon^2)$ for the O(4) model in $d=3$ \cite{Brezin:1972fb}:
\be
\beta = \frac{16}{41} \approx 0.3902, \quad \quad \gamma = \frac{133}{96} \approx 1.3854, \quad \quad \delta = \frac{107}{24} \approx 4.458. \nonumber
\ee
These values satisfy the scaling law $\gamma=\beta (\delta -1)$ to within $3 \%$

We observe that an ansatz $y(x) = c_0 (1+x)^\tau$ reproduces the first terms of the $\epsilon$-expansion around $x=-1$, if we assume that the exponent can be expanded as $\tau = 1 + \tau_1 \epsilon + \ldots$ in $\epsilon$:
\be
y(x) &=& = c_0 (1+x)^{1+\tau_1 \epsilon +\ldots} = c_0 (1+x) (1+ \tau_1 \epsilon \log(1+x) + \dots). 
\label{eq:phenomansatz}
\ee
For this reason we use such an ansatz with the exponent $\tau$ as a free parameter to fit our results, and we find that such a parameterization works surprisingly well. In addition, for large $x \gg 1$ it coincides with the leading term of Griffiths' expression if $\tau = \gamma$.

From Monte-Carlo lattice simulations of an O(4) spin model, the parameterization for the equation of state from Engels and Mendes \cite{Engels:1999wf} is also provided in the inverted form $x=x(y)$. For small values of $x$, the parameterization $x_s(y)$ is taken from the $\epsilon$-expansion \eqref{eq:epsiloninverted}, with coefficients determined non-perturbatively from a fit to the lattice results \cite{Engels:1999wf,Engels:2001bq}:
\be
& x_s(y) + 1 =  \tilde{c}_2 y^{1/2} + ( \tilde{c}_1 + \tilde{d}_3) y + \tilde{d}_2 y^{3/2}&\mathrm{with}\\
 &\tilde{c}_2 =0.674(08), \;\;\tilde{c}_1 + \tilde{d}_3 =0.345(12),\;\;  \tilde{d}_2 = -0.023(5).\nonumber
\ee
For large values of $x$, the first two terms of an inverted form of Griffiths' expression \eqref{eq:Griffithsasymptotic} are used \cite{Engels:2001bq}:
\be
&x_l(y) = a y^{1/\gamma} + b y^{(1-2\beta)/\gamma}&\mathrm{with} \nonumber\\
&a =1.084(6), \;\; b=-0.994(109).&
\ee
The values of the critical exponents are here
\be
&\beta=0.380, \;\; \delta = 4.86\;\; \gamma= 1.4668, \;\; \nu=0.7423.& \nonumber
\ee
The authors also formulate an interpolating expression for the equation of state
\be
x(y) = x_s(y) \frac{y_0^3}{y_0^3+ y^3} + x_l(y) \frac{y^3}{y^3+ y_0^3}, \;\;\; y_0=10.0
\label{eq:Mendesinterpolation}
\ee
which describes the results well for a range of at least $-1< x <30$ and $0<y<150$. Inspired by this idea, we will use a similar interpolation for the equation of state $y=y(x)$ to describe our results.

In \cite{Cucchieri:2004xg}, in addition to a parametric description, the results for the scaling function is also given in the above form. The coefficients found there are
\be
\tilde{c}_2 =0.746(3), \quad \tilde{c}_1 + \tilde{d}_3 =0.19(1), \quad \tilde{d}_2 = 0.061(8)
\ee
for the small-$x$ region ($x<1.5$), and 
\be
a=1.07(1), \quad b=-0.95(3) 
\ee 
for the large-$x$ region. The same expression is used for an interpolation.
The differences to the earlier results are attributed to slightly different values of the critical exponents used in the calculation.

\section{Renormalization Group approach to scaling}
\label{sec:RG}
In this section we discuss our Renormalization Group (RG) approach to scaling behavior in the presence of an external symmetry-breaking field and the derivation of the
proper-time flow equations for the $O(4)$-potential in infinite volume. We also
give a detailed discussion of the approximations that we have used throughout
this work. Reviews of and introductions to functional RG methods can be found in e.g. 
 \cite{Litim:1998nf,Bagnuls:2000ae,Berges:2000ew,Polonyi:2001se,Pawlowski:2005xe,
 Gies:2006wv,Schaefer:2006sr,Delamotte:2007pf}.
The effective action of the $O(4)$ model in $d$ space-time 
dimension is given by
\begin{equation}
\Gamma [\phi]=
\int d^{d}x \left\{
\frac{1}{2}Z_{\phi}(\partial_{\mu}\vec{\phi})^{2} + U(\phi)\right\}.
\end{equation}
Here we have neglected possible kinetic terms of higher order 
such as $Y_{\phi}(\vec{\phi} \partial _{\mu} \vec{\phi})^2$.
The components of the vector $\phi$ are labeled according the role that the
corresponding fields are playing in the spontaneously broken regime, $\vec{\phi}^{\mathrm{{T}}}
=(\sigma,\pi^{1},\pi^{2},\pi^{2})$. We choose the first component to be the radial
mode in the regime where the ground state of the theory is not symmetric 
under $O(4)$ transformations:
\be
\langle \vec{\phi} \rangle = \vec{\phi_{0}} ^{\mathrm{{T}}} =(\sigma_0,0,0,0)
\ee
The potential $U(\phi)$ in its present form depends only 
on $\phi^2=\sigma ^2 + \vec{\pi}^2$. As 
discussed in \cite{Braun:2004yk,Braun:2005gy,Braun:2005fj} such an ansatz for the 
potential is not appropriate if one is interested in a study of phase transitions in a finite 
volume, since strictly speaking spontaneous symmetry breaking does not occur
in a finite volume. Therefore the presence of an non-vanishing external source in the
ansatz for the effective action is indispensable for our study.
Moreover we do not solve the RG flow for the full potential but expand the potential
in a Taylor series around the vacuum expectation value and solve then
the RG flow for the expansion coefficients of this series. Before we discuss our
ansatz for the potential in more detail, we briefly sketch the derivation of proper-time
RG equations in general.

In a Gaussian approximation, we can perform the functional integration
of the bosonic fields and obtain the one-loop effective
action for the scalar fields $\phi$,
\begin{equation}
\Gamma[\phi] = \Gamma_{\Lambda_{UV}}[\phi] +
\frac{1}{2}\Tr\log\left(\Gamma_{B}^{(2)}[\phi]\right) \label{eq:1loop}
\end{equation}
where $\Gamma_{B}^{(2)}[\phi]$ is the inverse two-point function evaluated
at the background-field and $\Gamma_{\Lambda}$ contains the initial
values of the RG flow at ultraviolet (UV) scale $\Lambda$. 
From now on, we neglect a possible space dependence of the expectation
value and take the wave-function renormalization $Z_{\phi}$
to be constant, $Z_{\phi}=1$. Therefore the anomalous dimension of the
scalar field is zero, $\eta=0$. We have to keep this approximation in mind for 
the interpretation of our results, since the anomalous dimension of the 
scalar fields influences the critical exponents which in turn are the
key ingredients for finite-size scaling. However, the scalar anomalous 
dimension is small compared to one, see e. g. \cite{Tetradis:1993ts}, and therefore our 
approximation to neglect the running of the wave-function renormalization is 
well-justified. 

In order to regulate the infrared (IR) divergences
in Eq.~\eqref{eq:1loop} we use the Schwinger proper-time representation of
the logarithm and introduce an IR cutoff
function\footnote{On the one hand physical quantities
  calculated from the RG flow should not depend on the choice of the
  cutoff function in the limit $k \rightarrow 0$. On the other hand, a study of the regulator 
  dependence of the results in this limit allows us to check the quality of
  our truncations. A detailed study of the 
  dependence of our results on 
the choice of the regulator function is in progress \cite{BraunKlein:2008}.} 
$f_{a} (\tau k^{2})$, where the variable
$\tau$ denotes Schwinger's proper time and $k$ is a cutoff scale which has
mass dimension one. 
The derivative of the cutoff function with respect to the scale $k$ is
given by (see e. g. Refs.~\cite{Liao:1994cm,Schaefer:1999em})
\begin{equation}
k\frac{\partial}{\partial k}f_{a}(\tau k^{2}) =
-\frac{2}{\Gamma(a+1)}(\tau k^{2})^{a+1}e^{-\tau k^{2}} \,.
\label{eq:cutoff-fct}
\end{equation}
In the following we choose $a=\frac{d}{2}$. The relation between
the so-called proper-time RG and the Functional RG has been
worked out in detail in \cite{Litim:2001ky,Litim:2002hj,Litim:2002xm}. As was found 
in Refs. \cite{Litim:2001up,Litim:2001hk}, the flow equation which results from this choice for 
the proper-time cutoff 
function coincides with the flow equation obtained in the Functional RG
framework with an optimized cutoff. 

The inverse two-point function $\Gamma_{B}^{(2)}[\phi]$ 
in Eq.~\eqref{eq:1loop} depends on the second derivatives of the effective potential $U$, 
\be
M_{ij}=\frac{\partial ^2 \bar{U}(\phi)}{\partial \phi_i \partial \phi _j}\,.
\ee
The eigenvalues  of this matrix evaluated at the minimum of the 
potential are the masses of the fields. 
 By replacing the bare masses and couplings in the
inverse two-point functions with the scale-dependent quantities, we
obtain the so-called renormalization group improved flow equation
for the effective potential $\bar{U}_{k}$ in infinite volume (see also 
e. g. \cite{Liao:1994cm, Litim:2001up,Litim:2001hk, Bohr:2000gp}):
\begin{eqnarray}
k \frac{\partial}{\partial k} \bar{U}_k(\phi) &=&   \frac{(k^2)^{d/2+1}}{(4 \pi)^{d/2}}\frac{1}{\Gamma(d/2+1)} \left( \frac{3}{k^2 + M^2_{\pi,k}(\phi)} + \frac{1}{k^2 +M^2_{\sigma,k}(\phi)} \right).
\label{eq:floweq_dim}
\end{eqnarray}
The quantities $M_{\sigma}$ and $M_{\pi}$ are the eigenvalues of the 
the second-derivative matrix of the potential. Note that
these qunatities still depend on the background field $\phi$.

For a study of critical exponents and scaling it is convenient
to deal with dimensionless quantities rather than dimensionful
quantities. Therefore we introduce the dimensionless potential $u$, the
dimensionless masses $m_{\sigma}$ and $m_{\pi}$ as well as 
the dimensionless field $\varphi$ by
\be
u_k (\varphi)=k^{-d} U_k (\phi)\,\quad m^2 _{i,k}(\varphi)=k^{-2} M^2 _{i,k}(\phi)
\quad\text{and}\quad \varphi=k^{-\frac{d-2}{2}}\phi\,.
\label{eq:dimquant}
\ee
Applying these definitions to the flow equation~\eqref{eq:floweq_dim}, we obtain
\be
\partial_t u_t(\varphi) &=&  -d u(\varphi)  + \frac{1}{(4 \pi)^{d/2}}\frac{1}{\Gamma(d/2+1)} \left( \frac{3}{1 
+ m^2 _{\pi}(\varphi)} + \frac{1}{1 + m^2_{\sigma}(\varphi)} \right)\,,
\label{eq:floweq_dimless}
\ee
where the dimensionless flow variable $t$ is given by $t=\ln (k/\Lambda)$.
By integrating the flow equation (either the dimensionless or the dimensionful formulation)
from the UV scale $\Lambda$ to $k \to 0$, we obtain an effective potential in which quantum 
corrections from all scales have been systematically included.

We now discuss the ansatz for the effective potential $U_k(\phi)$. Since we are 
interested in a study of phase transitions with an external symmetry-breaking field, we have to introduce a 
corresponding linear term in the field into our ansatz
for the effective action. In order to solve the RG flow for the effective potential $\bar{U}$, we
expand the potential in local $n$-point couplings around its minimum $\sigma_0(k)$
\be
U_k(\sigma, \vec{\pi}^2) = a_0(k) + a_1(k) (\sigma^2 + \vec{\pi}^2 - \sigma_0(k)^2) + a_2(k) (\sigma^2 + \vec{\pi}^2 - \sigma_0(k)^2)^2 + \ldots - H \sigma
\label{eq:pot_ansatz}
\ee
where $H$ is the fixed, external symmetry-breaking field. The minimum $\sigma_0(k)$ 
is the order parameter of the system which can be identified with the 
magnetization $M$ in an Ising model, 
\be
M=\sigma_0. 
\ee
Since we have absorbed the
symmetry-breaking linear term in the effective action into the ansatz for the potential, $U_k$
depends now on the fields $\sigma$ and $\pi$ separately. The condition
\be
\frac{\partial U_k(\sigma,\vec{\pi}^2)}{\partial \sigma}\Bigg|_{\sigma=\sigma_0,\vec{\pi}^2=0} 
\stackrel{!}{=} 0
\label{eq:min_cond}
\ee
ensures that we are expanding around the actual physical minimum. From Eq. \eqref{eq:min_cond}, we find that the
RG flow of the coupling $a_1$ and the minimum $\sigma_0$ are related by the condition
\be
2 a_1(k) \sigma_0(k) = H\,.\label{eq:min_cond2}
\ee 
This condition keeps the minimum at $(\sigma, \vec{\pi}) = (\sigma_0(k), \vec{0})$. The flow
equation of the minimum is thus related to the flow of the coupling $a_1$ in a simple way.
Taking the derivate with respect to $k$ of Eq.~\eqref{eq:pot_ansatz}, we obtain
\be
k\frac{\partial U_k(\sigma, \vec{\pi}^2)}{k} = k\frac{\partial  a_0(k)}{\partial k} -
2 a_1(k) \sigma _0 (k) k\frac{\partial \sigma_0 (k)}{\partial k} + 
k\frac{\partial a_1(k)}{\partial k} (\sigma^2 + \vec{\pi}^2 - \sigma_0(k)^2) + \ldots 
\label{eq:pot_ansatz_dt}
\ee
Note that the derivative of the potential with respect to the regulator scale does not contain
any contributions linear in the fields. Of particular interest for our studies in this work is, 
apart from the order parameter $\sigma_0(k)$, the longitudinal as well as the transversal susceptibility.
Taking the total derivative of the minimum condition \eqref{eq:min_cond} with 
respect to $H$, one finds that the longitudinal susceptibility is inverse
proportional to inverse mass squared of the radial mode:
\be
\chi _L=\frac{1}{m_{\sigma}^2}=\frac{1}{2 a_1 + 4 a_2 {\sigma_0}^2}\,.
\ee
The transverse susceptibility is defined as the ratio of the order parameter $M=\sigma_0$
(magnetization) to the external field $H$ and is inverse proportional to the mass of the
(pseudo-) Goldstone particles
\be
\chi _T=\frac{M}{H}=\frac{1}{m_{\pi}^2}=\frac{1}{2 a_1}\,.
\ee

The RG flow equation for the couplings $a_i$ can now be obtained 
straightforwardly by expanding the flow equation \eqref{eq:floweq_dim} around
the minimum $\sigma_0 (k)$ and projecting it on Eq.~\eqref{eq:pot_ansatz_dt}.
The RG flow of the minimum of the potential is
determined by Eq.~\eqref{eq:min_cond2}. Apparently, this procedure for a derivation of the flow
of the potential $u$ results in an infinite
set of flow equations for the couplings $a_{i}$. In order to solve the set of equations
for the couplings, we have to truncate our ansatz~\eqref{eq:pot_ansatz} 
for the potential. In the following, we include fluctuations around the minimum up
to eighth order in the fields, i. e. we keep track of the running of the couplings 
$a_1$, $a_2$, $a_3$, and $a_4$. The resulting finite set of coupled
first-order differential equations are then solved numerically.
We have checked that our results do not change
significantly by including couplings of higher order.  We have also checked
numerically that our results coincide with earlier results (e. g. \cite{Litim:2001hk, Bohr:2000gp}) 
in the limit of vanishing external source $H$.

For our calculations, we have chosen a cutoff scale $\Lambda =1000\;\text{MeV}$, which is
comparable to a typical lattice cutoff ($\pi/a\approx 1500\;\text{MeV}$ with $a$ being the
lattice spacing) in current finite-temperature lattice simulations.
The initial values for the differential equations of the couplings 
are given by the effective potential at the UV scale in the form $U_\Lambda(\sigma, \vec{\pi}^2) = a_2(\Lambda) (\phi^2 - \phi_0^2(\Lambda))^2 - H \sigma$, where we choose 
$a_2(\Lambda)/\Lambda=0.025$. The flow of $a_1$ is determined by the flow of the
minimum $\sigma_0(k)$. In $d=3$ dimensions, the initial value of the minimum of the symmetric potential at the cutoff scale $\phi_0(\Lambda)$ serves as proxy for the 
temperature, and we assume that an expansion $(\phi_0(\Lambda) - \phi_0^{\mathrm{critical}}(\Lambda)) \sim (T - T_c)$ is possible.
This is to be seen in analogy to lattice simulations of spin models, where the temperature is absorbed into the spin-spin-coupling, which can then be tuned such that the system is
at its critical point.  Similarly, in Landau effective theory it is assumed that the couplings have an expansion in $T-T_c$ around the critical values.

\section{Critical exponents}
\label{sec:critex}

Critical exponents for the O(4)-model in $d=3$ have been determined by many different methods to a high degree of accuracy. Theoretical results are available from perturbative field-theoretical calculations \cite{Antonenko:1998es,Butera:1997ak,Guida:1998bx}, from lattice Monte-Carlo simulations of spin models \cite{Kanaya:1994qe, Ballesteros:1996bd} and $\phi^4$ theory \cite{Hasenbusch:2000ph}, and from non-perturbative RG calculations \cite{VonGersdorff:2000kp, Tetradis:1993ts, Bohr:2000gp,Litim:2001up,Litim:2001hk,Bervillier:2007rc}. (See e.g. \cite{Pelissetto:2000ek} for an overview.) It is not the purpose of the present paper to add to this list, in particular since our results coincide with \cite{Litim:2001hk} and since we work in the local-potential approximation where the anomalous dimension vanishes, $\eta=0$. More accurate results from functional RG methods beyond this approximation are available \cite{VonGersdorff:2000kp, Tetradis:1993ts, Bohr:2000gp}. But in order to present a consistent evaluation and to make contact with these calculations, we determine critical exponents from our results and use these values in the analysis.

Since we are working in $d=3$, temperature cannot be defined field-theoretically.
However, the relevant coupling for approaching the critical point is the initial value of the expectation value of the field $\phi$ at the ultraviolet scale $\Lambda$, $\phi_0(\Lambda)$, and we assume that an expansion  $(\phi_0(\Lambda) - \phi_0^{\mathrm{critical}}(\Lambda)) \sim (T - T_c)$ exists.

With our choice of scale $\Lambda=1000$ MeV and initial parameters, we determine the non-universal critical temperature of our model in the absence of an external symmetry-breaking field to be
\be
T_c /\Lambda^{1/2} = \phi_0^{\mathrm{critical}}(\Lambda) = 13.682\,368\;165\;072\;75(1)\; \mathrm{MeV}^{1/2}.\nonumber
\ee
(We find that this accuracy is actually necessary in the choice of the initial conditions to see the fixed point behavior clearly in the RG flow.)

We determine the critical exponents $\beta$ and $\delta$ from the order parameter $M$, and the critical exponent $\nu$ from the correlation length $\xi = 1/m_\sigma$ in the phase with restored O(N) symmetry. The critical critical exponent $\beta$ can be determined  from the order parameter at $H=0$ from 
\be
\log M &=& \beta \log(-t) = - \beta \log(T_0) + \beta \log(T_c - T),
\ee
by regarding $\log M$ as a function of the variable $\log(T-T_c)$ and either fitting this directly as a linear expression in a region close to $T_c$, or by numerically taking the derivative of the function in the limit $T \to T_c$. 
From an RG point of view, the critical exponents are defined from the eigenvalues of the linearized flow equations at the critical point. Methodically, it is therefore more sound to use the limit of the derivative for the determination. We do indeed find good agreement with the results of 
\cite{Litim:2001hk,Litim:2002cf}, where the critical exponents were determined in local potential approximation from the same RG flow equations by diagonalizing the stability matrix at the fixed point.

In order to estimate how well the critical exponents can be determined from our calculation, we also fit the linear expression below $T_c$. We repeat the determination from the derivative and from the fit in the presence of a small external field $H=1.0 \times 10^{-9}$ MeV$^{5/2}$.
The results are given in Tab.~\ref{tab:betanudet}. While the original determination from the derivative gives the most exact value, we use the results from the other determinations in order to estimate an error for this determination. We observe that the values from a fit tend to be larger for $\beta$ and smaller for $\nu$, compared to the values from the derivative.  
The values for the critical exponents $\beta$ and $\nu$ satisfy the scaling law $ \beta = \frac{1}{2} (d-2+\eta) \nu$ with $\eta=0$ and $d=3$ with better than $0.2 \%$.

\begin{table}
\begin{tabular}{l c c c c c }
\hline\hline
 & \multicolumn{2}{c}{$H=0$} &  \multicolumn{2}{c}{$H=1.0 \times 10^{-9}$ MeV$^{5/2}$}&\\
 & derivative & fit & derivative & fit & quoted\\
 \hline
 $\beta$ & $0.4030$ & $0.4051$ & $0.4033$ & $0.4108$ & $0.4030(30)$\\
 $\nu$ & $0.8053$ & $0.7953$ & $0.8007$ & $0.7920$ & $0.8053(60)$\\
 \hline\hline
 \end{tabular}
\caption[Critical exponents $\beta$ and $\nu$ from our results.]{  \label{tab:betanudet} Critical exponents $\beta$ and $\nu$ determined from our results. For the evaluation, we use the quoted values. The errors reflect the uncertainty of the determination, in addition there is a systematic error for the calculation which is not quoted and which cannot be determined from the present calculation taken by itself.}
\end{table}

The critical exponent $\delta$ can be determined from the order parameter $M$ at the critical temperature $T_c$, taken as a function of $H$
\be
\log M &=& - \frac{1}{\delta} \log H_0 + \frac{1}{\delta} \log H
\ee
again by either calculating the derivative in the limit $H \to 0$ or by fitting this as a linear expression in $\log H$. We expect the result from the limit of the derivative to be more reliable, but also calculate $\delta$ from a fit to estimate the accuracy of our determination.
The results are given in Tab.~\ref{tab:deltadet}. In contrast to our direct determination, in most functional RG calculations \cite{Tetradis:1993ts, Bohr:2000gp}, the critical exponent $\delta$ is calculated from the anomalous dimension $\eta$ with the scaling law
\be
\delta = \frac{d+2-\eta}{d-2+\eta}
\ee
or determined from the asymptotic behavior of the effective potential. Since $\eta=0$ in the local potential approximation, we expect $\delta=5$ from the scaling law. In addition, we can also use the values for the critical exponents $\beta$ and $\nu$ to calculate $\delta$ with the scaling law $d \nu = \beta (1 + \delta)$. The results are also given in the table and are used in the estimation of the error of the determination.

\begin{table}
\begin{tabular}{l c c c c c}
\hline\hline
 & derivative & fit & $\nu d/\beta -1$ & $(d+2-\eta)/(d-2+\eta)$ & quoted \\
 \hline
 $\delta$ & $4.9727$ & $4.8409$ & $4.995(90)$ & $5.0000$ & $4.973(30)$\\
 \hline\hline
\end{tabular}
\caption[Critical exponent $\delta$ from our results.]{\label{tab:deltadet} The critical exponent $\delta$, determined directly from the derivative and from a fit to the order parameter at the critical temperature as a function of $H$, and indirectly from the other critical exponents by means of the scaling laws. In our calculation, the anomalous dimension $\eta=0$, and thus $\delta=5$ is theoretically expected.}
\end{table}

The total error for the values of the critical exponents is larger than the uncertainty from the determination which we give explicitly.
It is dominated by the larger systematic error from the approximations that are necessary to solve the RG flow equations. 
This systematic error cannot be estimated by looking at the current calculation in isolation.
It is generally difficult to assess these systematic errors in non-perturbative functional RG calculation, but
the errors due to the necessary truncations can be estimated by comparing different RG cutoff schemes for the same approximation, and by systematically improving the approximations. 
The two main approximations in the present calculation are a restriction to only a finite number of $n$-point couplings in the effective potential, and a truncation in the momentum dependence to local couplings. 

The restriction to a finite number of $n$-point couplings is not very severe, and its effects can be estimated rather reliably by including a larger number of couplings. The results in \cite{Litim:2001hk, Tetradis:1993ts} show that the convergence is relatively fast, and the inclusion of a small number of 
$n$-point couplings is usually sufficient in the case of O(N) models. Our own observations confirm these findings. 

The effects of the truncation in the momentum dependence of the couplings (derivative expansion) are more difficult to assess. In the current calculation, the vanishing anomalous dimension $\eta$, which is due to the restriction to local couplings, affects all coefficients, as can be seen from the scaling laws. For this reason we do not expect to obtain values for the critical exponents comparable to those from simulations, where the anomalous dimension is non-zero. 

For our results, the effects from this truncation can be best estimated by comparing to other results from functional RG calculations. We list a number of results in Tab.~\ref{tab:critex}. We expect our results to coincide with those of \cite{Litim:2001hk}, since this determination of the critical exponents from the fixed point is performed in the same RG cutoff scheme and in local potential approximation, and we do indeed observe good agreement. The FRG calculations \cite{VonGersdorff:2000kp, Tetradis:1993ts, Bohr:2000gp} all include a wave function renormalization, which 
leads to a non-zero anomalous dimension. As one can see from the table, the resulting critical exponents are much closer to the values observed in lattice Monte-Carlo simulations. The differences to our results are a measure for the systematic error due to the local potential approximation.

Nevertheless, the values obtained here are the ones needed for a scaling analysis of the results from our calculation, since the effects that are responsible for the non-zero anomalous dimension are absent in the effective potential and our observables as well. For this reason we must use these values for a consistent evaluation of the scaling behavior.

\begin{table}
\begin{tabular}{l l l l l l l}
\hline\hline
 &  & method & \quad \quad$\nu$ & \quad \quad \, $\beta$ & \quad \quad \, $\eta$ & \quad \quad $\delta$ \\
 \hline
R. Guida, and J. Zinn-Justin & \cite{Guida:1998bx} & FT & $0.741(6)$ & $0.3830(45)$ & $0.0350(45)$ & $4.797(25)^*$\\
R. Guida, and J. Zinn-Justin & \cite{Guida:1998bx} & $\epsilon$-exp & $0.737(8)$ & $0.3820(25)$ & $0.036(4)$ & $4.792(22)^*$\\
\hline
K. Kanaya, and S. Kaya & \cite{Kanaya:1994qe} & MC & $0.7479(90)$ & $0.3836(46)$ & $0.0254(38)^*$ & $4.851(22)$ \\
H.~G.~Ballesteros {\it et al.} &  \cite{Ballesteros:1996bd} & MC & $0.7525(10)$ & $0.3907(10)^*$ & $0.0384(12)$ & $4.778(8)^* $ \\ 
 M. Hasenbusch & \cite{Hasenbusch:2000ph} & MC $\phi^4$ & $0.749(2)$ & $0.388(2)^*$ & $0.0365(10)$ & $4.789(6)^*$\\
\hline  
G. v. Gersdorff, and C. Wetterich & \cite{VonGersdorff:2000kp} & FRG & $0.739$ & $0.387^*$ & $0.047$ & $4.73^*$\\
N. Tetradis, and C. Wetterich &\cite{Tetradis:1993ts} & FRG & $0.791$ & $0.409 $ & $0.034$ & $4.80^*$\\
O. Bohr {\it et al.} & \cite{Bohr:2000gp} & FRG & $0.78$ & $0.40$ & $0.037$ & $4.80$ \\
\hline
D.~F.~Litim, and J.~M.~Pawlowski& \cite{Litim:2001hk} &FRG & $0.8043$ & $0.4022^*$ & $-$ &$5.00^*$\\
our result & & FRG & $0.8053(60)$ & $0.4030(30)$ & $-$ &$4.973(30)$ \\
 \hline\hline
\end{tabular}
\caption[Comparison of critical exponents for the O(4) model in $d=3$ dimensions.]{\label{tab:critex} Comparison of critical exponents for the O(4) model in $d=3$ from different methods. Values with an asterisk are not obtained independently, but calculated from other exponents with the scaling laws. This table is by no means comprehensive, see e.g. \cite{Pelissetto:2000ek} for a more complete listing. In addition to the functional RG (FRG) results, we list results from lattice Monte-Carlo (MC) calculations, from perturbative field-theoretical (FT) calculations in $d=3$, and from the $\epsilon$-expansion.}
\end{table}

Once the values for the critical exponents $\beta$ and $\delta$ are known, we use these values to determine the normalization constants $T_0$ and $H_0$ from fits to the results. For $T_0$, we do this by fitting with fixed $\beta$ to the order parameter for small values of the external symmetry breaking field $H=1.0 \times 10^{-10}$ MeV$^{5/2}$ and $H=1.0 \times 10^{-11}$ MeV$^{5/2}$. For $H_0$, we fit results obtained exactly at $T_c$, using the fixed value of $\delta$. We find for the non-universal normalization constants in our calculation
\be
T_0/\Lambda^{1/2} = 0.014916(5) \;\mathrm{MeV}^{1/2}, \;\;\; H_0 = 6.032(10)  \;\mathrm{MeV}^{5/2}.
\ee
In the following, we use these values in our analysis.

\section{Order parameter}
\label{sec:orderparameter}

For the determination of the scaling function, we have calculated results for the order parameter $M$, the longitudinal susceptibility (or mass $m_\sigma$ of the radial excitations), and the transverse susceptibility (or mass $m_\pi$ of the transverse excitations), over a wide range of values for $T$ and $H$. For $H$, the values span a range of seven orders of magnitude, from $H=1.0 \times 10^{-4}$ MeV$^{5/2}$ to $H=1.0 \times 10^{3}$ MeV$^{5/2}$. For each value, we have chosen appropriate ranges for $T$ around $T_c$, so that we always cover a similar range of values for the scaling variable $z=-10\ldots 10$, or $z=-30 \ldots 30$ for the larger values of $H$. 

In this way we can determine the scaling functions with high accuracy from the results with small symmetry-breaking fields, where scaling corrections remain small. At the same time, we can assess scaling violations from the results at large fields and determine where these corrections become large. 
This is relevant for the application of a scaling analysis to a physical system, and our range of values extends beyond the currently available results from spin model lattice simulations.

In the following, we use the normalization constants and critical exponents determined in section~\ref{sec:critex}. Since we work in an approximation with local couplings the anomalous dimension in our calculation is zero, $\eta=0$. This introduces a systematic error into critical exponents and observables which cannot be estimated from our calculation taken by itself. Within our approach, the systematic error can only be estimated by comparing to other RG calculations which go beyond this approximation or which use a different cutoff scheme. Work on a comparison of different RG schemes and on an improvement of the approximation are in progress. It should be kept in mind that for now this systematic error remains unquantified in our results. Of course, it can also be estimated by comparing directly to spin model lattice simulations. 

However, despite the observed deviation from the lattice in the current approximation, we will see below that the results are internally consistent: The scaling laws among the critical exponents are satisfied very well, and the expected scaling behavior is observed to an impressive accuracy. 
The same values of the critical exponents are also recovered from the asymptotic behavior of the scaling functions, as expected.
In addition, the scaling functions for the order parameter and for the susceptibility are well consistent with each other, as we will demonstrate below.
These observations serve as a cross-check of our results and confirm that the critical exponents have been determined correctly for our calculation. 

For our determination of the scaling function, we only consider the leading-order scaling behavior and do not quantify the observed corrections to scaling. For the purpose of judging the quality of a fit to the leading-order scaling behavior, we treat scaling deviations due to a small variation of the external field as an error. This is suitable for weighing how well a single scaling function can fit the results after rescaling. We estimate the scaling corrections by comparing the rescaled results for different values of $H$.

In the following, we first determine a fit for the scaling function in Widom-Griffiths parameterization. 
We found this to be the most convenient parametrization, and also the one most sensitive to deviations from scaling, so that it provides the most stringent test. We will then compare the results to those from the lattice and the $\epsilon$-expansion. After translating the results into the more intuitive scaling form $f(z)$ as a function of the scaling variable $z$, we compare our parameterization to results at large values of the external field and demonstrate deviations from scaling.  

\subsection{Scaling function in Widom-Griffiths parameterization}

\subsubsection{Phenomenological ansatz for the scaling function}

\begin{table}
\begin{tabular}{l l l r r l }
\hline\hline
\quad \quad $x$-range &\quad \quad  \quad $c$ &\quad \quad $\tau$ & \quad $\chi^2$ \quad & $\#$ d.o.f. & \quad $\chi^2/\#$ d.o.f. \\
\hline
$-1 < x < 1$ &  \quad $0.9928(14)$ & $1.6712(27)$ & $568$ & $838$ & \quad $0.6778$\\
\hline
$-1 < x < 5.3 \times 10^4$ &  \quad $0.9928(21)$ &  $1.5941(8)$ &$3682$ & $1551$ & \quad $2.374$\\
\hline\hline
\end{tabular}
\caption{Fit to the rescaled order parameter for $H=1.0 \times 10^{-4}$, $1.0 \times 10^{-3}$, $1.0 \times 10^{-2}$ MeV$^{5/2}$ with the ansatz  $y(x)=c (1+x)^\tau$. The scaling deviation is estimated from the spread of the rescaled susceptibility. Shown are separate fits for the range $-1<x<1$ and for the global fit $-1< x < 5.3 \times 10^{4}$.}
\label{tab:pheneos}
\end{table}
\begin{figure}
\includegraphics[scale=0.5]{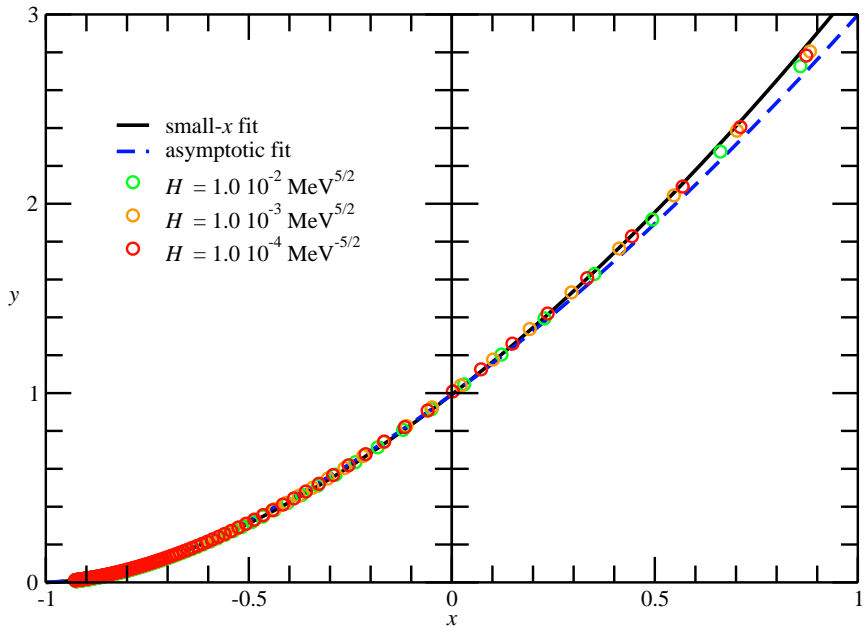}
\caption{Scaled RG results $y(x)$ for $H=1.0 \times 10^{-4}$, $1.0 \times 10^{-3}$, and $1.0 \times 10^{-2}$ MeV$^{5/2}$. Errors are estimated from the spread of the data points after rescaling and are due only to scaling corrections. Not all data points are displayed. Shown are further fits to the small-$x$ values $x <1$ (solid black line, small-$x$ fit) and to all $x$-values up to approximately $x=5.3 \times 10^4$ (dashed blue line, asymptotic fit).}
\label{fig:griffithssmallx}
\end{figure}
\begin{figure}
\includegraphics[scale=0.5]{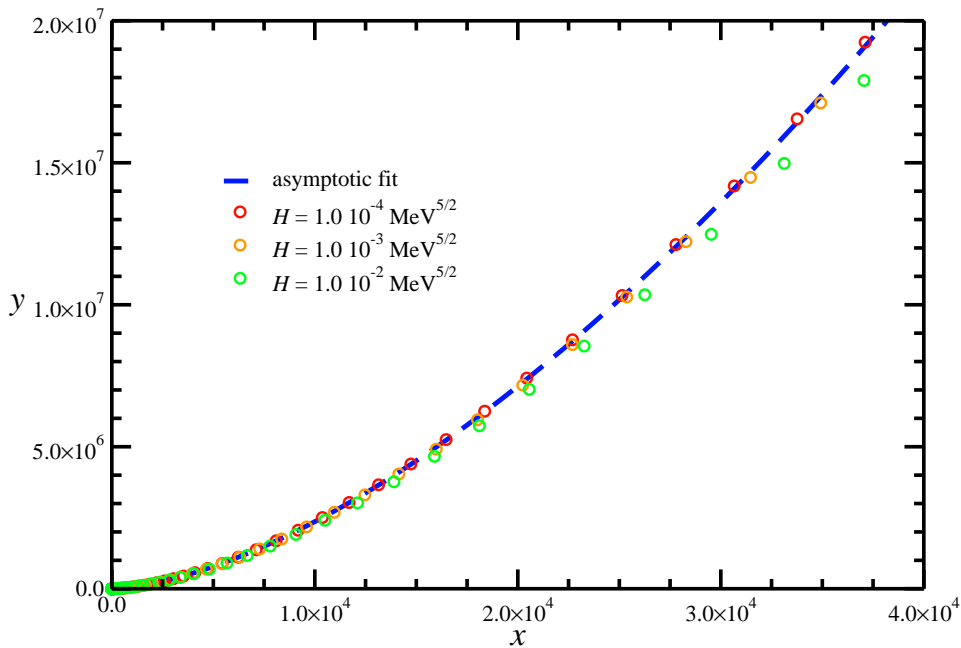}
\caption{Scaled RG results $y(x)$ for $H=1.0 \times 10^{-4}$, $1.0 \times 10^{-3}$, and $1.0 \times 10^{-2}$ MeV$^{5/2}$. Errors are estimated from the spread of the data points after rescaling (so they are mainly due to the appearance of scaling corrections). Not all data points are displayed. The fit to all $x$-values up to approximately $x=5.3 \times 10^4$ is also shown (black dashed line, asymptotic fit).}
\label{fig:griffithslargex}
\end{figure}
\begin{figure}
\includegraphics[scale=0.5]{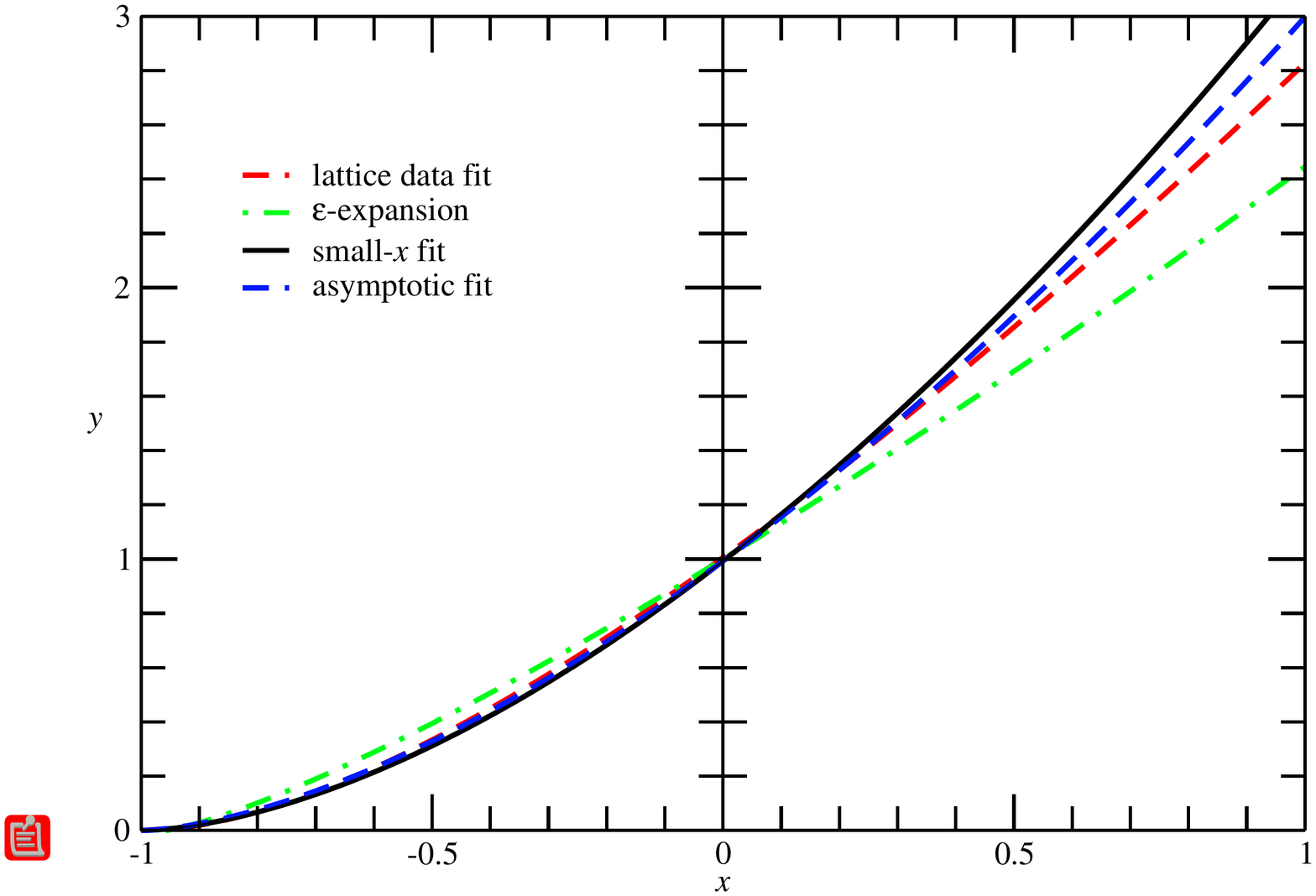}
\caption{Comparison of the results for the Widom-Griffiths scaling function $y(x)$, where $x=t/M^{1/\beta}$ and $y=h/M^\delta$ ($M>0$). Shown is the result from our fit to the RG results for $H=1.0 \times  10^{-4}$ MeV$^{5/2}$, $1.0 \times 10^{-3}$ MeV$^{5/2}$, and $1.0 \times 10^{-2}$ MeV$^{5/2}$ for $x$-values $x<1$ (black solid line, small-$x$ fit) and for all $x$ values (up to approximately $x = 5.3 \times 10^4$) (blue dashed line, asymptotic fit). There is little difference between these fits for $x<0$, but for asymptotically  large $x$-values the difference between these curves due to the difference in the exponents becomes quite large. The fit to the O(4) lattice data from \cite{Engels:1999wf} (red dashed line, lattice data fit) differs from our results for large $x$-values due to the different values for the critical exponents. The result from the $\epsilon$-expansion \cite{Brezin:1972fb} (green dot-dash line) is valid only for $x<0$ and displays a significantly smaller curvature than both our RG result and the O(4) lattice simulation result.}
\label{fig:griffithssmallxcomparison}
\end{figure}

We start our analysis with the results for $H=1.0 \times 10^{-4}$ MeV$^{5/2}$, $H=1.0 \times 10^{-3}$ MeV$^{5/2}$, and $H=1.0 \times 10^{-2}$ MeV$^{5/2}$, since corrections to the leading-order scaling behavior are small in this range. This allows us to obtain a simple but quite decent parameterization of the scaling function $y=y(x)$. 
For the fitting procedure, we estimate errors due to corrections to scaling by comparing the results after rescaling for the different values of $H$.

For small values of $x$ close to the coexistence point $x=-1$, it is natural to expand the equation of state in terms of $(1+x)$. Inspired by eq. \eqref{eq:phenomansatz}, we fit the phenomenological expression
\be
y(x) &=& c(1+x)^\tau
\ee
to our results, which provides a surprisingly good fit with only two parameters. Since this expression behaves as $y(x) \simeq c x^\tau$ for large values of $x \gg 1$, it reproduces the leading term of the expansion \eqref{eq:Griffithsasymptotic}, provided $\tau = \gamma$. Therefore this form also suggests itself for a global fit of the results.

We first restrict ourselves to the range $-1<x<1$.
The fit to the data for $H=1.0 \times 10^{-4}$ MeV$^{5/2}$, $H=1.0 \times 10^{-3}$ MeV$^{5/2}$, and $H=1.0 \times 10^{-2}$ MeV$^{5/2}$ is shown in Fig.~\ref{fig:griffithssmallx}. The field $H$ varies over two orders of magnitude, yet the rescaled results fall neatly onto a single scaling curve. For these extremely small values of $H$, corrections to the leading scaling behavior are very small. In the plot, for $x<1$ the differences between the scaled results are smaller than the width of the symbols. 

For a global fit, we use all points in the range $-1<x<5.3 \times 10^4$. The result is dominated by the behavior at asymptotically large values of $x$. Using the scaling laws, we find from our values for the critical exponents $\beta$, $\delta$ and $\nu$ for the critical exponent $\gamma$
\be
\gamma = (2 - \eta) \nu = 1.611(12), \quad \gamma=\beta(\delta-1) = 1.601(24), 
\ee
which is still compatible with the result expected from the asymptotic behavior when we identify
\be
\gamma = \tau = 1.594(1).
\ee
Below we also perform a fit to the exact expression of Griffiths' expansion, including the next correction term. This leads to the value $\gamma=1.5997(1)$, in perfect agreement with the result from the scaling laws.

The values for the fit parameters for both the range $-1<x<1$ and the global range $-1<x<5.3 \times 10^4$ are given in Tab.~\ref{tab:pheneos}. The quality of the fit for small $x$ is better, and it appears that the correction terms to the leading behavior in Griffiths' expansion for large $x$ become important.
Incidentally, the fits to the scaling function for large and small $x$-values differ only in the value of the exponent $\tau$, and have the same value for the coefficient $c$. 

In Fig.~\ref{fig:griffithssmallxcomparison}, our results for the scaling function for both small and large $x$ are compared to the fit to lattice Monte-Carlo results from \cite{Engels:1999wf} and to the result from $\epsilon$-expansion \cite{Brezin:1972fb} over the interval $-1<x<1$. For $x<0$, our fits and the results from the lattice simulations are very close together and have a significantly larger curvature than the $\epsilon$-expansion result. For $x>0$, all curves diverge quickly: The asymptotic behavior is determined by the value of the critical exponent $\gamma$, and the values in our calculation and the lattice simulation differ. The result  \cite{Brezin:1972fb} from the $\epsilon$-expansion satisfies the asymptotic behavior only order by order in $\epsilon$, but not explicitely, and is not a good description of the scaling function for $x>0$. 

We adapt the interpolation idea of eq.~\eqref{eq:Mendesinterpolation} from \cite{Engels:1999wf} to parameterize the scaling function $y(x)$ over the full range of $x$-values by combining the two-parameter fit for $x<1$ and the two-parameter fit for large $x$-values which captures the asymptotic behavior determined by the exponent $\gamma$. 
We find that the result is not very sensitive to the exact point $x_0$ at which we switch from one functional from to the other, as long as $0 \le x_0 \lesssim 10$, but smaller values give a slightly better interpolation. A suitable parameterization is
\be
y(x) &=&  \frac{(1+x_0)^2}{(1+x_0)^2+(1+x)^2}c_s (1+x)^{\tau_s} + \frac{(1+x)^2}{(1+x_0)^2+(1+x)^2}c_l (1+x)^{\tau_l}\quad \mathrm{with}\nonumber \\
&& c_s =0.9928, \;\; \tau_s = 1.6712, \;\; c_l=0.9928, \;\; \tau_l=1.5941, \quad x_0=0.
\label{eq:interpolation}
\ee
We will use this expression together with the relations \eqref{eq:translation} to obtain the scaling function in the form $f(z)$ for later comparison with additional results.

\subsubsection{Scaling behavior for small $x$}

\begin{table}
\begin{tabular}{l l l l l l}
\hline \hline
\quad $l_0$ &  \quad $l_1$ & \quad \quad $l_2$ & \quad $\chi^2$  & $\#$ d.o.f. & $\chi^2/ \#$d.o.f. \\
\hline
$0.980(5)$ & $0.506(16)$ & $\phantom{-}-$ & $2.7 \times 10^6$ & $282$ & $9586$ \\
$1.0058(6)$ & $0.642(3)$ & $\phantom{-}0.135(2)$ & $39963$ & $281$ & $142.3$\\
\hline
$1.11142$ & $0.273902$ & $-0.015625$ & \multicolumn{3}{c}{D. J. Wallace and R. K. P. Zia \cite{Wallace:1975vi}} \\
\hline\hline
\end{tabular}
\caption{\label{tab:Wallacefit}Fit of the form $y(x) = (1+x) \left[l_0 + l_1 \log(1 +x) +l_2 \log^2(1+x)  \right]$. The ansatz proves to be not suitable to fit our results over the range $-1<x<1$.}
\end{table}
\begin{table}
\begin{tabular}{l l l l l l}
\hline
\hline
\quad $a_3$  &  \quad $b_1$ & \quad $a_1$ & \quad $\chi^2$ & \# d.o.f. & $\chi^2$/(\# d.o.f.)\\
\hline
$0.9661(13)$ & $\phantom{-}-$ & $-$ & $2.3 \times 10^6$ & $283$ & $8153$ \\
$1.2619(52)$ & $-0.2601(45)$ & $-$   & $5.0 \times 10^4$ & $282$ & $178.1$\\
$1.3457(8)$   & $-0.3977(12)$ & $0.0561(5)$ & $277.6$ & $281$ & $0.9879$\\ 
\hline
\hline
\end{tabular}
\caption{Coefficients for the fit $y(x) =  a_3 (1+x)^{3/2} + b_1 (1+x) + a_1(1+x)^{1/2}$ to the $284$ points in the region $-1 < x < 1$. The fit with 3 parameters provides a reasonable description of our results for $H = 1.0 \times 10^{-4}$ MeV$^{5/2}$.}
\label{tab:4xsmall1}
\end{table}
\begin{table}
\begin{tabular}{l l l l l  l l }
\hline
\hline
\quad $c$  & \quad  $\tau$ & \quad $d_1$ & \quad $d_2$ & \quad $\chi^2$ & \# d.o.f. & $\chi^2$/(\# d.o.f.)\\
\hline
$1.0031(5)$ & $1.6472(20)$ & $\phantom{-}-$ & $-$ & $3.0 \times 10^4$ & $282$ & $106.3$ \\
$1.2242(27)$ & $1.7542(14)$ & $-0.179(2) $ & $-$&  $431.4$ & $281$ & $1.536$\\
$1.390(9)$ & $1.797(2)$ & $-0.332(7)$ & $0.055(3)$  & $77.18$ & $280$& $0.2756$\\
\hline
\hline
\end{tabular}
\caption{Coefficients for the fit $y(x) =  c (1+x)^\tau (1+d_1 (1+x)^{1/2}+d_2 (1+x) )$ to the $284$ points in the region $-1 < x < 1$. Both the 3- and the 4-parameter fits are a reasonable description of the results for $H = 1.0 \times 10^{-4}$ MeV$^{5/2}$.}
\label{tab:-4xsmall2}
\end{table}

For a determination of the scaling function for small $x$, we will use only the results  for $H = 1.0 \times 10^{-4}$ MeV$^{5/2}$ in the range  $-1 < x <1$. The corrections to the scaling behavior are small only if the external symmetry-breaking field is small, and this value is the smallest one for which we calculated results over a reasonably large range of values for the scaling variable $z$.

Here we estimate the scaling corrections by comparing to the results with $H=2.0 \times 10^{-4}$ MeV$^{5/2}$. 
Due to the normalization $y(0) =1$, the error is minimal at $x=0$. It is largest close to $x=-1$. The relative error due to scaling corrections is less than $0.04$ in the limit $x \to -1$, and for $x>-0.75$ it drops to less than $0.01$. In the region $0<x<1$, the relative error is less than $0.0005$.

In the absence of analytical results, different ans\"atze for the scaling function are possible. The results \cite{Brezin:1972fb, Wallace:1975vi, Pelissetto:1999cq} from the $\epsilon$-expansion suggest from the expansion around $x=-1$ to attempt a fit of the form
\be
y(x) &=&  (1+x) \left[l_0 + l_1 \log(1 +x) +l_2 \log^2(1+x)  + \ldots \right],
\label{eq:logfit}
\ee
which is only expected to hold for $(1+x)\ll 1$, or from the proposed expansion \eqref{eq:generalepsiloninverted} 
\be
y(x) &=& a_3 (1+x)^{3/2} + b_1 (1+x) + a_1(1+x)^{1/2} + \dots.
\label{eq:afit}
\ee
Assuming that the $\epsilon$-expansion can be resummed, a third possibility is our phenomenological expression
\be
y(x) =  c (1+x)^\tau (1+d_1 (1+x)^{1/2}+d_2 (1+x) +\ldots).
\label{eq:expcorfit}
\ee
We will take these possible expansions in turn. The results for the logarithmic expression \eqref{eq:logfit}
are given in Tab.~\ref{tab:Wallacefit}. We find that this expression is not suitable to fit our results over the range we consider here. This is not unexpected, since the expansion is only valid close to $x=-1$.
Both the expression \eqref{eq:afit} and \eqref{eq:expcorfit} describe the results about equally well, the first one perhaps a little better. The coefficients are given in Tables~\ref{tab:4xsmall1} and \ref{tab:-4xsmall2}. For practical purposes, both descriptions are equivalent.

\begin{table}
\begin{tabular}{l l l l}
\hline\hline
\quad $\tilde{c}_2$ & $\tilde{c}_1+\tilde{d}_3$ & \quad  $\tilde{d}_2$ & \\
\hline
$0.681(4)$ & $0.391(8)$ &  $-0.074(4)$ & our result \\
\hline
$0.674(8)$ & $0.345(12)$ &  $-0.023(5)$ & J. Engels and T. Mendes \cite{Engels:1999wf, Engels:2001bq}\\
$0.746(3)$ & $0.19(1)$ & $\phantom{-}0.061(8)$ & A. Cucchieri and T.Mendes \cite{Cucchieri:2004xg}\\
$0.530$ & $0.528$ &  $\phantom{-}-$ &D.  J. Wallace and R. K. P. Zia \cite{Wallace:1975vi}\\ 
\hline
\end{tabular}
\caption{Fit to the equation of state in inverted form, $1+x =  \tilde{c}_2 y^{1/2} + (\tilde{c}_1 + \tilde{d}_3) y + \tilde{d}_2 y^{3/2} $, from our results for $H=1.0 \times 10^{-4}$ MeV$^{5/2}$ in the region $0<y<1.5$, compared results from the $\epsilon$-expansion and from O(4) spin model lattice calculations.}
\label{tab:inverted}
\end{table} 
In \cite{Engels:1999wf} and \cite{Wallace:1975vi}, the equation of state is parameterized in inverted form as $x=x(y)$, and  the  terms expected from the $\epsilon$-expansion  in $d=3$ ($\epsilon =1$) are
\be
1+x &=& \tilde{c}_2 y^{1/2} + (\tilde{c}_1 + \tilde{d}_3) y + \tilde{d}_2 y^{3/2} + \ldots.
\ee 
To make a direct comparison to our results easier, we also used this form of the equation for a fit to the results for $H= 1.0 \times 10^{-4}$ MeV$^{5/2}$ in the range $0<y<1.5$.

The coefficients are given in Table~\ref{tab:inverted}. The coefficient $\tilde{d}_2$ has not been calculated in the $\epsilon$-expansion. For the comparison to the lattice results, please keep in mind that our results have an additional systematic error not included in the fit uncertainty given in the table.
While strictly speaking the results from the spin model lattice calculation \cite{Engels:1999wf, Engels:2001bq} and from our results do not agree within the errors, they are remarkably consistent with each other, in particular when compared with the perturbative results from the $\epsilon$-expansion:
The values for the leading coefficient, $\tilde{c}_2$, agree within $1\% $. The values for the sub-leading coefficent, $\tilde{c}_1 + \tilde{d}_3$, differ only by about $10 \%$. In contrast, both our results and the lattice results differ from the perturbatively calculated coefficients from the $\epsilon$-expansion by more than $20-30 \%$.

\subsubsection{Asymptotic scaling behavior} 

\begin{table}
\begin{tabular}{ l l l l  l l }
\hline\hline
$\gamma$ & $c_1$ & \quad $c_2$ & \quad $\chi^2$ & \# d.o.f. & $\chi^2$/\# d.o.f. \\
\hline
$1.5948(4)$ & $0.9855(25)$ & $-$ &  $2144$ & $204$ & $10.51$ \\
$1.59973(4)$ & $0.94294(28)$ & $1.196(8)$ &  $4.186$ & $203$&  $0.02062$\\
\hline\hline
\end{tabular}
\caption{\label{tab:coeffasyeos}Fit to Griffiths' expansion  $y(x) =  x^\gamma ( c_1 + c_2 x^{-2\beta} + c_3 x^{-4 \beta} +\ldots )$, for $x>0$, $M>M_0>0$. We fit only to the results for $H =1.0 \times 10^{-4}$ MeV$^{5/2}$ and only to points with $x>100$.}  
\end{table}
Griffiths' expansion \eqref{eq:Griffithsasymptotic} can be used to describe the results for large values of $x$, and to determine the critical exponent $\gamma$ to high accuracy. A comparison to the value obtained with the scaling laws from the other exponents serves as a consistency check for our results.
We use for the fit
\be
y(x) &=& c_1 x^\gamma + c_2 x^{\gamma - 2\beta} +c_3 x^{\gamma - 4 \beta} + \ldots 
 = x^\gamma \left( c_1 + c_2 x^{-2\beta} + c_3 x^{-4 \beta} + \ldots \right)
\ee
with $\gamma$ as a free parameter. In order to retain the hierarchy of the corrections in the expansion, we use the previously determined value $\beta=0.4030(30)$ and keep it fixed.
Since $\gamma - 4 \beta \approx \gamma - 2 \nu \approx 0$ according to the $d=3$ scaling laws, only the exponents of the first two terms in Griffiths' expansion are positive and contribute significantly for large $x$. The additional terms with $n>3$ are small corrections for large $x$, but diverge for $x \to 0$ where the expansion is no longer valid.

We fit again only to the results for $H= 1.0 \times 10^{-4}$ MeV$^{5/2}$ and estimate scaling corrections from a comparison to the results for $H= 2.0 \times 10^{-4}$ MeV$^{5/2}$. Only points with $x>100$ are included, so we can be certain to be in a region where the expansion is valid.
The results for the coefficients and the exponent $\gamma$ are given in Tab.~\ref{tab:coeffasyeos}. 
The inclusion of more than the first two terms does not lead to a meaningful improvement of the fit quality. As already noted above, the results for the critical exponent $\gamma=1.5997(1)$ is in perfect agreement with the determination from the scaling laws.

\subsection{Scaling behavior for large fields}

\begin{figure}
\includegraphics[scale=0.3]{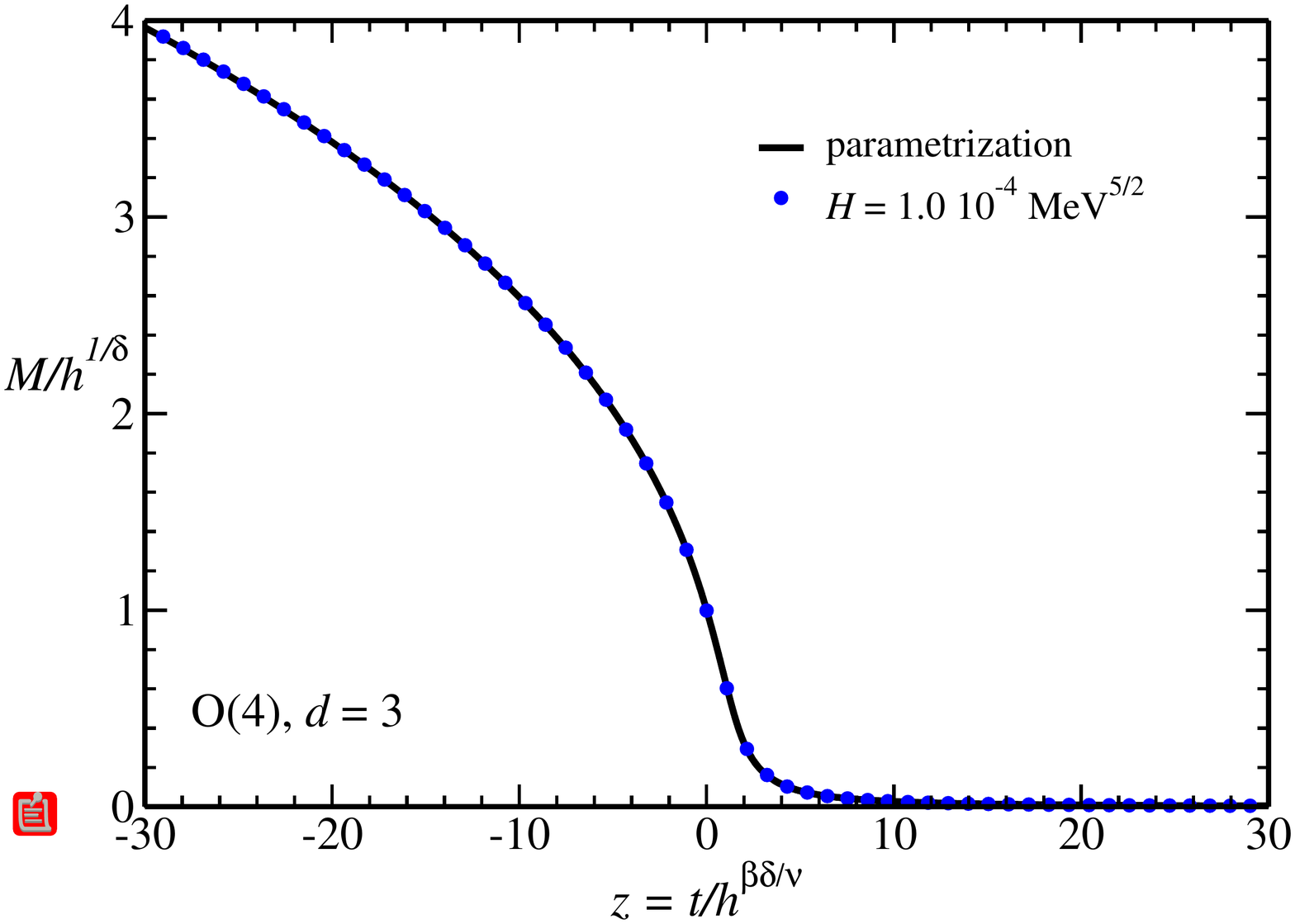}
\includegraphics[scale=0.3]{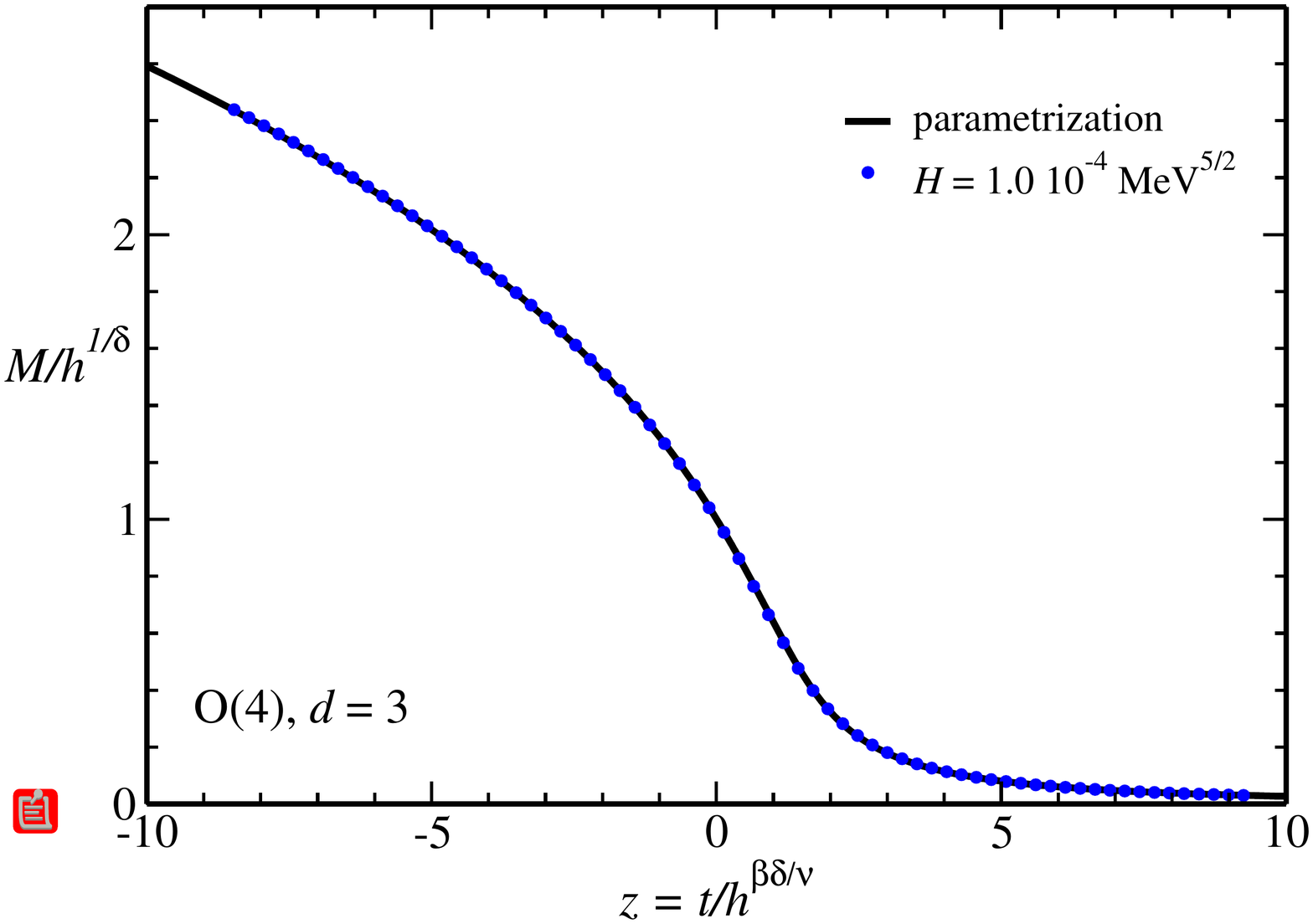}
\caption{The parameterization of the scaling function $y(x)$ in terms of the scaling variables $x$ and $y$ provides an implicit parameterization of the scaling function $f(z)$. Shown is a comparison of $f(z)$ from the interpolated fit eq.(\ref{eq:interpolation}) with the rescaled results for $H=1.0 \times 10^{-4}$ MeV$^{5/2}$ (blue circles, for clarity not all points are shown). The fit is perfect on the scale of the plot. The asymptotic behavior of $y(x)$ for $x \to \infty$ determines the behavior of the fit for $z \gtrsim 0.5$. Using the parameterization in terms of only the small-$x$ behavior underestimates $f(z)$ in this region. In this region, the results from the $\epsilon$-expansion show the largest deviation.}
\label{fig:interpolation}
\end{figure}
\begin{figure}
\includegraphics[scale=0.5]{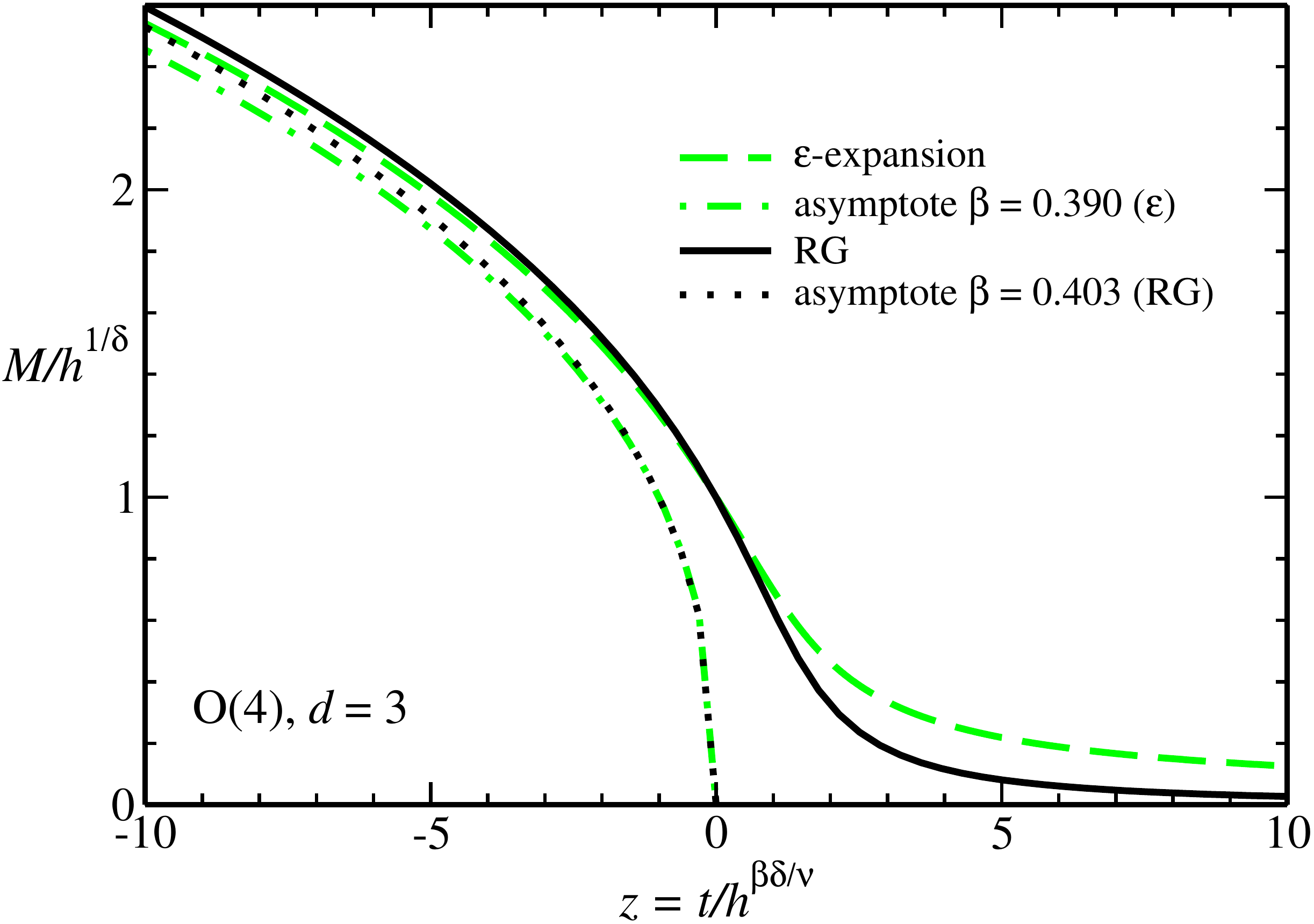}
\caption{Plot of the scaling function $f(z)$ as a function of $z$. Shown are results from our calculation for  $H=1.0 \times 10^{-4}$ MeV$^{5/2}$ (black solid line) and from the $\epsilon$-expansion calculation  to ${\mathcal O}(\varepsilon^2)$ of Br\'ezin {\it et al.} \cite{Brezin:1972fb} (green dashed line). For comparison, we also plot the expected asymptotic behavior $f(z) \to (-z)^\beta$ for $\beta = 0.390$ ($\epsilon$-expansion) (dot-dashed green line) and for $\beta = 0.403$ (our result) (dotted black line). The behavior of the scaling functions for large negative values of $z$ is again determined by the different values of the critical exponent $\beta$. The $\epsilon$-expansion result is expected to provide a good description only for $t<0$. We find that indeed the deviation from our scaling function is large for $z \gtrsim 0.5$.}
\label{fig:fzscalingepsilon}
\end{figure}
\begin{figure}
\includegraphics[scale=0.5]{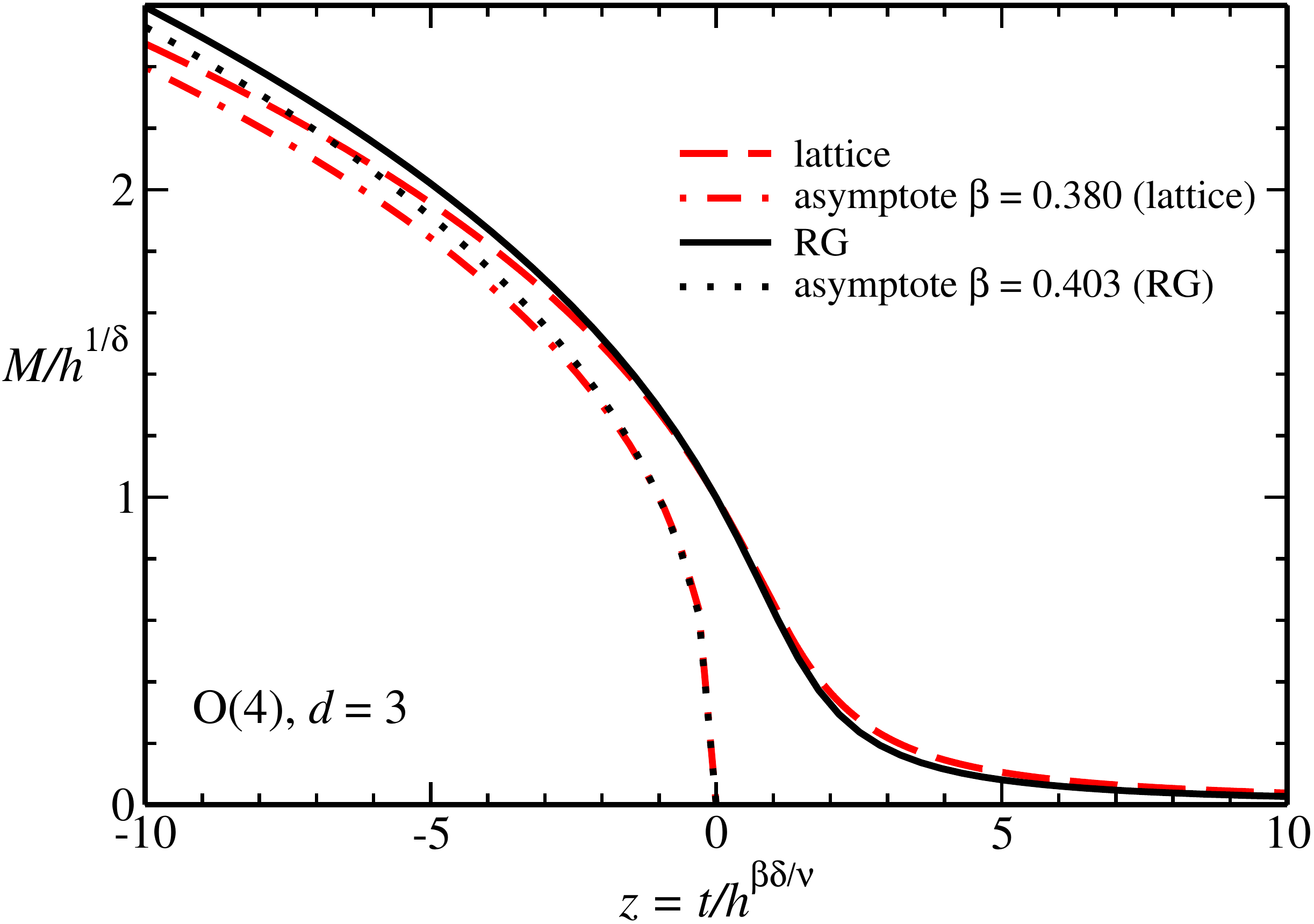}
\caption{Plot of the scaling function $f(z)$ as a function of $z$. Shown are results from our calculation for  $H=1.0 \times 10^{-4}$ MeV$^{5/2}$ (black solid line) and from the calculation of Mendes and Engels \cite{Engels:1999wf} (using the parameterization as given in \cite{Engels:2001bq}) (red dashed line). For comparison, we also plot the expected asymptotic behavior $f(z) \to (-z)^\beta$ for $\beta = 0.380$ (Mendes and Engels) (dot-dashed red line) and for $\beta = 0.403$ (our result) (dotted black line). As expected, the behavior of the scaling functions for large negative values of $z$ is determined by the different values of the critical exponent $\beta$.}
\label{fig:fzscalingmendes}
\end{figure}
After having determined the scaling function from results with small values of the symmetry-breaking field $H$, we now compare these results to the scaling behavior at much larger values of $H$. For this purpose, we also switch to the more intuitive description by means of the scaling function $f(z)$. 

With the help of Eqns.~\eqref{eq:translation} and the critical exponents, the parametrization Eq.~\eqref{eq:interpolation} implicitly also  provides a parameterization of the scaling function in the form $f(z)$. 
In Fig.~\ref{fig:interpolation}, the interpolation function $f(z)$ obtained from the parameterization Eq.~(\ref{eq:interpolation}) is compared to the rescaled results for the order parameter $M/h^{1/\delta}$  for $H= 1.0 \times 10^{-4}$ MeV$^{5/2}$ as a function of $z$. On the scale of the plot, the agreement is perfect. 
The scaling form $f(z)$ is less sensitive to small changes in $M$ than the Widom-Griffiths scaling form $y(x)$ and thus the agreement appears to be even better than for the fit in the form $y(x)$. 

Using the appropriate values for the critical exponents, we can also translate the scaling functions from the $\epsilon$-expansion \cite{Brezin:1972fb} and from the lattice simulations \cite{Engels:1999wf} into this form.
The comparison to our results for $H= 1.0 \times 10^{-4}$ MeV$^{5/2}$ is shown in Fig.~\ref{fig:fzscalingepsilon} for the $\epsilon$-expansion and in Fig.~\ref{fig:fzscalingmendes} for the lattice results. Since $f(z) \to (-z)^\beta$ for $z \to -\infty$, the differences in the values of the critical exponents lead to different behavior of the scaling functions. To illustrate this, the asymptotic behavior for the different values of $\beta$ is also shown in the plots, and the scaling functions can be seen to approach the asymptotic functions. 

The result from the $\epsilon$-expansion shows the largest deviation from our results for $z\gtrsim 0.5$, which corresponds to the deviation observed in $y(x)$ at large $x$-values, where the expansion is not a good description of the scaling function.

The agreement between our results and those from the lattice simulations is much better. Most of the difference at large negative $z$-values can be attributed to the different values of the critical exponents. Overall we consider the agreement quite satisfactory.

The scaling behavior of the order parameter can be demonstrated very nicely in this scaling form with our results. Having first confirmed the agreement between our scaling function $f(z)$ and the rescaled results for $H=1.0 \times 10^{-4}$ MeV$^{5/2}$ in Fig.~\ref{fig:interpolation}, we now compare these results to results for a wide range of values for $H$.

\begin{figure}
\includegraphics[scale=0.3]{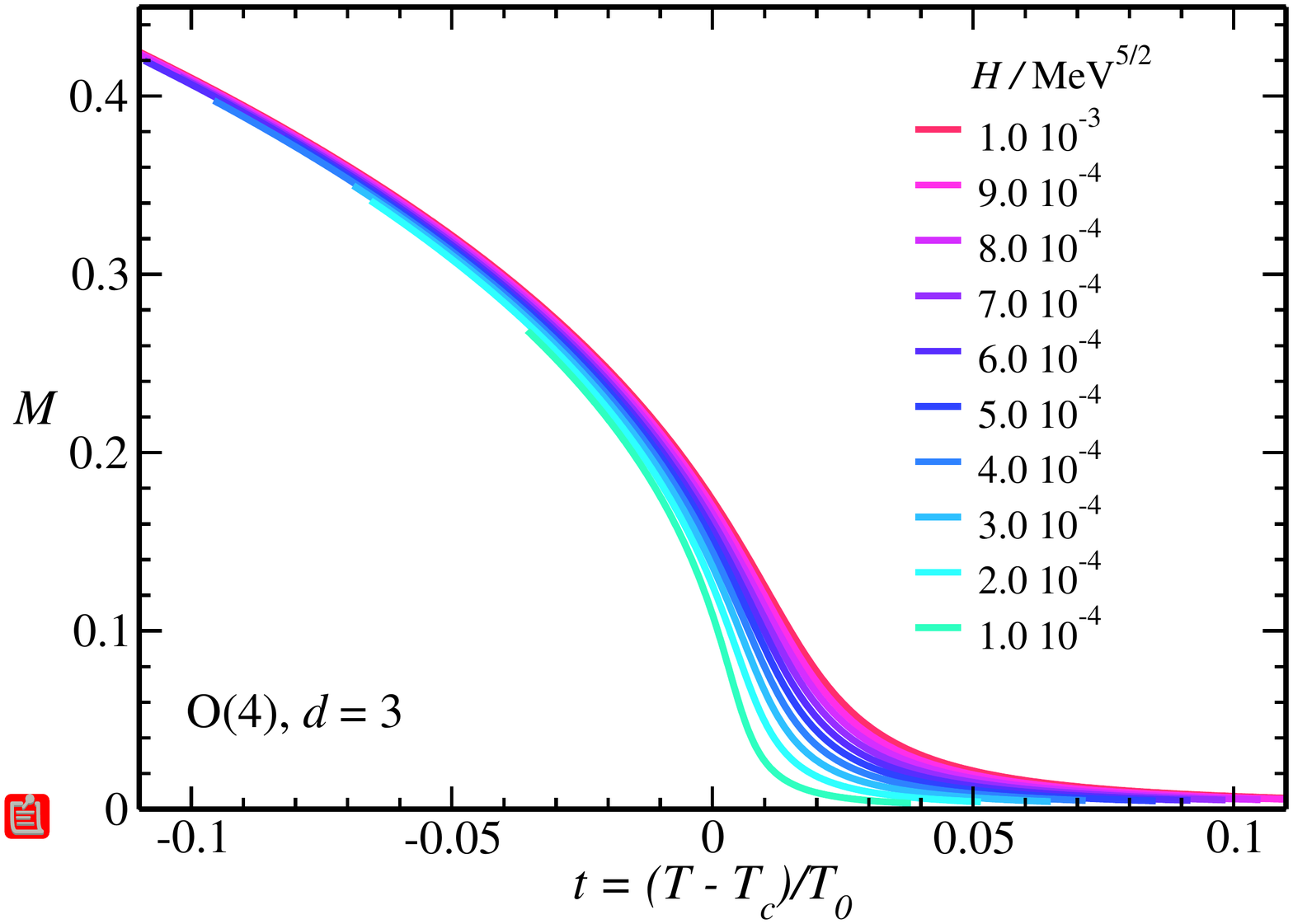}
\includegraphics[scale=0.3]{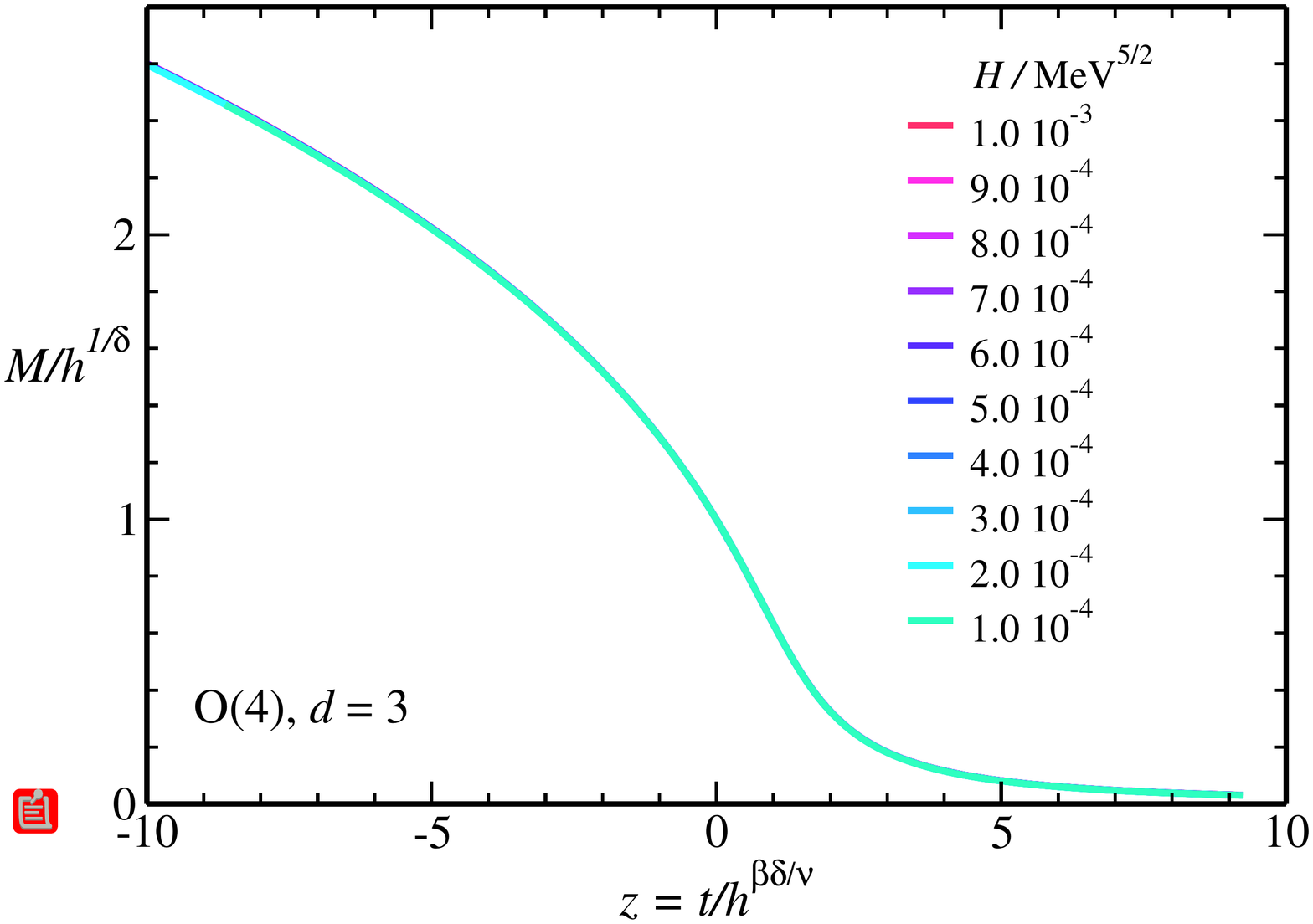}
\caption{Order parameter as a function of the reduced temperature $t$ for very small values of the external field $H$ from $1.0 \times 10^{-4}$ to $1.0 \times 10^{-3}$ MeV$^{5/2}$ (left panel), and for the rescaled order parameter $M/h^{1/\delta}$ as a function of $z=t/h^{1/(\beta\delta)}$ (right panel). The $t$-ranges for the different values of $H$ are chosen such that $z$ covers the range $-10\ldots 10$ after rescaling. On the scale of the plot in the left-hand panel, a deviation from the scaling behavior is not visible in this range of values.}
\label{fig:zscalingsmallH}
\end{figure}

In Fig.~\ref{fig:zscalingsmallH}, we consider small values of $H$ in the range from $H= 1.0 \times 10^{-4}$ to $H=1.0 \times 10^{-3}$ MeV$^{5/2}$. In the left-hand panel, the order parameter $M$ is plotted as a function of the reduced temperature $t$ for the different values of $H$. In the right-hand panel, the rescaled order parameter $M/h^{1/\delta}$ is plotted as a function of $z=t/h^{1/(\beta\delta)}$
for the same $H$ values. After rescaling, the curves for all values of $H$ fall onto a single line and are indistinguishable at the scale of the plot. This agrees with our observation in the Wisdom-Griffiths scaling analysis, where scaling corrections were also negligible in this $H$-range. 

\begin{figure}
\includegraphics[scale=0.3]{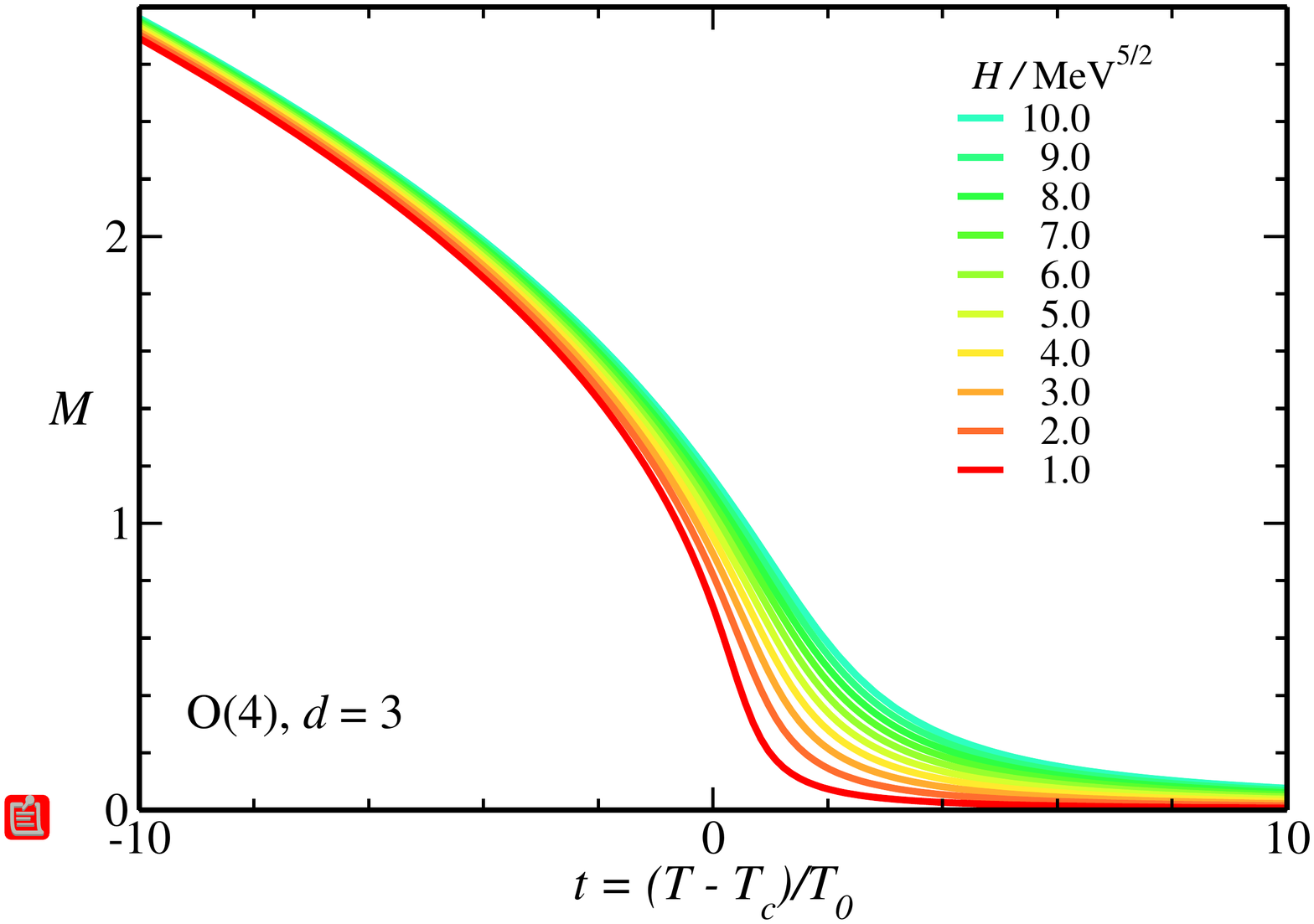}
\includegraphics[scale=0.3]{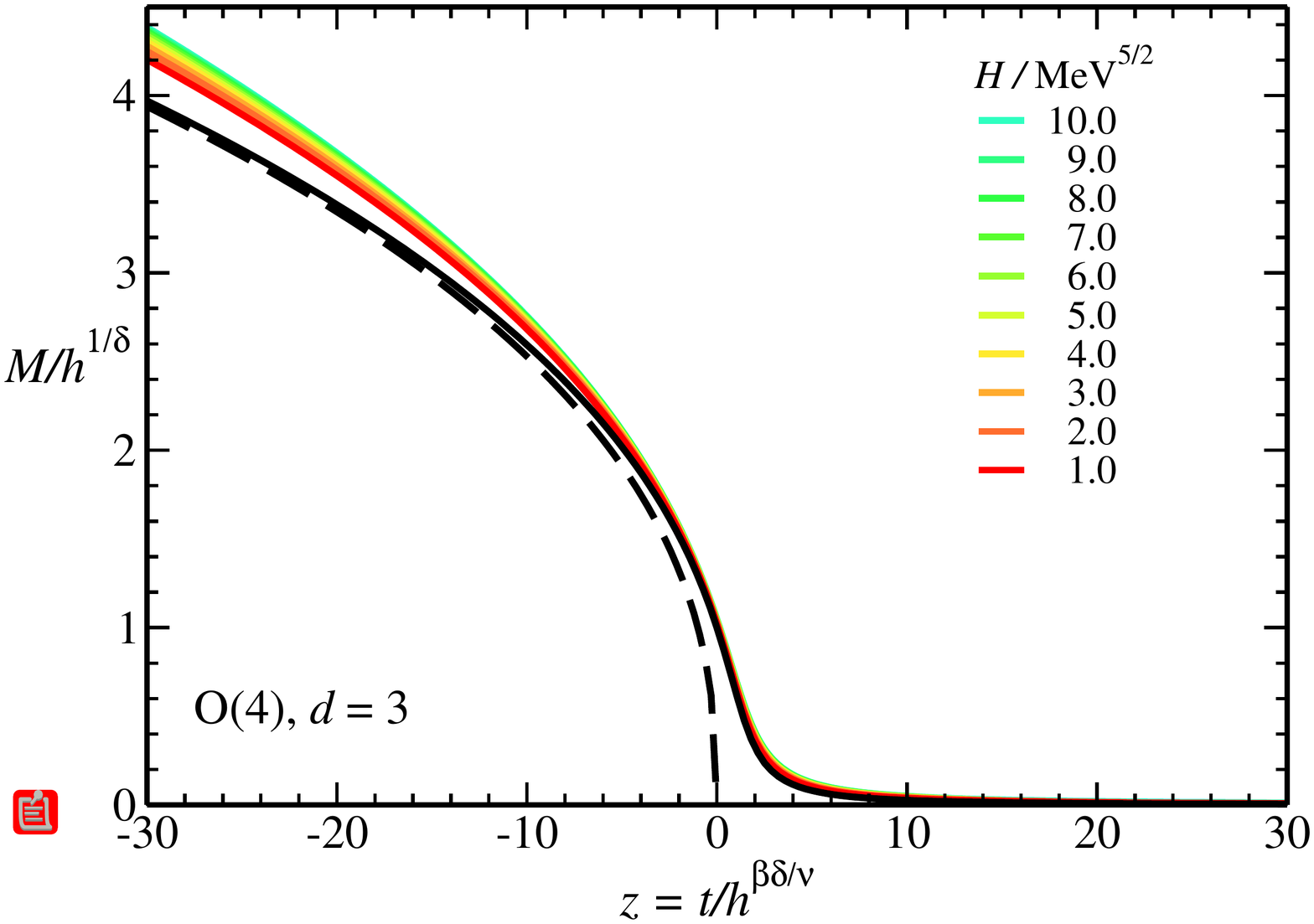}\\
\includegraphics[scale=0.3]{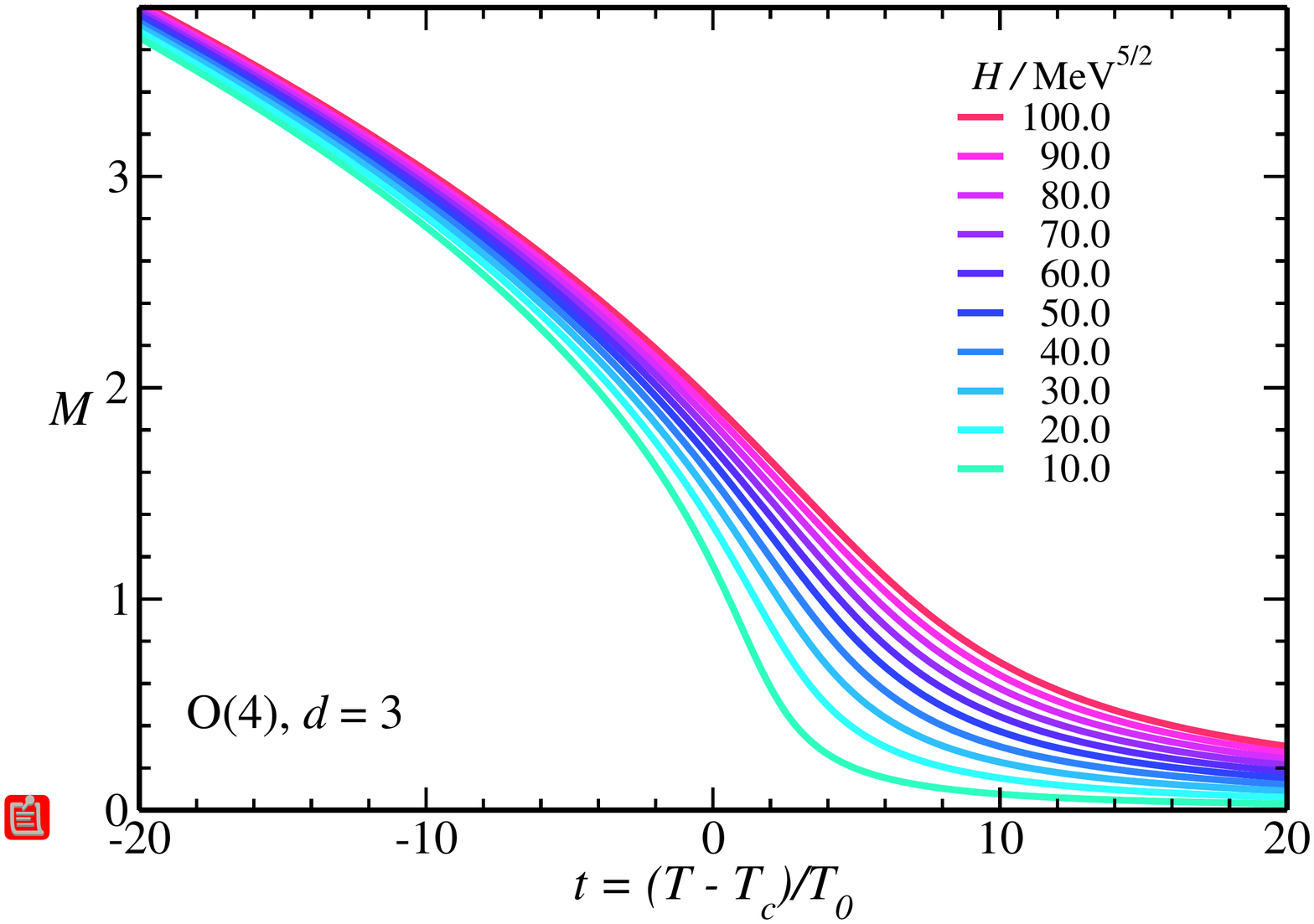}
\includegraphics[scale=0.3]{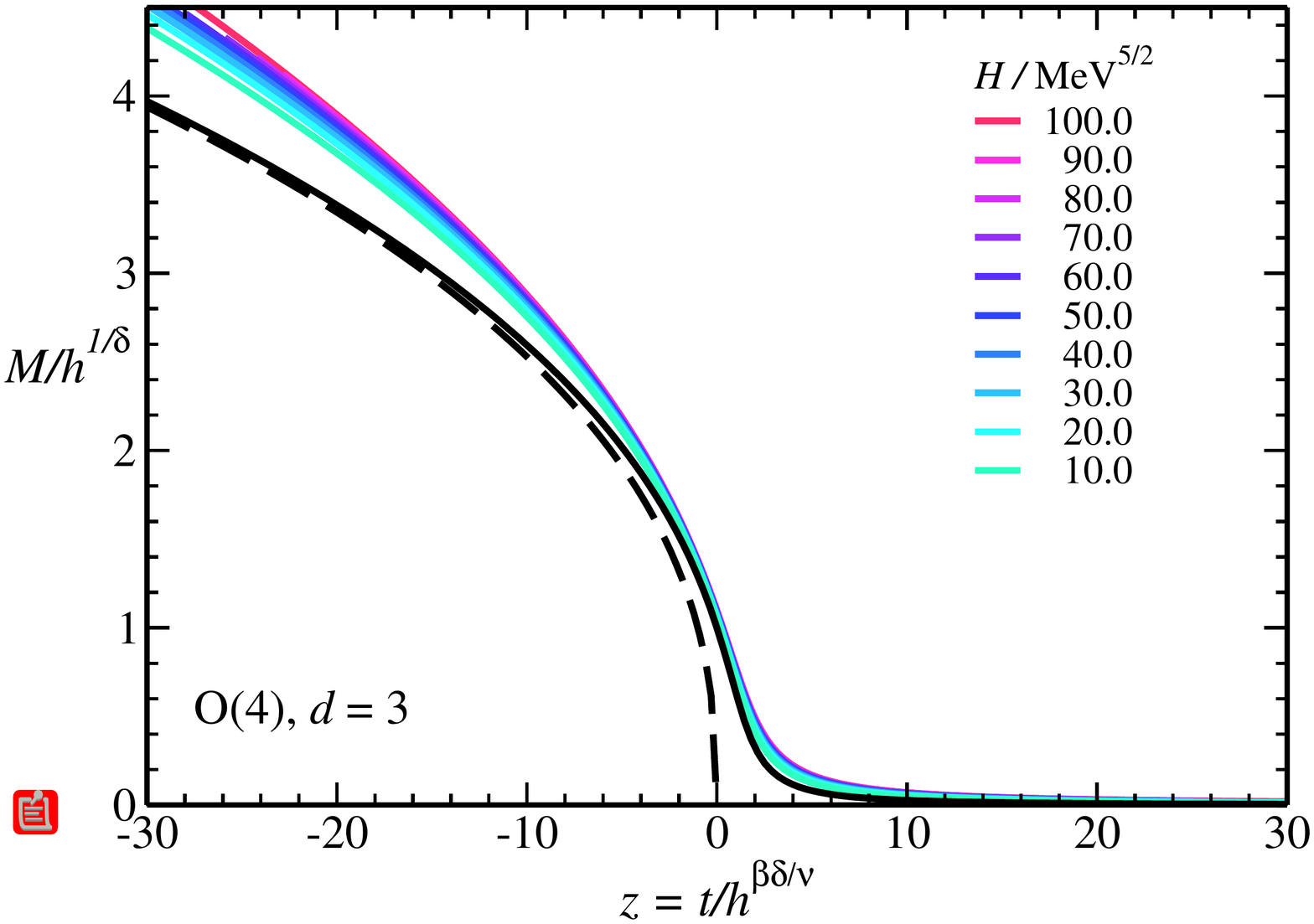}
\\
\includegraphics[scale=0.3]{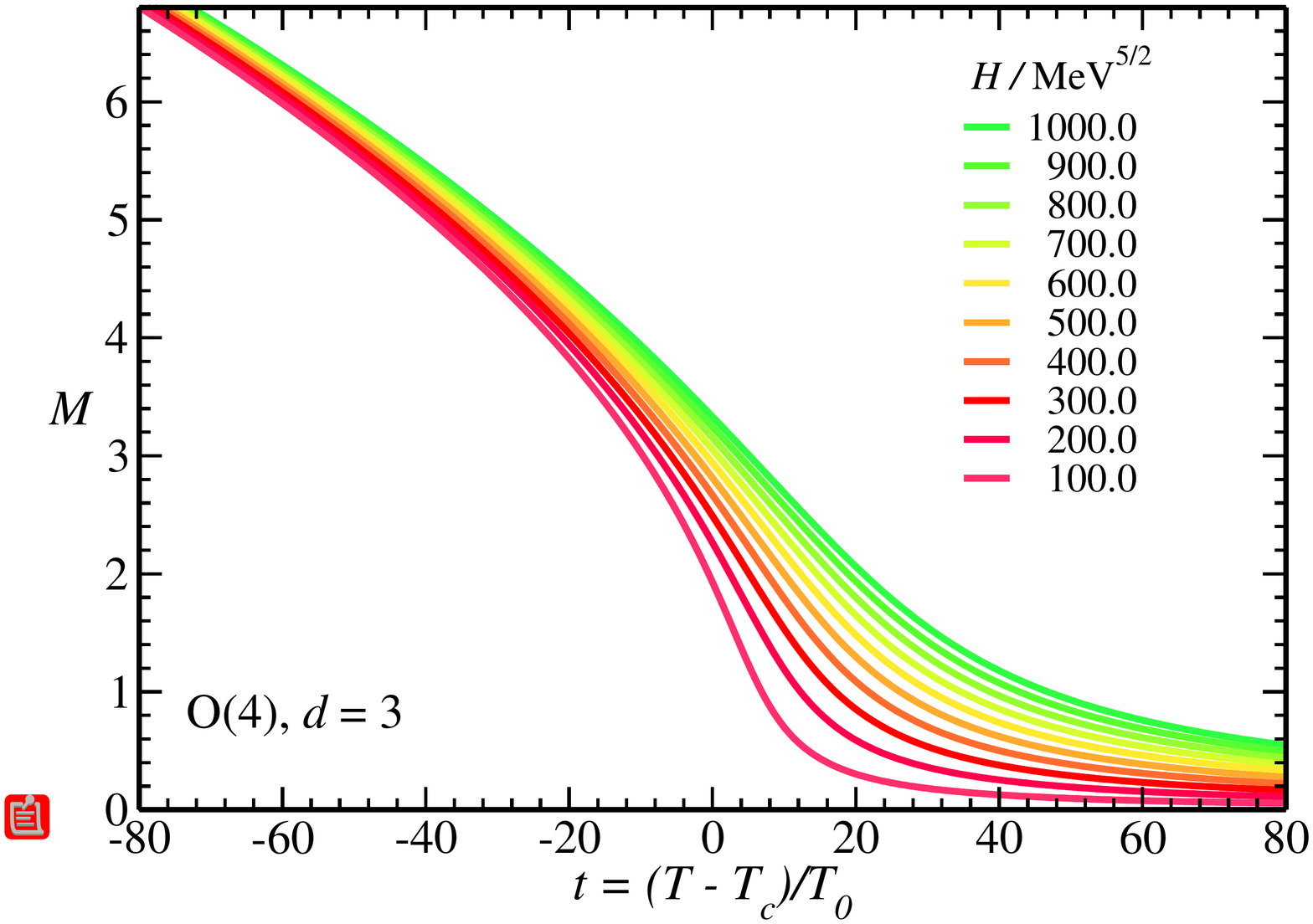}
\includegraphics[scale=0.3]{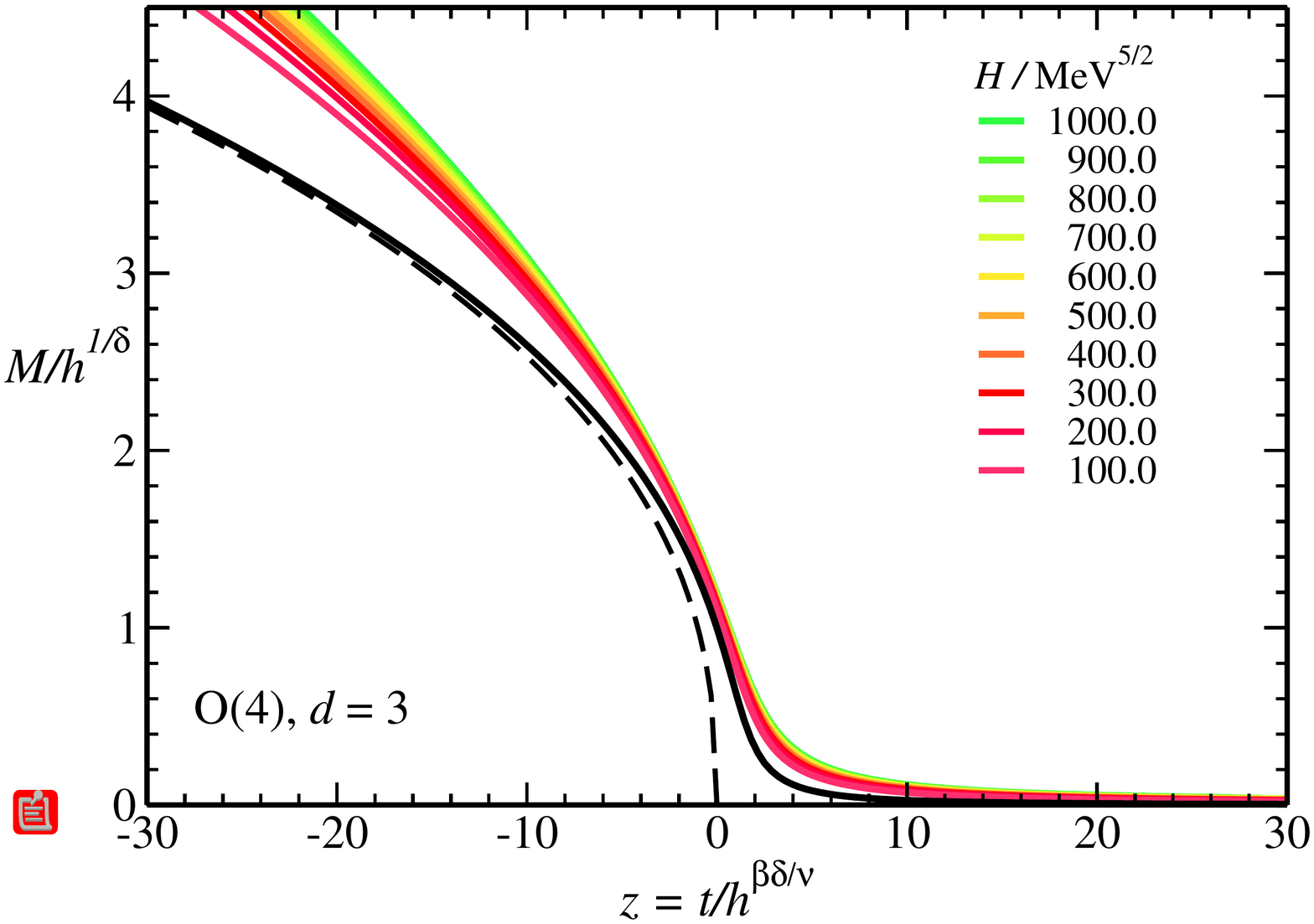}
\caption{Results for the order parameter as a function of the reduced temperature $t$ for different values of the field $H$ (left-hand column), and for the rescaled order parameter $M/h^{1/\delta}$ as a function of the scaling variable $z=t/h^{1/(\beta\delta)}$ (right-hand column). 
For comparison, the scaling function obtained from the result for $H=1.0 \times 10^{-4}$ MeV$^{5/2}$ (solid black line) and the asymptotic behavior of the scaling function $(-z)^\beta$ (dashed black line) are shown with the rescaled results.
Note that the axes on the figures for the rescaled order parameter in the right-hand column all have the same scales, while the scales of the axes on the plots with the unscaled results  differ widely for the different $H$-ranges.}
\label{fig:zscalinglargeH}
\end{figure}

However, for larger values of $H$, corrections to the scaling behavior soon become apparent. In Fig.~\ref{fig:zscalinglargeH}, we show results in the range from $H = 1.0 $ MeV$^{5/2}$ to $H=1.0 \times 10^3$ MeV$^{5/2}$. Again both the order parameter as a function of temperature and the rescaled order parameter as a function of $z$ are shown. For comparsion, the rescaled results for $H=1.0 \times 10^{-4} $ MeV$^{5/2}$ and the asymptote $(-z)^\beta$ are shown with the rescaled results.
While the curves for different values of $H$ still collapse after rescaling in the vicinity of the critical temperature, the deviations from the scaling function become quite large, and they are already plainly visible for fields of the order $H=10.0$ MeV$^{5/2}$. 

This result is actually not very surprising: For small values of the field $H$, only a small region in $t$ around the critical temperature $t=0$ is rescaled and becomes a sizable region in $z$. 
Therefore we can expect that all results are in the scaling region around $T_c$ and scaling deviations are small over a large region in $z$. 
For example, for $H=1.0 \times 10^{-4}$ MeV$^{5/2}$, the rescaled results approach the expected asymptotic behavior at least down to $z=-30$. In contrast, for large values of $H$, a much wider region in $t$ is mapped onto the same range of $z$-values, and corrections to the scaling behavior are bound to appear, since points far away from the critical point are included. In Fig.~\ref{fig:zscalinglargeH}, one can see how the $t$-region has to be increased with $H$ to keep the $z$-range constant after rescaling. 
Results from the boundary of the scaling region at large $-t$ then lead to deviations from the expected asymptotic scaling behavior at large $-z$.

\begin{figure}
\includegraphics[scale=0.5]{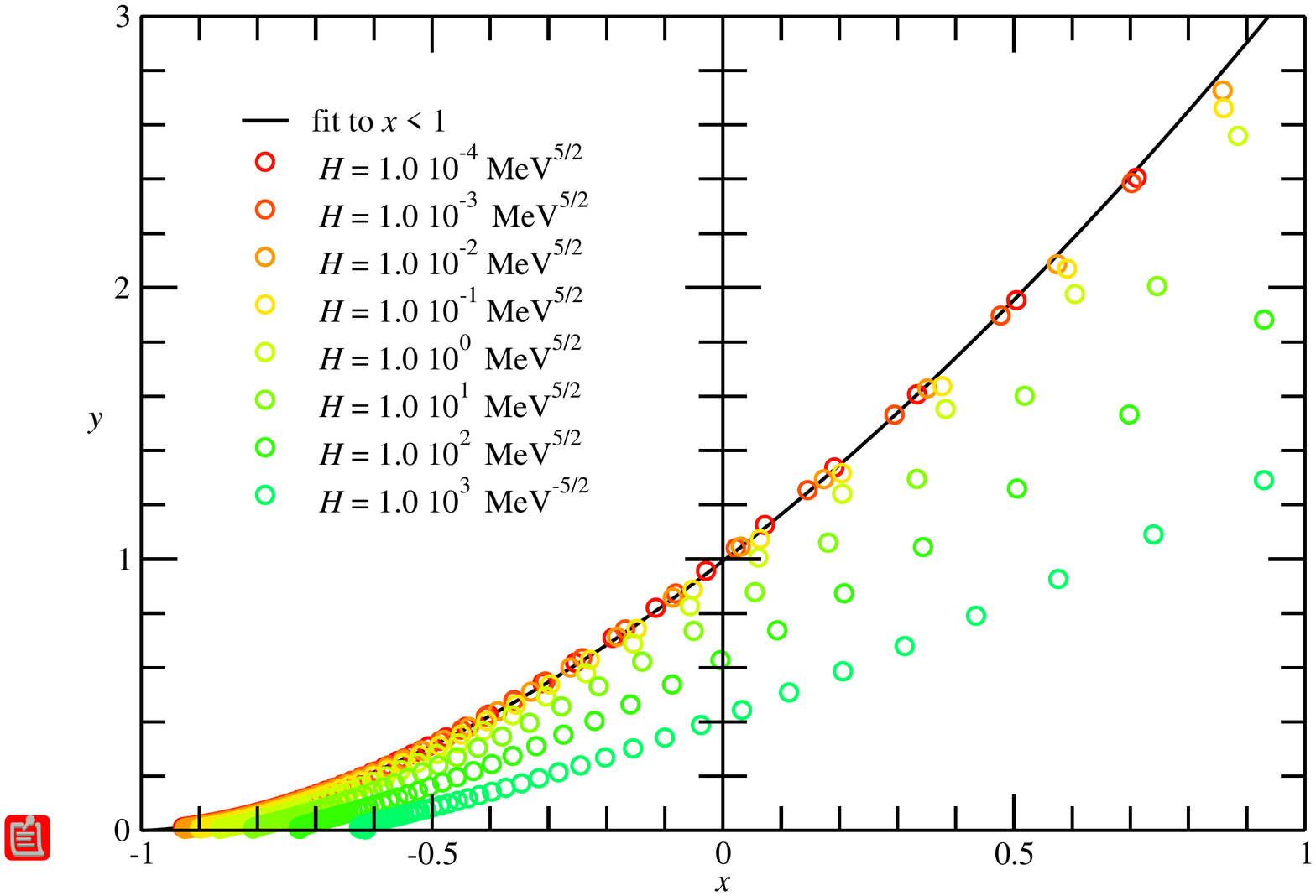}
\caption{Comparison of scaling results over a wide range of values (7 orders of magnitude) for the symmetry-breaking field $H$. On this scale, deviations from the scaling behavior become readily apparent for fields of the order of $H=1.0 \times 10^{-1}$ MeV$^{5/2}$. For comparison, the fit 
$y(x)= c (1+x)^\tau$ with $c=0.9228(24)$ and $\tau =1.6712(27)$ to the results for $H=1.0 \times 10^{-4}$, $1.0 \times 10^{-3}$, and $1.0 \times 10^{-2}$ MeV$^{5/2}$ for $x<1$ is also shown (solid black line). For clarity, not all points are shown.}
\label{fig:Griffithsdeviationsmallx}
\end{figure}
\begin{figure}
\includegraphics[scale=0.5]{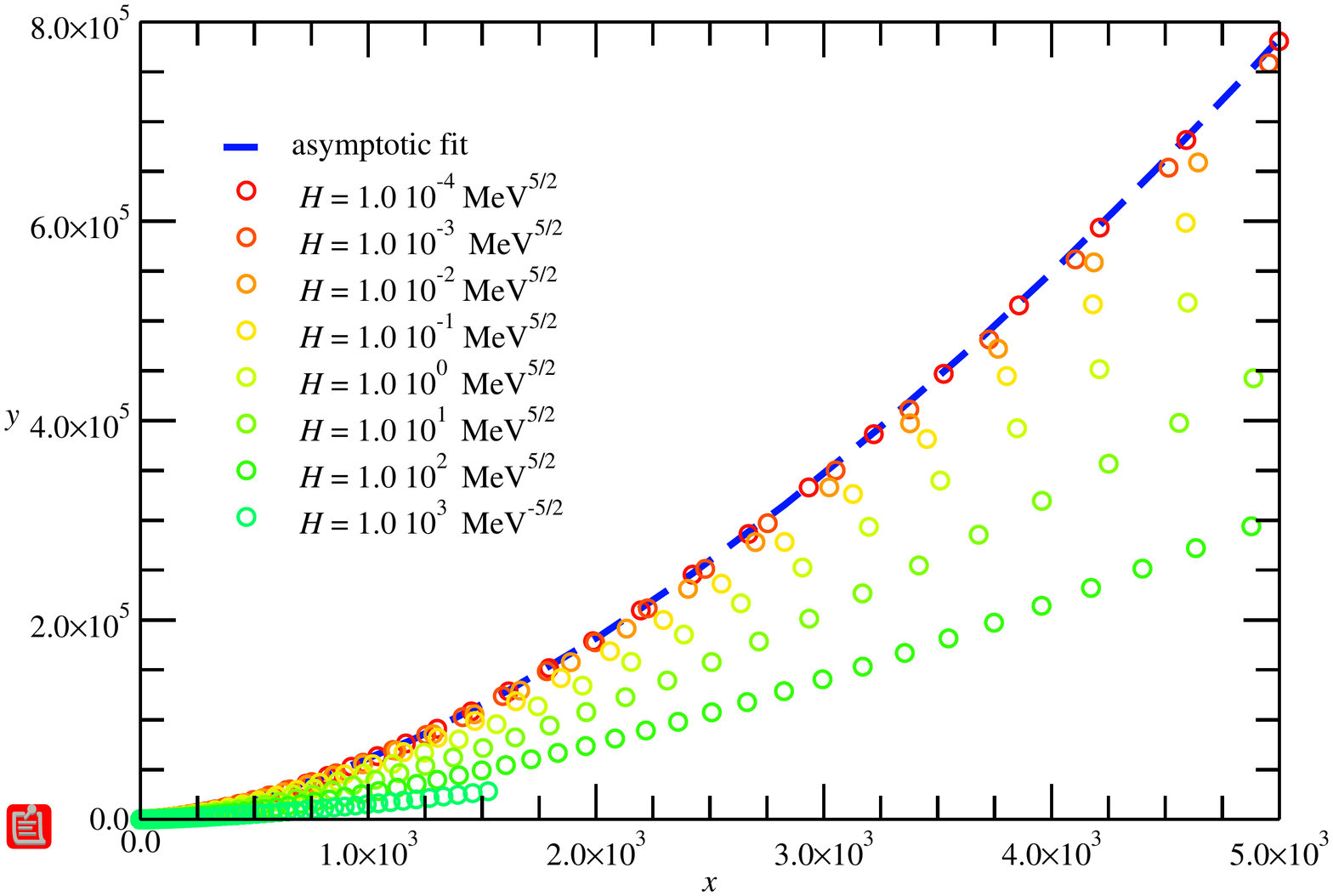}
\caption{Comparison of scaling results over a wide range of values (7 orders of magnitude) for the symmetry-breaking field $H$. Deviations from the scaling behavior are clearly visible for fields of the order of $H=1.0 \times 10^{-1}$ MeV$^{5/2}$. For comparison, the asymptotic fit $y(x)= c (1+x)^\tau$ with $c=0.9228(24)$ and $\tau =1.5941(8)$ to the results for $H=1.0 \times 10^{-4}$, $1.0 \times 10^{-3}$, and $1.0 \times 10^{-2}$ MeV$^{5/2}$ is also shown (dashed blue line). For clarity, not all points are shown.}
\label{fig:Griffithsdeviationlargex}
\end{figure}

The deviations from the leading-order scaling behavior are even more starkly visible from a plot of the results in the Widom-Griffiths scaling form as $y(x)$. This is shown for values of $H$ over 7 orders of magnitude, from $H=1.0 \times 10^{-4}$ MeV$^{5/2}$ to $H=1.0 \times 10^{3}$ MeV$^{5/2}$, in Fig.~\ref{fig:Griffithsdeviationsmallx} for the range $-1<x<1$, and in Fig.~\ref{fig:Griffithsdeviationlargex} for the range $-1<x<5.0 \times 10^3$. On the scale of these plots, corrections become significant for fields $H> 1.0 \times 10^{-2}$ MeV$^{5/2}$.

In conclusion, we find that we can extract a parameterization of the scaling function in Widom-Griffiths form $y=y(x)$ from our results for the order parameter at small values of the field $H$. The scaling function $f(z)$ can be obtained from this parameterization, and the agreement with the order parameter in this scaling form is also very good. 
These results are also in satisfactory agreement with the results for the scaling function from the O(4) spin model lattice simulations \cite{Engels:1999wf}, and the principal differences for large negative values of the scaling variable $z$ can be explained by the difference in the values of the critical exponents. 

Beyond small values of the external field $H$, we find consistently in both scaling forms, $y(x)$ and $f(z)$, that corrections to scaling become large and the results deviate from the scaling function away from the critical temperature. The scaling deviations can be discerned more readily in the Widom-Griffiths scaling form.

\section{Susceptibility}
\label{sec:susceptibility}

The susceptibility affords us an additional opportunity to test our results for the scaling function. 
As outlined in the discussion of the scaling behavior in Sec.~\ref{sec:scaling}, the scaling function for the susceptibility is completely determined by the scaling function for the order parameter and by the critical exponents:
\be
H_0\, h^{1-1/\delta} \chi  = \frac{1}{\delta} \left[f(z) -\frac{z}{\beta} f^\prime(z)\right],  \quad \quad \mathrm{or} \quad \quad \left[ H_0 M^{\delta -1} \chi \right]^{-1} = \delta \, y(x) -\frac{1}{\beta} x\, y^\prime(x). \nonumber
\ee
From a view focussing on the effective potential, this relation is \emph{a priori} far from obvious: the order parameter is determined as the minimum from the first derivative of the effective potential, the susceptibility from the second derivative, which is related to the masses of the longitudinal fluctuations and the four-point interaction. For this reason we consider this to be a significant test not only of our result for the scaling function, but also of our approach in general.

In the following, we will proceed in the same way as for the order parameter. We will first determine the scaling function for the susceptibility in Widom-Griffiths form from small values of the eternal field $H$, and then proceed to larger values. We will compare the determination of the scaling function from the susceptibility to the determination from the order parameter. For completeness, we also compare to the scaling function obtained from the lattice scaling function for the order parameter.

\subsection{Scaling function in Widom-Griffiths parameterization}

\subsubsection{Phenomenological ansatz for the scaling function}

\begin{table}
\begin{tabular}{l  l l l r l }
\hline\hline
\quad \quad $x$-range &  \quad  \quad \quad $c$ &\quad \quad $\tau$ & $\chi^2$ & $\#$ d.o.f. & \quad $\chi^2/\#$ d.o.f. \\
\hline
$-1 < x < 1$ & \quad $1.0009(19)$ & $1.7018(33)$ & $371.1$ & $838$ & \quad $0.4428$\\
\hline
$-1 < x < 5.3 \times 10^4$ &  \quad $0.9802(27)$ &  $1.5954(12)$ &$3754$ & $1551$ & \quad $2.420$\\
\hline\hline
\end{tabular}
\caption{\label{tab:suscphenlarge} Fit to the inverse rescaled susceptibility  for $H=1.0 \times 10^{-4}$, $1.0 \times 10^{-3}$, $1.0 \times 10^{-2}$ MeV$^{5/2}$ with ansatz for the equation of state of the form $y(x)=c (1+x)^\tau$. The scaling corrections are estimated from the spread of the rescaled susceptibility. Shown are separate fits for the range $-1<x<1$ and for $-1< x < 5.3 \times 10^{4}$.}
\end{table}
\begin{figure}
\includegraphics[scale=0.5]{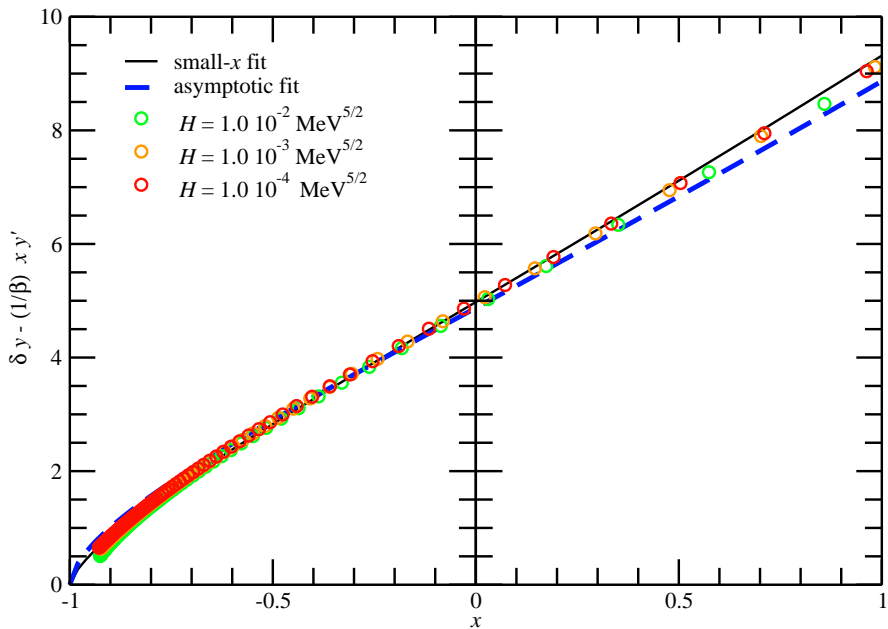}
\caption{\label{fig:inversesuscscalingsmallx} Results for the inverse scaled susceptibility $\left[H_0 M^{\delta-1} \chi \right]^{-1}$ for $H=1.0 \times 10^{-4}$, $1.0 \times 10^{-3}$, and $1.0 \times 10^{-2}$ MeV$^{5/2}$ (not all points are shown). We estimate an error due to the scaling violations from the spread of the results. 
Shown are in addition the fits for small $x$-values $x<1$ (solid black line), and from all values up to approximately $x=5.3 \times 10^4$ (dashed blue line, asymptotic fit).}
\end{figure}
\begin{figure}
\includegraphics[scale=0.5]{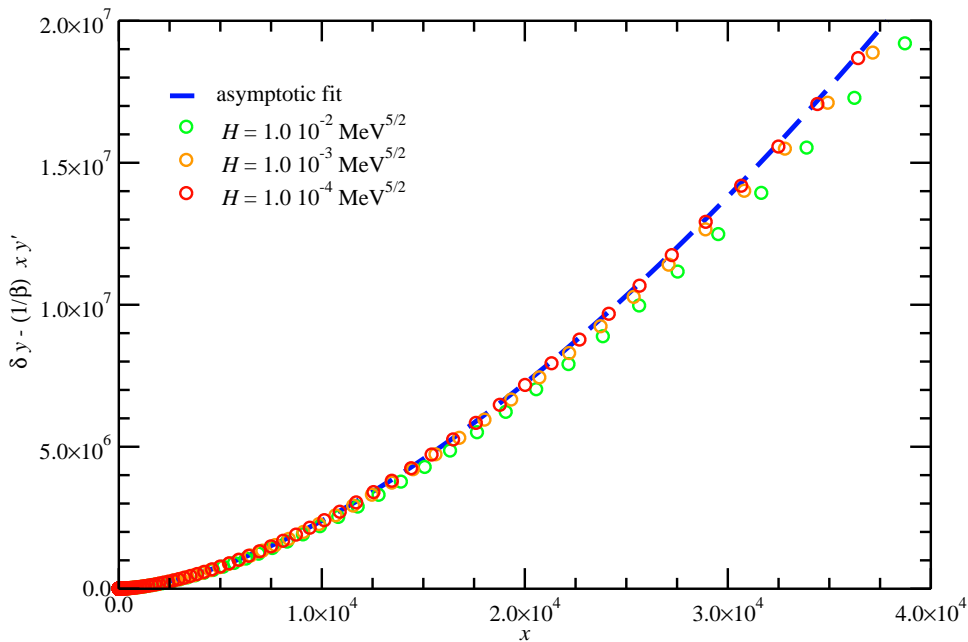}
\caption{\label{fig:inversesuscscalinglargex} Results for the inverse scaled susceptibility $\left[H_0 M^{\delta-1} \chi \right]^{-1}$ for $H=1.0 \times 10^{-4}$, $1.0 \times 10^{-3}$, and $1.0 \times 10^{-2}$ MeV$^{5/2}$ (not all points are shown). The fit to all data points is also shown (asymptotic fit, blue dashed line). From comparison to the equation of state $y(x)$ in Fig.~\ref{fig:griffithslargex} it is apparent that for large $x$ indeed $\delta y(x) - \frac{1}{\beta} x y(x)^\prime \simeq y(x)$ as expected from Griffiths' expansion and the scaling laws (see discussion of the asymptotic behavior below).}
\end{figure}
\begin{figure}
\includegraphics[scale=0.6]{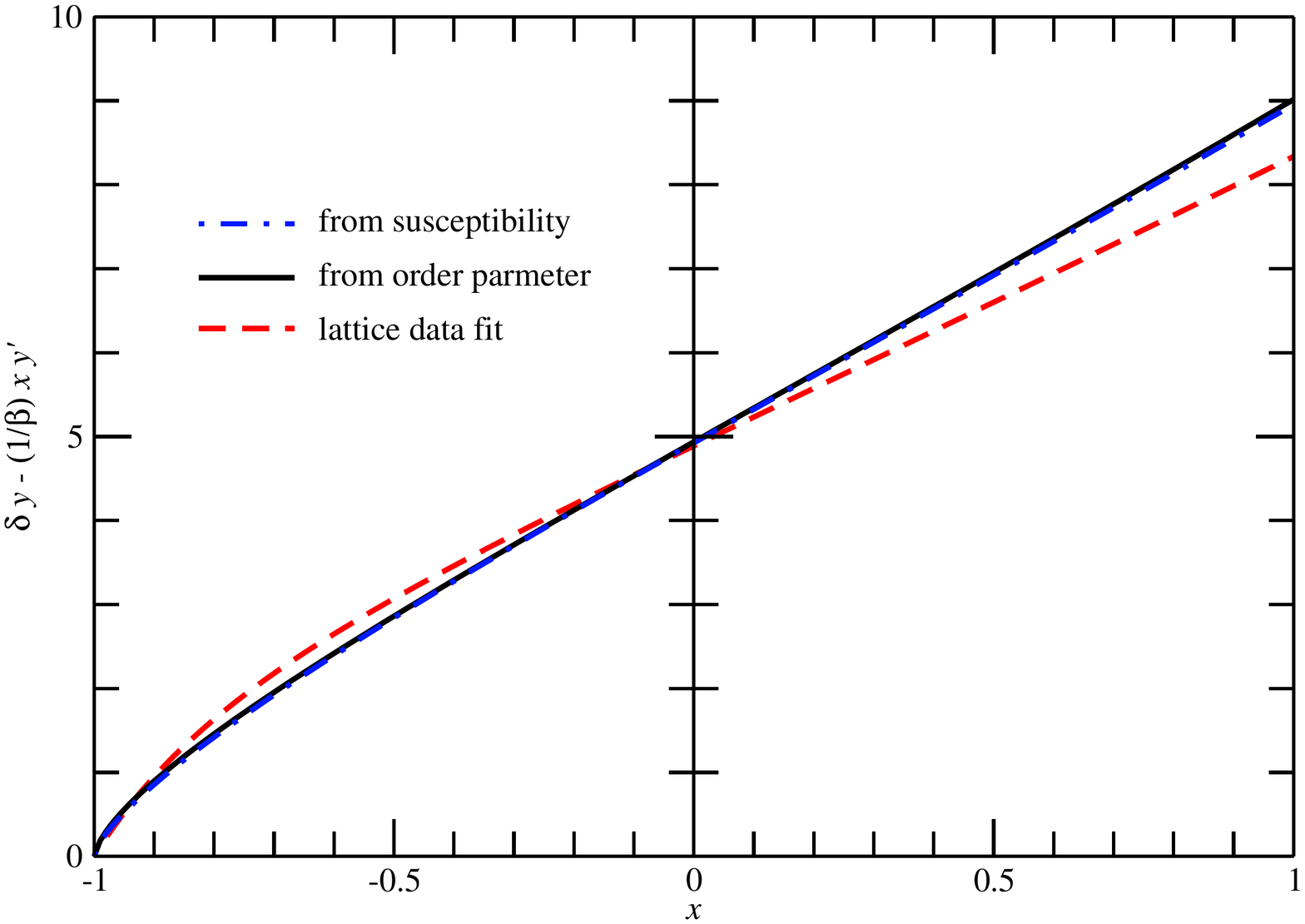}
\caption{\label{fig:suscgscaling} Comparison of different fits for the universal function $\delta y(x) -\frac{1}{\beta} x y^\prime(x)$ which describes the inverse rescaled susceptibility. We show the function with parameters obtained directly from a fit to our results for the rescaled susceptibility (blue dot-dashed line), from our results for the order parameter (black solid line), and from the order parameter in the lattice simulation \cite{Engels:1999wf} (note that the values for the critical exponents are different).
For all practical purposes, there is complete agreement between our curves with parameters from the order parameter and with those from the susceptibility. }
\end{figure}

As for the order parameter, we fit a two-parameter ansatz $y(x)=c (1+x)^\tau$ for the equation of state to the results for $H=1.0 \times 10^{-4}$ MeV$^{5/2}$,  $H=1.0 \times 10^{-3}$ MeV$^{5/2}$, and $H=1.0 \times 10^{-2}$ MeV$^{5/2}$. The size of the scaling corrections is estimated from the spread of the rescaled results for these values of $H$.
Inserting the ansatz in the expression for the rescaled susceptibility, one obtains
\be
\delta y(x) -\frac{1}{\beta} x y^\prime(x) &=& c (1+x)^\tau \left(\delta -\frac{\tau}{\beta} \frac{x}{1+x}  \right),
\ee
which we use to fit the inverse rescaled susceptibility. The critical exponents $\beta$ and $\delta$ are kept fixed at the previously determined values. 

The results for the separate fits in the range $-1<x<1$ and in the range  $-1< x < 5.3 \times 10^{4}$ are shown in Fig.~\ref{fig:inversesuscscalingsmallx} and Fig.~\ref{fig:inversesuscscalinglargex}. Scaling for the susceptibility, here over two orders of magnitude for the field $H$, is clearly observed, with small corrections at these values of $H$.
By comparing the results at large $x$ in Fig.~\ref{fig:inversesuscscalinglargex} for the susceptibility to the results in Fig.~\ref{fig:griffithslargex} for the order parameter, one can already see an indication that $\delta y(x) - \frac{1}{\beta} x y(x)^\prime \simeq y(x)$ as expected from Griffiths' expansion. We will check this more rigorously below when we analyze the asymptotic behavior.

The coefficients for the two-parameter fits are shown in Table~\ref{tab:suscphenlarge}. They can be compared directly to the coefficients from the fit to the rescaled order parameter in Table~\ref{tab:pheneos}. While the agreement is perfect (better than $0.1 \%$) only for the exponent $\tau$ for the large-$x$ region, which corresponds to the critical exponent $\gamma$, all parameters agree to better  than $2 \%$.

We can find a global expression for the rescaled susceptibility that interpolates between the small-$x$ and the large-$x$ behavior, just as for the order parameter. We first form the combinations
\be
\delta y(x) - \frac{1}{\beta} x y^\prime(x) \nonumber
\ee
for the large-$x$ and the small-$x$ region separately, and then combine them with interpolation factors:
\be
\left[M^{\delta -1} H_0 \chi\right]^{-1} &=& \frac{(1+x_0)^2}{(1+x_0)^2+(1+x)^2}\; c_s (1+x)^{\tau_s} \delta \left(1-\frac{\tau_s}{\beta \delta} \frac{x}{1+x}\right) \nonumber \\
&&+ \frac{(1+x)^2}{(1+x_0)^2+(1+x)^2} \; c_l (1+x)^{\tau_l} \delta \left(1-\frac{\tau_l}{\beta \delta} \frac{x}{1+x}\right).
\label{eq:suscphenfit}
\ee
\begin{table}
\begin{tabular}{ l l l l l l}
\hline\hline
 \quad\quad$c_s$ & \quad\quad$\tau_s$ & \quad\quad$c_l$ & \quad\quad$\tau_l$ & \quad\quad$x_0$ & \\
\hline
\quad $1.0009$ & \quad $1.7018$ & \quad $0.9802$ & \quad $1.5954$ & \quad\quad$0$ & \quad \quad from $\chi$,  Tab.~\ref{tab:suscphenlarge}\\
\quad$0.9928$ & \quad$1.6712$ & \quad$0.9928$ & \quad$1.5941$ & \quad\quad$0$ &  \quad \quad from $M$, Tab.~\ref{tab:pheneos}\\
\hline
\hline
\end{tabular}
\caption{\label{tab:phenfitcomp}Coefficients for the phenomenological ansatz  for the interpolated equation of state. Values in the upper row are obtained from a fit to the susceptibility $\chi$, values in the lower row from the order parameter $M$.}
\end{table}
For the coefficients, we give both the values obtained directly from the susceptibility, and the values obtained from the order parameter in Tab.~\ref{tab:phenfitcomp}.
For all practical purposes, these functions are indistinguishable from one another. This is illustrated in Fig.~\ref{fig:suscgscaling}. Thus the predictions of 
scaling are borne out by our results, and the scaling function for the susceptibility is already predicted by the scaling function for the order parameter.

This result can be compared to the interpolated parameterization of the lattice results from Engels and Mendes. We use the parameterization $x(y)$ for small and large $x$-values obtained from the order parameter. For both regions we first form the combinations 
\be
\left[M^{\delta -1} H_0 \chi\right]^{-1} &=& \delta y - \frac{1}{\beta} \frac{x}{x^\prime(y)}, \nonumber
\ee  
where in this case $y$ is treated as the independent variable. We then use the interpolation prescription eq.~\eqref{eq:Mendesinterpolation} to obtain a result valid over a wide $y$-range.
For the comparison in terms of $x$ and $y$ and for small $x$-values, the resulting curve is also shown in Fig.~\ref{fig:suscgscaling} as the lattice data fit. The difference in the slope is to a large part once again due to the difference in the values for the critical exponents.

\subsubsection{Scaling behavior for small $x$}

\begin{table}
\begin{tabular}{l l l l l l l}
\hline\hline
\quad \quad$c$ & \quad\quad$\tau$ & \quad\quad$d_1$ & \quad\quad$d_2$ & \quad$\chi^2$ & \# d.o.f. & $\chi^2$/\# d.o.f. \\
\hline
$1.00324(26)$ & $1.6546(17)$ & $-$ & $-$ & $3.0 \times 10^4$ & $282$ & $109.2$\\
$1.299(10)$ & $1.7886(43)$ & $-0.229(4)$ & $-$ & $2762$ & $281$ & $9.827$\\
$1.5344(49)$ & $1.8472(13)$ & $-0.428(4)$ & $0.081(2)$ & $98.51$ & $280$ &  $0.3518$\\
\hline\hline 
\end{tabular}
\caption{\label{tab:suscfitphen}Coefficients for the fit to the inverse scaled susceptibility in the region $-1<x<1$, using the ansatz $y(x)= c (1+x)^\tau( 1+ d_1 (1+x)^{1/2} + d_2 (1+x))$ for the equation of state.}
\end{table}
\begin{table}
\begin{tabular}{l l l l l l}
\hline
\hline
\quad\quad$a_3$  & \quad\quad $a_2$ & \quad\quad$a_1$ & \quad$\chi^2$ & \# d.o.f. & $\chi^2$/(\# d.o.f.)\\
\hline
$0.98125(49)$ & $\phantom{-}-$ & $-$ & $3.4 \times 10^6$ & $283$ & $12100$ \\
$1.2855(17)$ & $-0.2839(16)$ & $-$   & $7746$ & $282$ & $27.47$\\
$1.3678(16)$   & $-0.4409(29)$ & $0.0751(15)$ & $196.9$ & $281$ & $0.7007$\\ 
\hline
\hline
\end{tabular}
\caption{Coefficients for the fit to the inverse scaled susceptibility in the region $-1 < x < 1$, using the ansatz $y(x) =  a_3 (1+x)^{3/2} + a_2 (1+x) + a_1(1+x)^{1/2}$ for the equation of state.}
\label{tab:suscfiteps}
\end{table}

We will again determine the scaling function for small $x$ as accurate as possible from the results for $H=1.0 \times 10^{-4}$ MeV$^{5/2}$ in the range $-1<x<1$, where scaling corrections are small.
We estimate the scaling corrections for the fit to the smallest-field results  $H=1.0 \times  10^{-4}$ MeV$^{5/2}$ by comparing the rescaled results to those of with $H=2.0 \times 10^{-4}$ MeV$^{5/2}$. Between $-0.5 < x< 1$, the scaling violation is less than $0.01 \%$, and becomes sizable ($\sim 5 \%$) only close to $x=-1$. Using these errors, we can judge how well the leading-order scaling behavior fits the results.

We use the same ans\"atze as for the order parameter. With the ansatz $y(x) = c(1+x)^\tau ( 1 + d_1 (1+x)^{1/2} +d_2 (1+x))$  for the equation of state, one finds for the scaling function of the susceptibility
\be
\delta \, y(x) -\frac{1}{\beta} x y^\prime(x) &=& \delta\, c (1+x)^\tau \left[ 1+ d_1 (1+x)^{1/2} + d_2 (1+x)\right] \nonumber \\
&&  -\delta\, c (1+x)^\tau \left[  \frac{1}{\beta \delta} \frac{x}{(1+x)}\left( \tau + (\tau + \frac{1}{2} ) d_1 (1+x)^{1/2} + (\tau +1) d_2 (1+x)\right) \right].
\ee
We take the critical exponents from our original determination, and thus fit the result in this functional from with the same number of parameters as for the order parameter. The results are given in Table~\ref{tab:suscfitphen}. They can be compared directly to the results from the order parameter in Tab.~\ref{tab:-4xsmall2}. The agreement with the parameters determined from the order parameter is best for the fit with only two parameters: we find $c=1.0032(3)$ and $\tau = 1.655(2)$ from the susceptibility, and  $c=1.0031(5)$ and $\tau=1.682(3)$ from the order parameter. The agreement for the coefficient $c$ is perfect, and the values for the exponent $\tau$ agree within $2 \%$. Even though the agreement of the coefficients becomes worse when additional corrections are included, the functions with parameters from the order parameter and from the susceptibility remain within $0.4\%$ of each other over the interval $-1<x<1$.

The ansatz $y(x) = a_3 (1+x)^{3/2} + a_2 (1+x) + a_1 (1+x)^{1/2}$ for the equation of state leads to the corresponding expression
\be
\delta \left[y(x) -\frac{1}{\beta \delta} x y^\prime(x)\right] &=& \delta a_3 (1+x)^{3/2} \left(1-\frac{3}{2} \frac{1}{\beta \delta} \frac{x}{1+x} \right) \nonumber \\
&& \;\;+ \delta a_2 (1+x) \left(1- \frac{1}{\beta \delta} \frac{x}{1+x} \right)  + \delta a_1 (1+x)^{1/2} \left(1-\frac{1}{2} \frac{1}{\beta \delta} \frac{x}{1+x} \right). 
\ee
The coefficients are given in Table~\ref{tab:suscfiteps}. They should be compared to those from the fit to the order parameter in Table~\ref{tab:4xsmall1}. Again, the leading-order coefficients agree within $ 2 \%$ in the one-parameter, and within $5\%$ in the two-parameter fit. While the agreement between the coefficients is not perfect, the agreement between the scaling functions as such is actually better: For the three-parameter fit, the scaling functions fitted directly to the susceptibility and the one calculated from the fit to the order parameter are within $0.01$ of one another over most of the interval ($-0.9<x<1$), i.e. the relative errors are less than $0.4 \%$ for most of the interval ($-0.9<x<1$). Since the function itself approaches $0$ for $x\to -1$, the relative errors become large close to $x=-1$.

Despite the differences in the coefficients that appear when additional corrections are included, the determination of the scaling function from the susceptibility and the one  from the order parameter are completely equivalent for all practical purposes.

\subsubsection{Scaling behavior for asymptotically large $x$}

\begin{table}
\begin{tabular}{l l l l l l l l}
\hline\hline
$\gamma$ & $c_1$ & $c_2$ & $c_3$ & $\chi^2$ & \# d.o.f. & $\chi^2$/\# d.o.f.  & $\delta(\gamma- 1/(\beta \delta))$\\
\hline
$1.5844(12)$ & $1.081(9)$ & $-$ & $-$ & $9918$ & $204$ & $48.62$ & $1.0411$\\
$1.59988(5)$ & $0.94158(42)$ & $1.2794(34)$ & $-$ & $4.2109$ & $203$ & $0.02074$ & $1.00277$\\
\hline\hline 
\end{tabular}
\caption{Coefficients for the fit to the inverse susceptibility in the region $x>100$, using the asymptotic expression $y(x)=x^\gamma(c_1 + c_2 x^{-2 \beta} + c_3 x^{-4 \beta} + \ldots)$ for the equation of state. The deviationfrom $1$ in the last column indicates the violation of the scaling law for the critical exponents with the value of $\gamma$ from the fit.}
\label{tab:suscfitasy}
\end{table}

Using Griffiths' expansion eq.~\eqref{eq:Griffithsasymptotic} in the scaling function of the susceptibility, one finds that the leading order behavior of the scaling functions for the order parameter and the susceptibility agree for large $x$ (see eq.~\eqref{eq:Griffithsasymptoticsusceptibility}), provided the  scaling law $\gamma = \beta(\delta -1)$ holds. In the following, we test this by fitting Griffiths' expansion to the results from $H =1.0 \times 10^{-4} $ for $x>100$, once again with the deviation from the scaling behavior estimated from the difference to the results for $H =2.0 \times 10^{-4} $ and used to assess the quality of the fit. As for the order parameter, we keep the values of the critical exponents $\beta$ and $\delta$ fixed to the previously determined values, and treat $\gamma$ as a free fit parameter. 

The coefficients of the fit are given in Table~\ref{tab:suscfitasy}. They should be compared to the results for the coefficients of the equation of state in Tab.~\ref{tab:coeffasyeos}. 
Because of the scaling relations between the critical exponents, the leading-order terms of the scaling form for the order parameter, $y(x) \simeq c_1 x^\gamma$, and for the inverse susceptibility, $\delta y(x) -\frac{1}{\beta} x y^\prime(x) \simeq c_1 x^\gamma$ are expected to coincide exactly.

With two fit parameters, the fit results for the exponent $\gamma$ agree to within $0.0004$, and the coefficients $c_1$ to within $9 \%$. Part of the difference for $c_1$ is explained by the fact that the exponent $\gamma$ satisfies the scaling law only to within $4 \%$, which translates directly into a deviation for the coefficient $c_1$ and explains about half of the difference.

For the three-parameter fit (with $\gamma$, $c_1$ and $c_2$ as fit parameters), the agreement between the equation of state $y(x)$ from the order parameter and from the susceptibility is perfect, the values for the critical exponent $\gamma$ agree to within $0.0002$, and the scaling law $\gamma = \beta(\delta-1)$ is satisfied to better than $0.3 \%$ (see deviation from $1$ in the last column of the table).

We find the degree to which the relations between the scaling functions are satisfied in our calculation truly remarkable. On the one hand, we calculate the order parameter from the minimum of the effective potential. On the other hand, the longitudinal susceptibility $\chi$ is calculated from the mass $m_\sigma$ of the longitudinal fluctuations, which involves the curvature of the effective potential and is \emph{a priori} independent of the order parameter. Nevertheless, we find that the values for the critical exponents and for the leading coefficients in the asymptotic expansion of the equation of state agree perfectly. 

In this region, this relation has been tested neither by results from the $\epsilon$-expansion, nor by results from lattice simulations. For the $\epsilon$-expansion, the large-$x$ region is not accessible, since the asymptotic behavior cannot be calculated explicitly, but only order by order in $\epsilon$. 
In spin model lattice calculations, the equation of state has also not been calculated this far in $x$ and the asymptotic behavior has not been tested to this extent. Engels and Mendes \cite{Engels:1999wf} confirm the large-$y$ behavior in the scaling form $x(y)$ up to $x \lesssim 150$, after inverting the equation of state $y(x)$ approximately. 

\subsection{Scaling for large fields}

\begin{figure}
\includegraphics[scale=0.3]{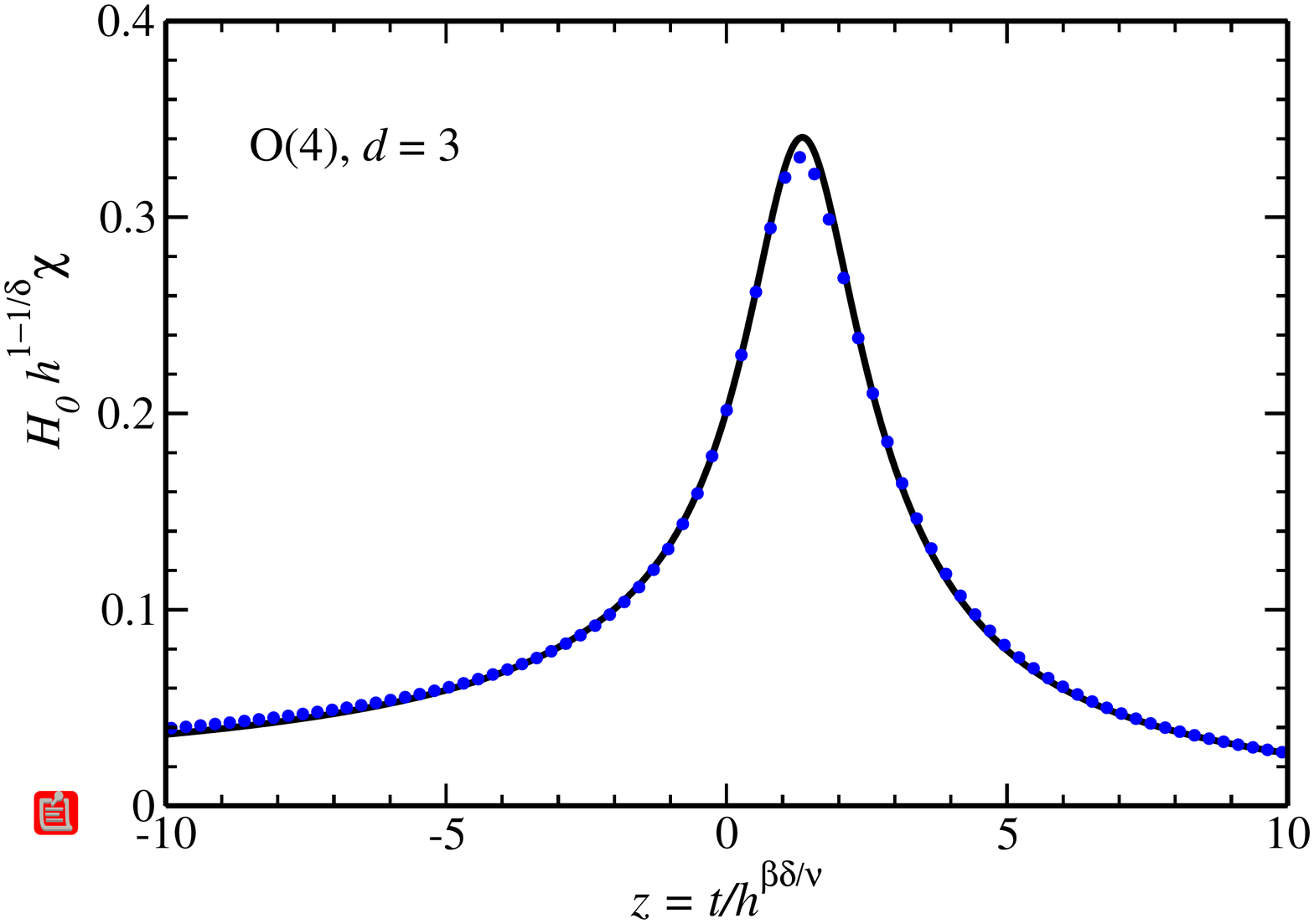}
\includegraphics[scale=0.3]{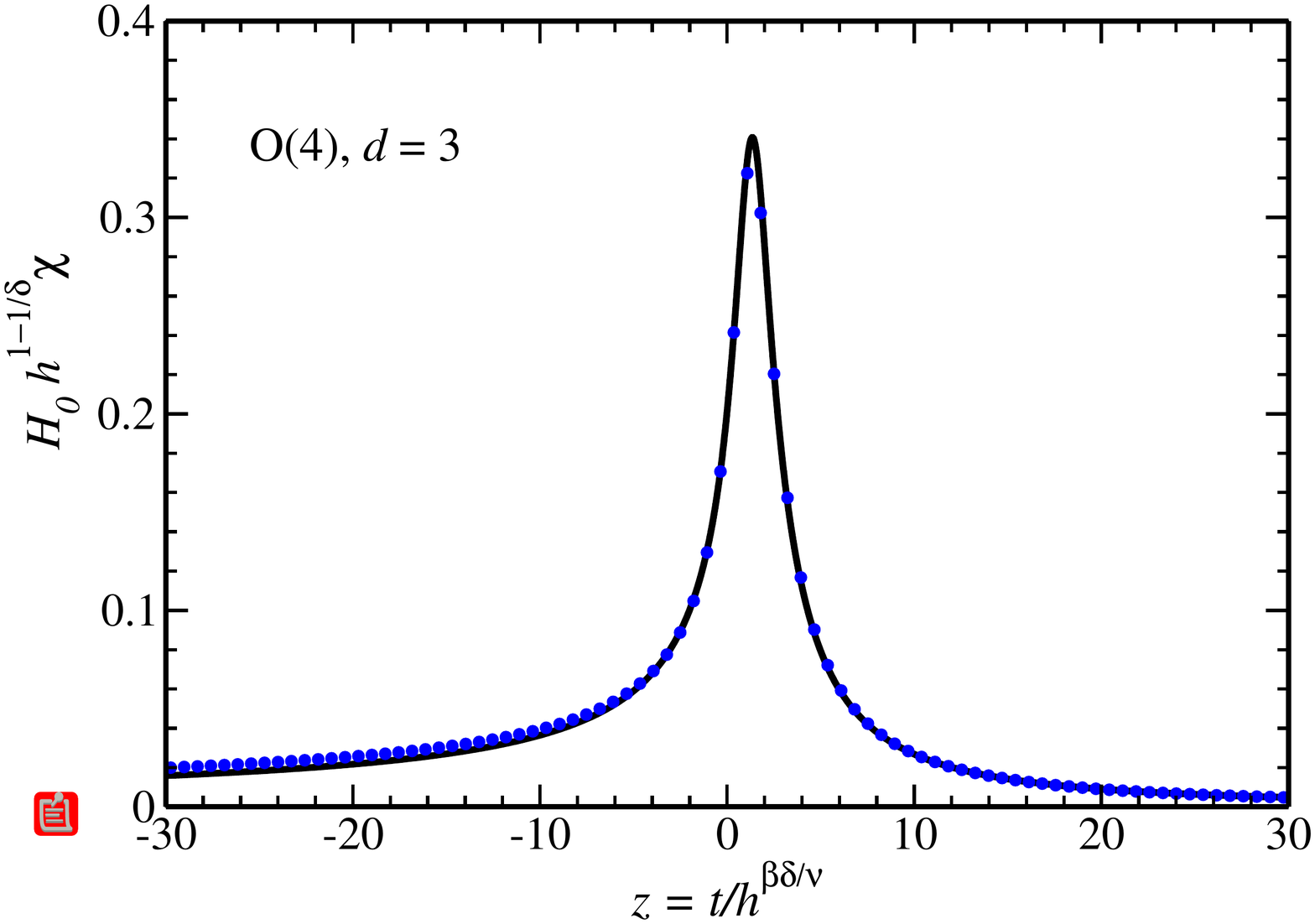}
\caption{\label{fig:suscfitvsz} The parameterization $y=y(x)$ of the equation of state in terms of the scaling variables $x$ and $y$ also provides implicitly a parameterization of the scaled susceptibility $H_0 h^{1-1/\delta} \chi = \frac{1}{\delta} \left[f(z) -\frac{z}{\beta} f^\prime(z)\right]$ in terms of the scaling variable $z$. The scaling 
function (black solid line) with parameters obtained from the order parameter, not the susceptibility itself, is compared to the results for $H=1.0 \times 10^{-4}$ MeV$^{5/2}$ (blue circles, for clarity not all points are shown). Apart from the slight difference at the peak the fit is almost perfect. Within the linewidth, the fit with parameters obtained from the order parameter and the one with parameters obtained from the susceptibility directly (not shown) are indistinguishable in this plot.}
\end{figure}
\begin{figure}
\includegraphics[scale=0.6]{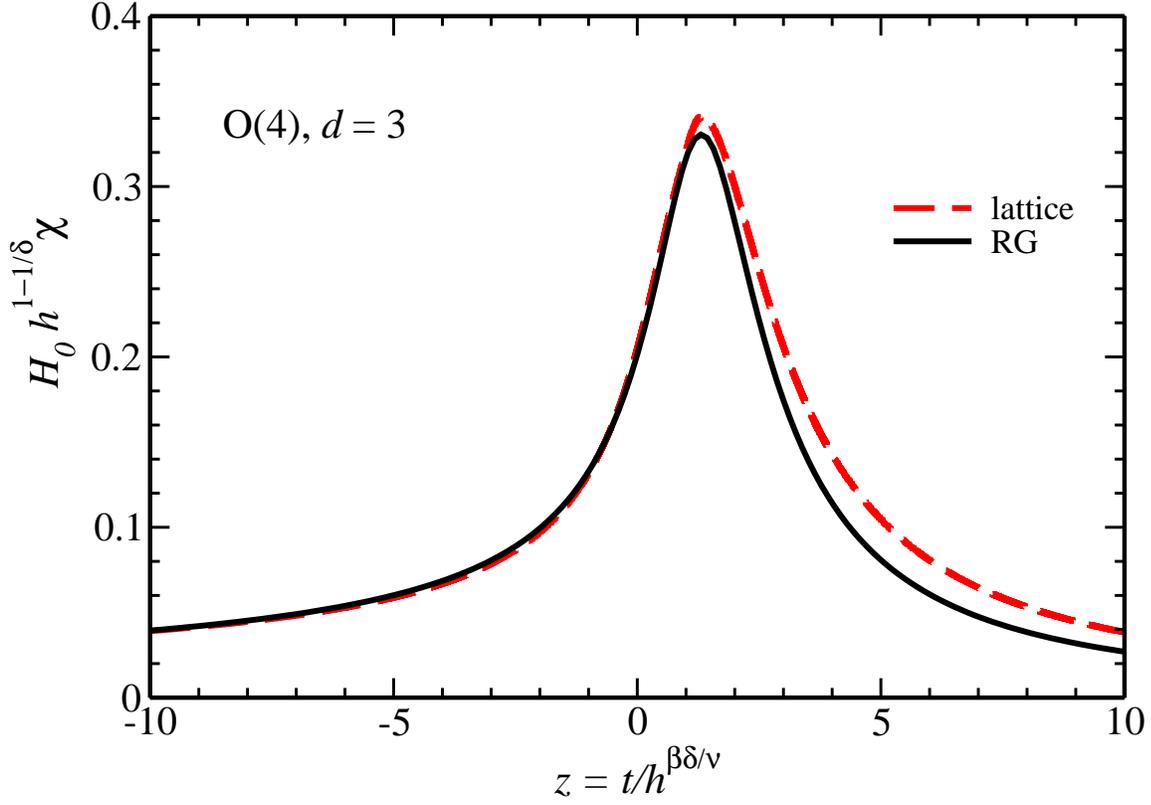}
\caption{\label{fig:suscmendesvsz} Comparison of the parameterization of the scaling function from the lattice results \cite{Engels:1999wf} (red dashed line) to our results for $H = 1.0 \times 10^{-4}$ MeV$^{5/2}$ (black solid line). The agreement is very good below the critical temperature, but deviations appear above $T_c$.}
\end{figure}

We have determined the scaling function for the susceptibility in the previous sections from results at small values of $H$, and we were able to confirm the expected relations with the scaling function for the order parameter. Proceeding as for the order parameter, we now turn to larger values of $H$ where scaling corrections become important.

The representation of the results in terms of Griffiths' scaling variables $x$ and $y$ is not very intuitive, and we therefore translate the results back into the scaling variable $z$ and the scaling function $f(z)$. In Fig.~\ref{fig:suscfitvsz}, the rescaled susceptibility $H_0 h^{1-1/\delta} \chi = \frac{1}{\delta} \left[f(z) -\frac{z}{\beta} f^\prime(z)\right]$ for $H=1.0 \times 10^{-4}$ MeV$^{5/2}$ is shown as a function of $z$, together with the fit with parameters obtained from the order parameter. Apart from the slight difference at the peak, which is most sensitive to scaling corrections, the scaling function agrees very well with the results for the susceptibility. This really confirms a prediction, since all parameters for the curve are already determined from the order parameter, and there is no freedom left. A direct fit to the susceptibility in a second step only confirms how well the parameters agree with each other.

A comparison of our results for $H = 1.0 \times 10^{-4}$ MeV$^{5/2}$ to the scaling function for the susceptibility predicted from the lattice spin model is shown in Fig.~\ref{fig:suscmendesvsz}. For $z<0$ (in the phase with large symmetry breaking), the agreement is very good, but in the phase with largely restored symmetry, the susceptibility predicted by the order parameter lattice scaling function is larger than the one we calculate in the RG. Overall, we consider the agreement still satisfactory. 

\begin{figure}
\includegraphics[scale=0.3]{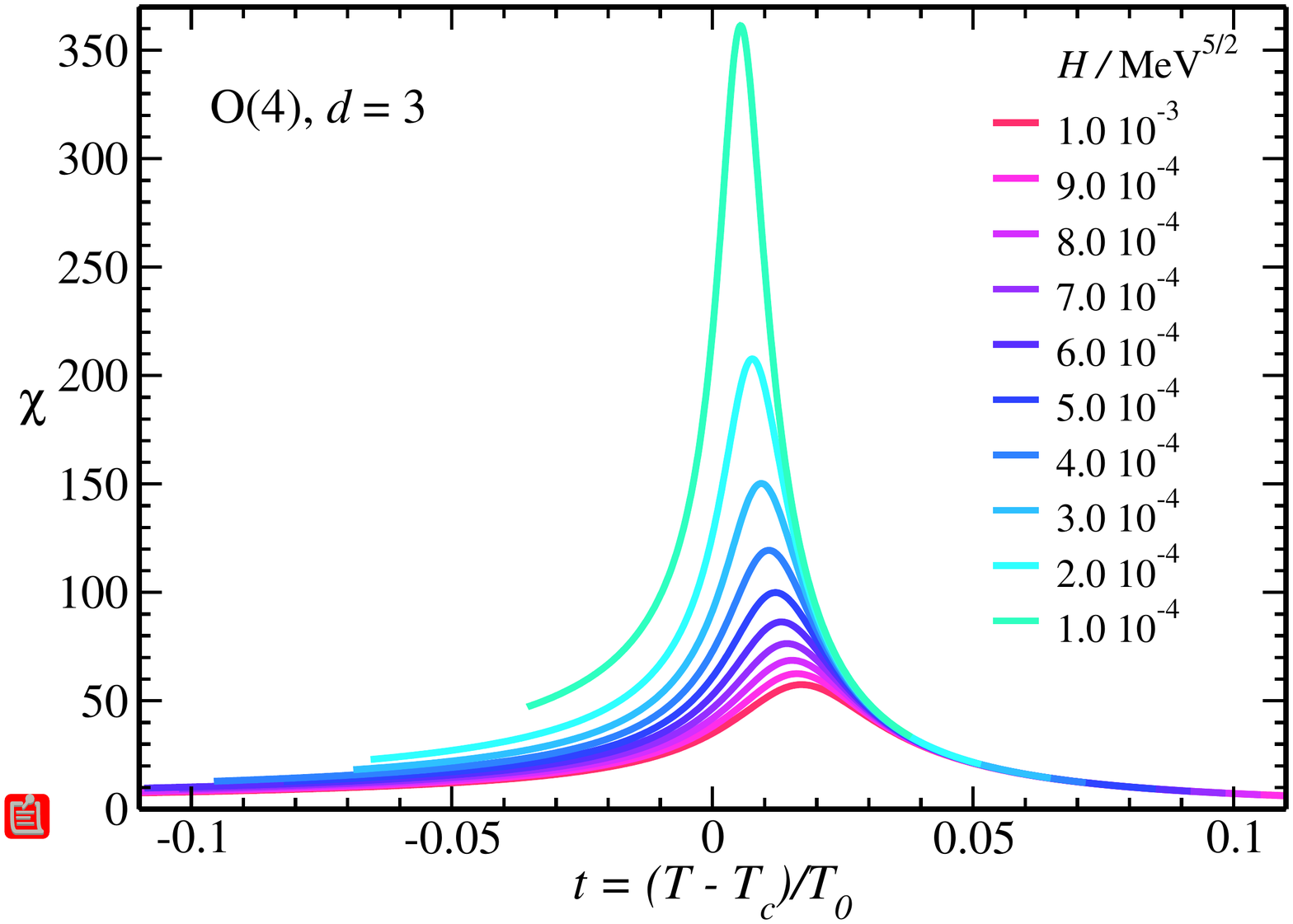}
\includegraphics[scale=0.3]{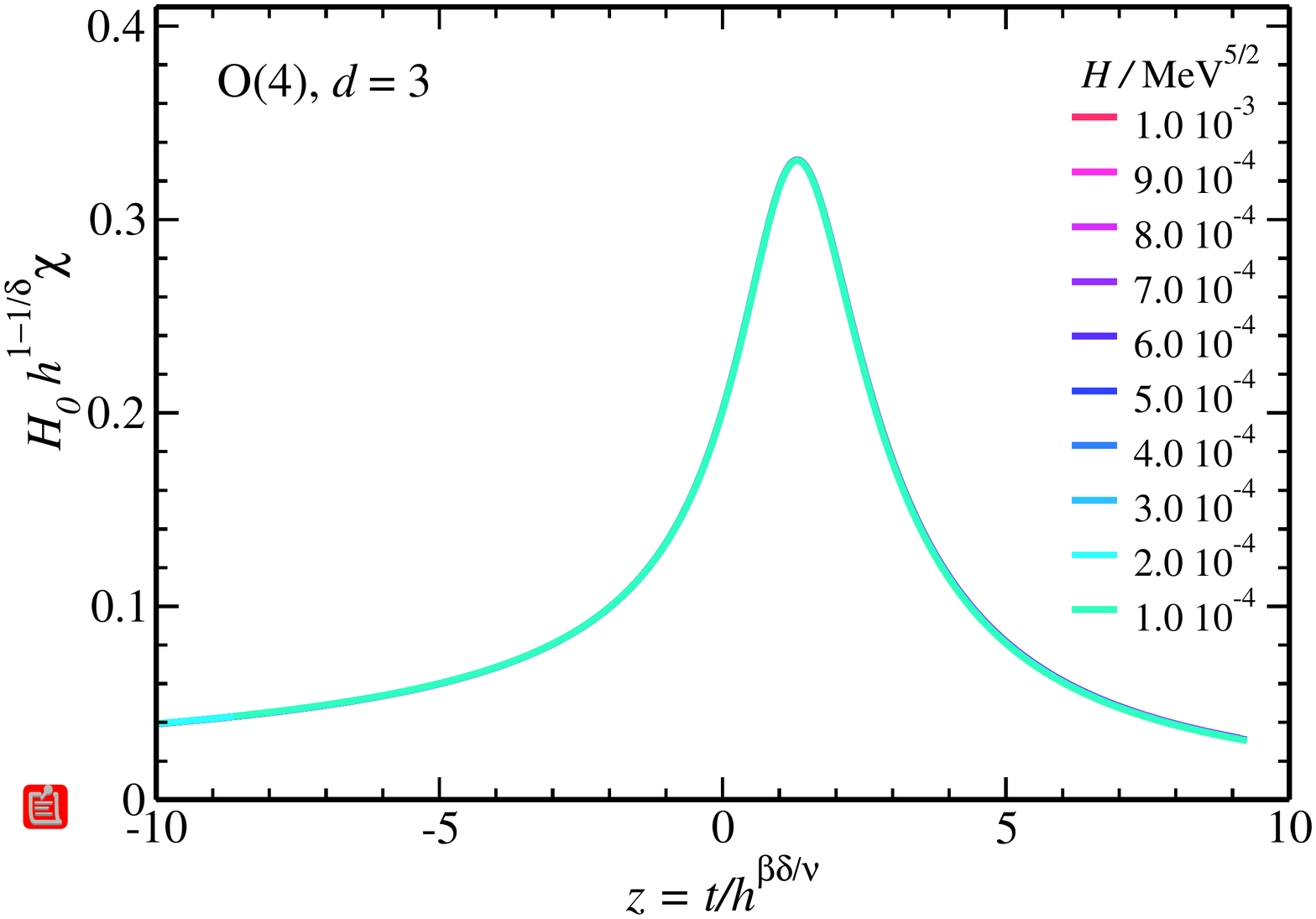}
\caption{\label{fig:susczscalingsmallH} Susceptibility as a function of the reduced temperature $t$ for very small values of the external field $H=1.0 \times 10^{-4}$ MeV$^{5/2}$ to $H=1.0 \times 10^{-3}$ MeV$^{5/2}$ (left panel), and rescaled susceptibility $H_0 h^{1-1/\delta} \chi$ as a function of $z=t/h^{1/(\beta \delta)}$ (right panel). The $t$-ranges for the different values of $H$ are chosen such that $z$ covers the range $-10\ldots 10$ after rescaling. Within the width of the symbols, the rescaled curves coincide exactly. Deviations from scaling are not visible at this scale.}
\end{figure}
\begin{figure}
\includegraphics[scale=0.3]{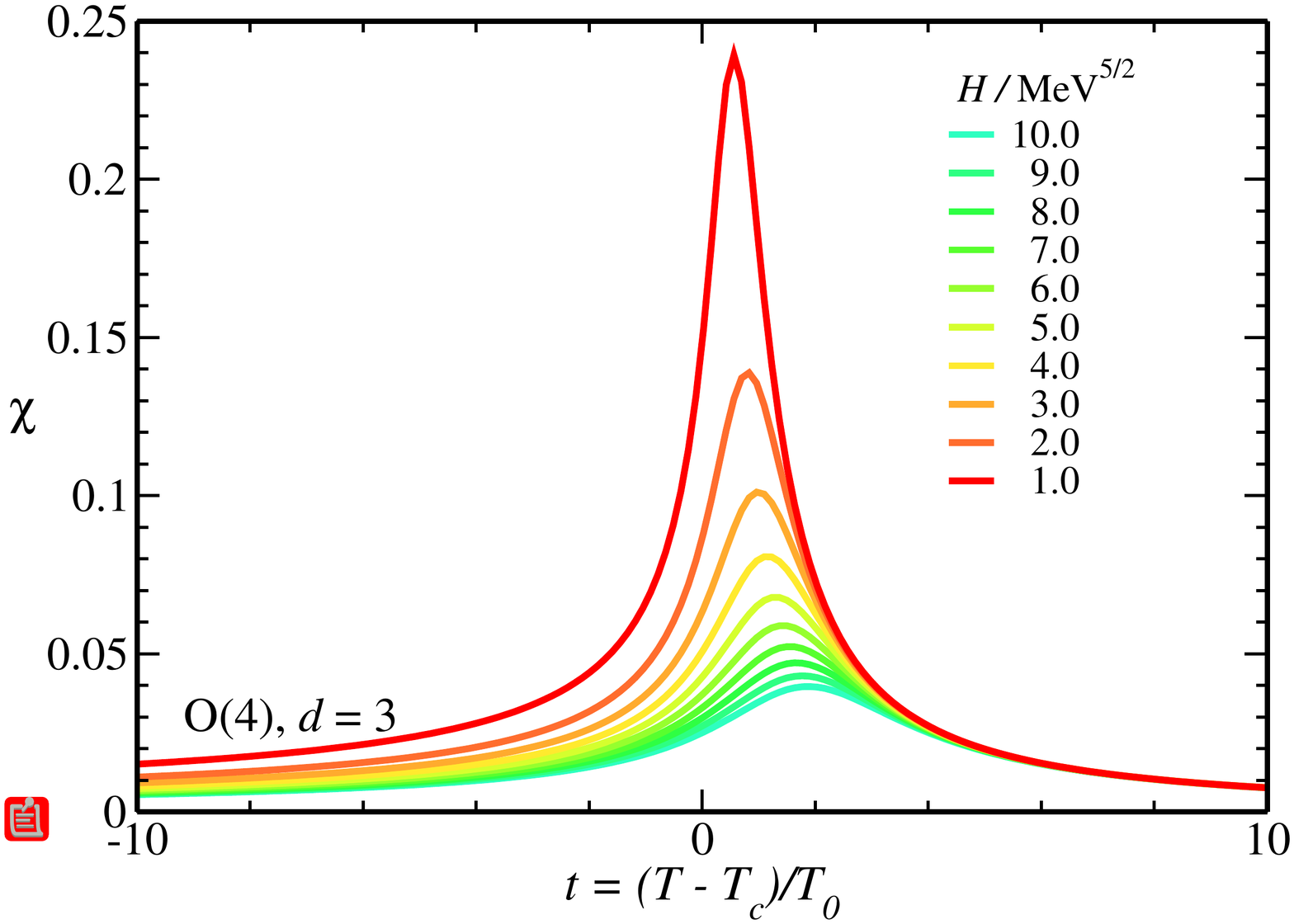}
\includegraphics[scale=0.3]{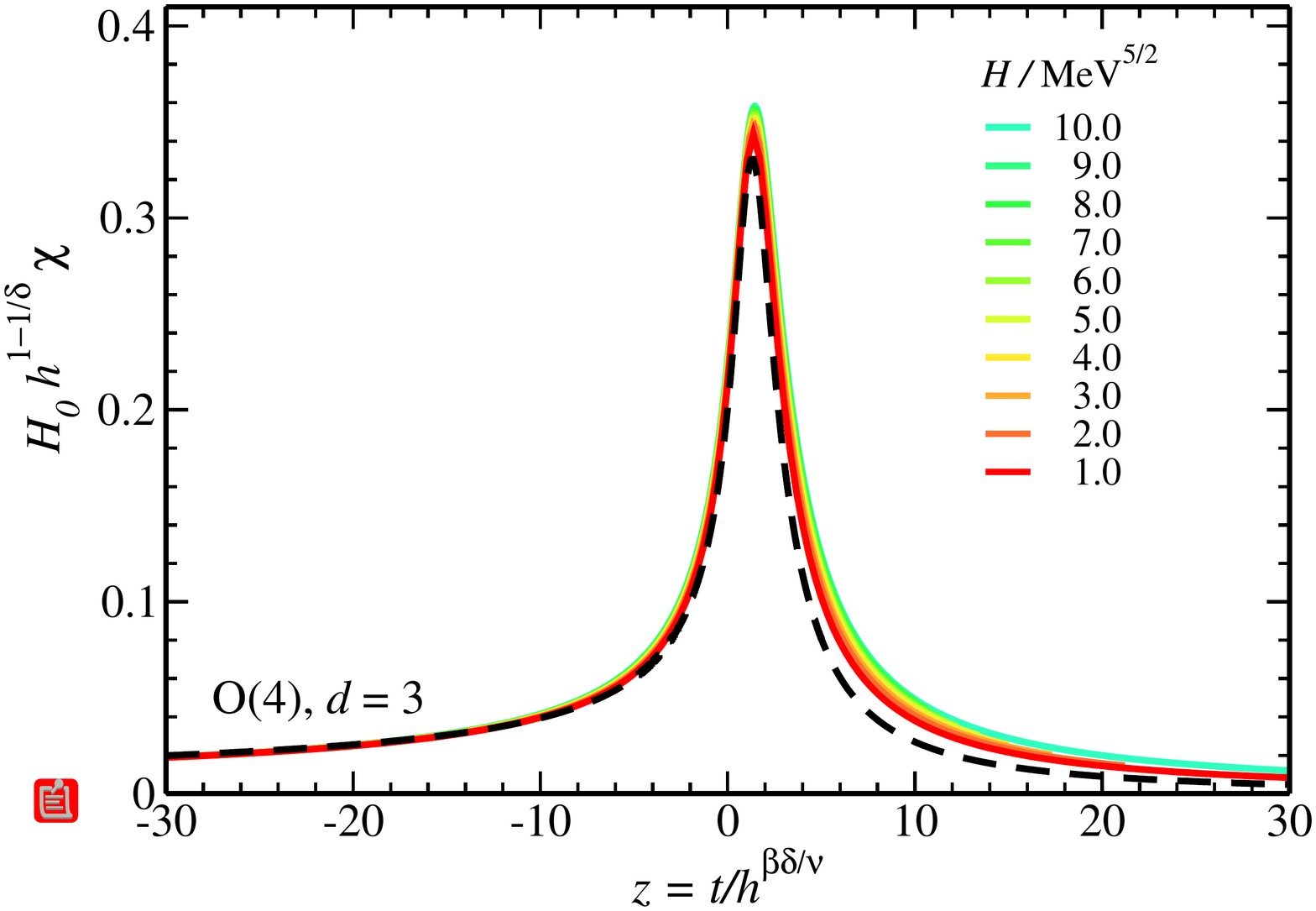}\\
\includegraphics[scale=0.3]{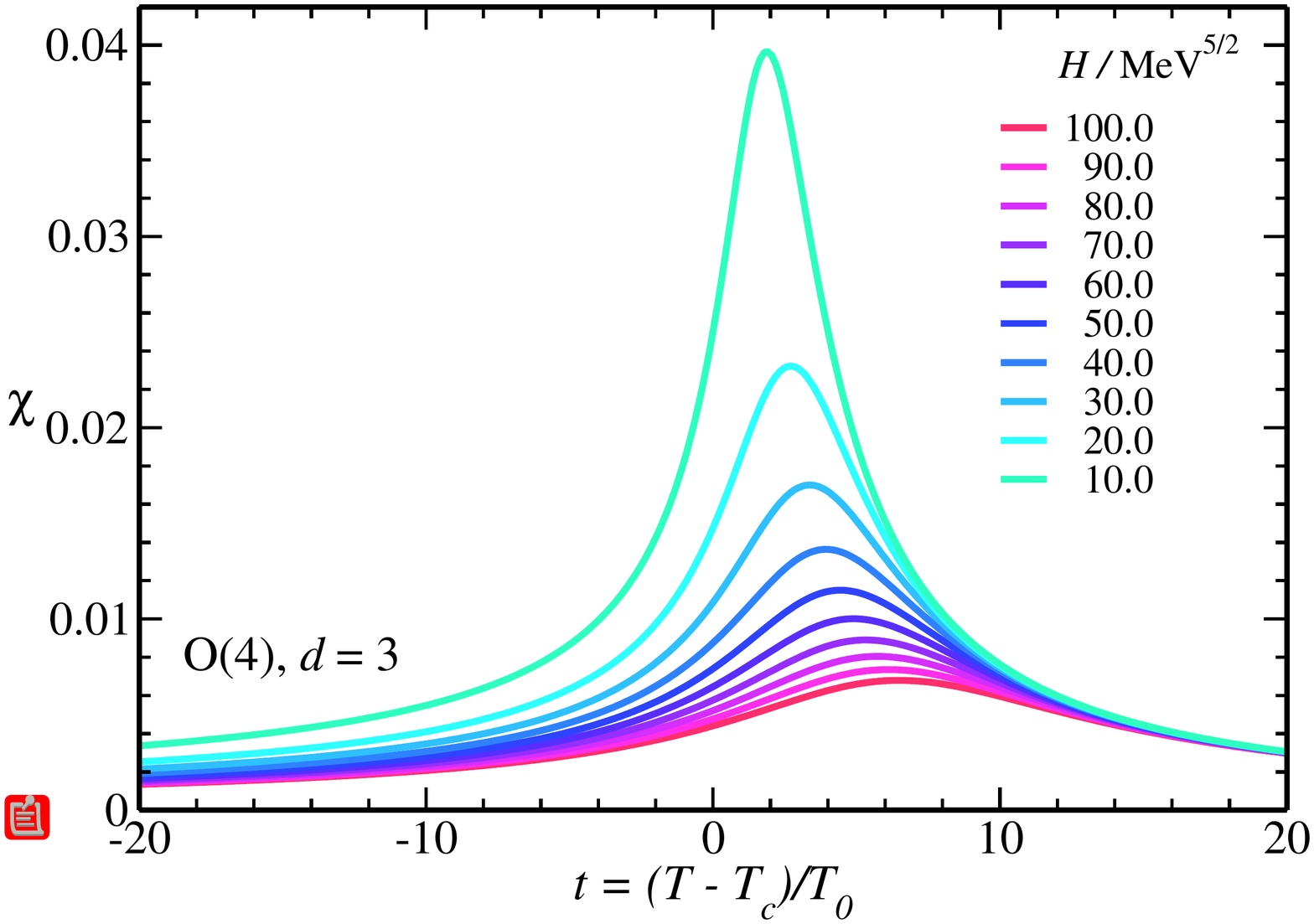}
\includegraphics[scale=0.3]{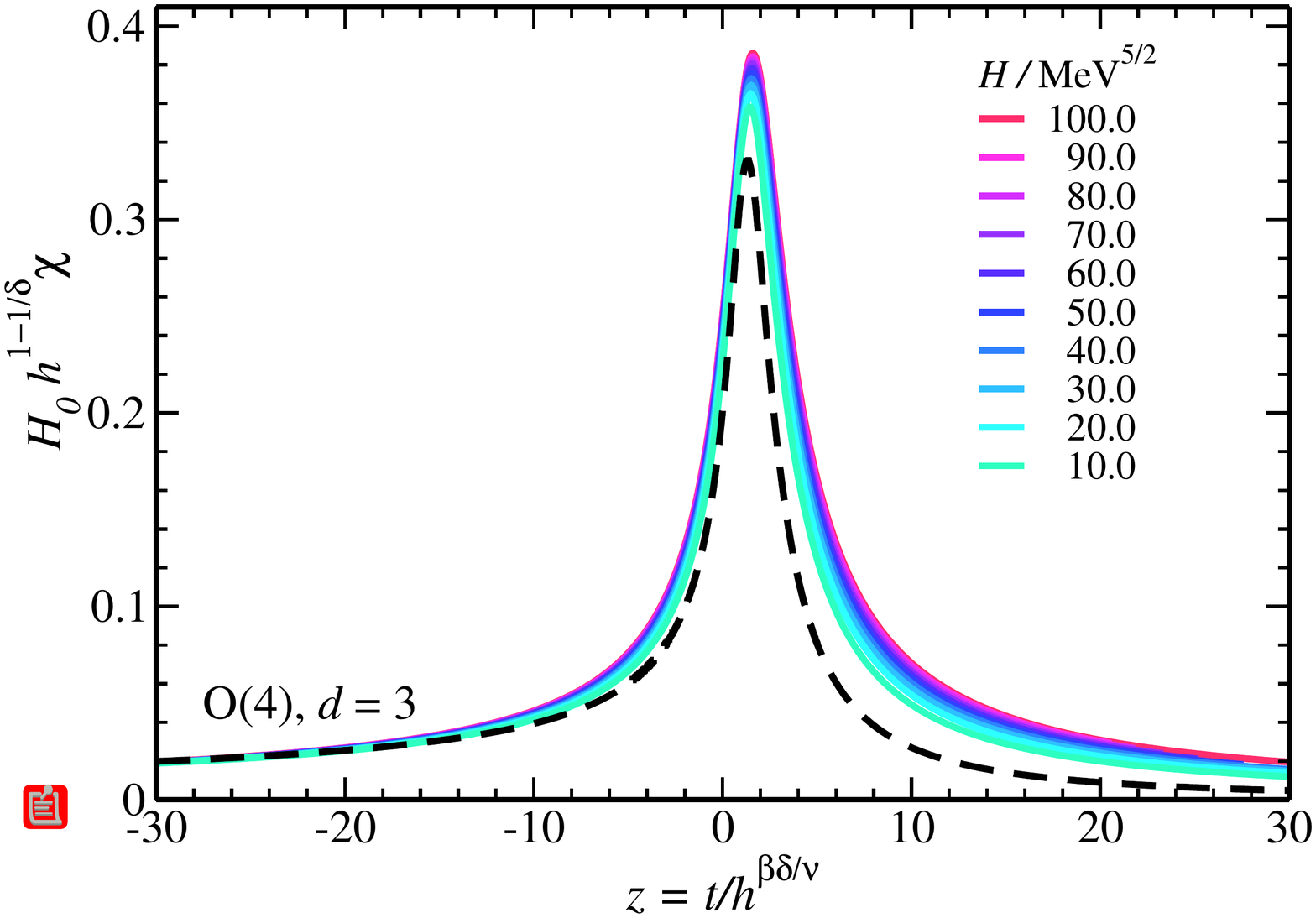}\\
\includegraphics[scale=0.3]{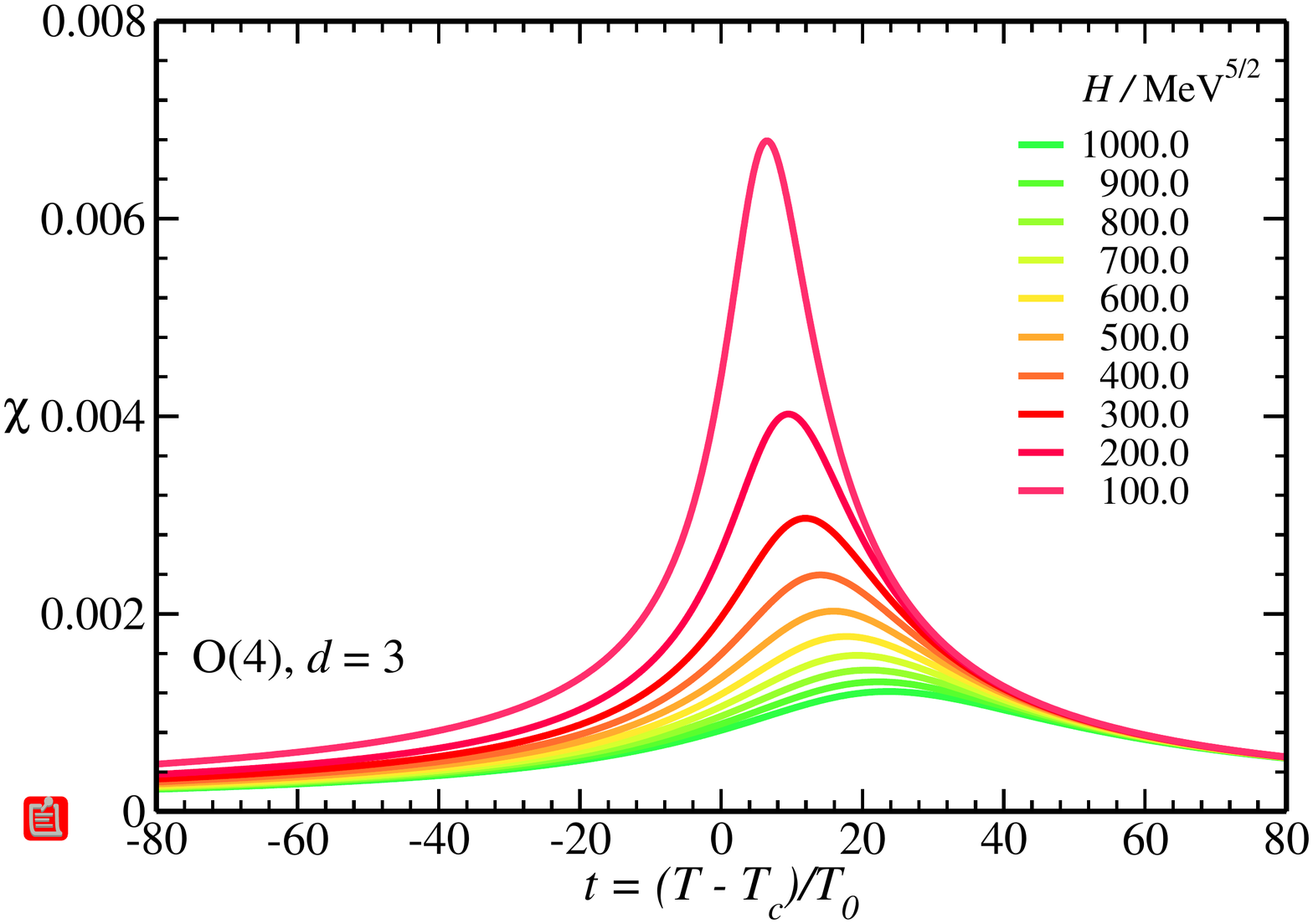}
\includegraphics[scale=0.3]{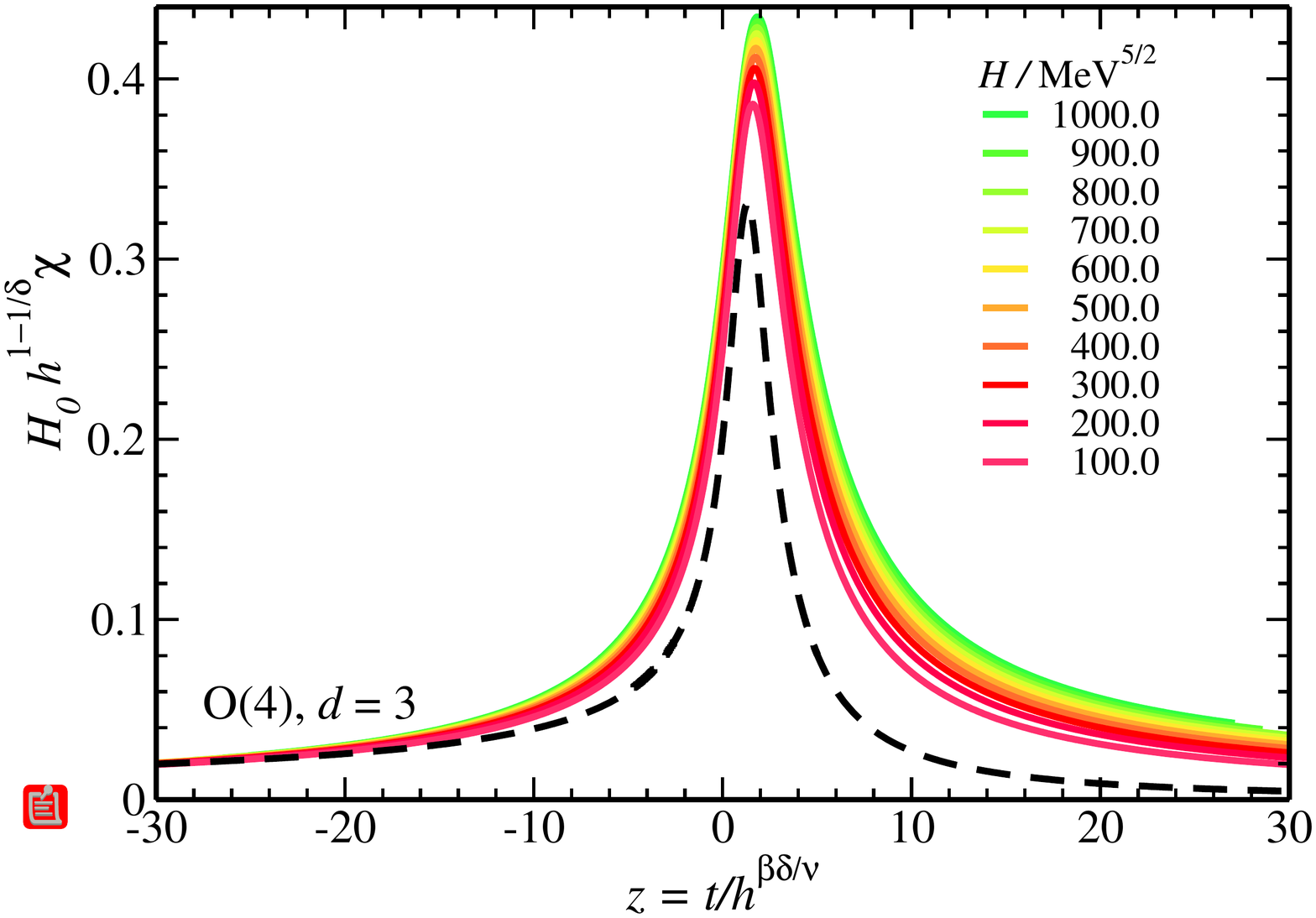}
\caption{\label{fig:suscscalingzlargeH}Results for the longitudinal susceptibility $\chi$ as a function of the reduced temperature $t$ (left-hand column), and for the rescaled susceptibility $H_0 h^{1-1/\delta} \chi$ as a function of $z=t/h^{1/(\beta \delta)}$ (right-hand column) for different values of the field $H$. For comparison, the scaling function obtained from the results for $H=1.0 \times 10^{-4}$ MeV$^{5/2}$ is plotted with the rescaled results (black dashed line). Note that the axes for the rescaled results in the left column are kept at the same scale, while the ranges in $t$ for the unscaled results are very different for different $H$-ranges.}
\end{figure}
\begin{figure}
\includegraphics[scale=0.6]{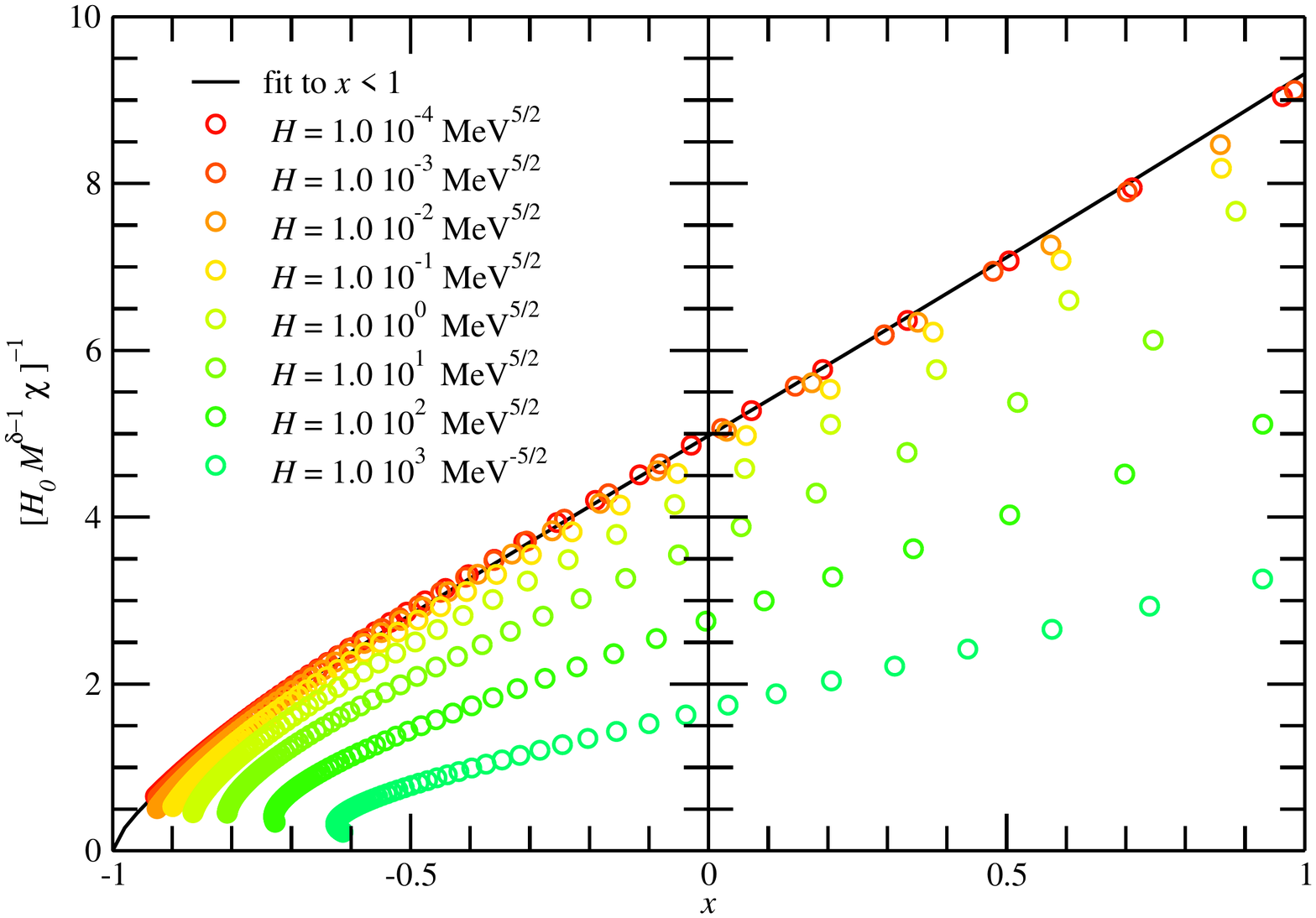}
\caption{\label{fig:suscGriffithslargeH} Comparison of the scaling results for the inverse susceptibility $\left[H_0 M^{\delta -1} \chi \right]^{-1}$ over a wide range of values for the symmetry-breaking field $H$ (7 orders of magnitude). The result of a fit with the ansatz eq.~(\ref{eq:suscphenfit}) to the susceptibility for $H=1.0 \times 10^{-4}$ MeV$^{5/2}$, $H=1.0 \times 10^{-3}$ MeV$^{5/2}$, and $H=1.0 \times 10^{-2}$ MeV$^{5/2}$ is shown as a solid line. For the larger values of $H$, the deviations from the leading-order scaling behavior are apparent.}
\end{figure}
We now demonstrate the scaling behavior of the susceptibility by plotting the susceptibility rescaled in the form $H_0 h^{1-1/\delta}\chi$ as a function of the scaling variable $z$.
In Fig.~\ref{fig:susczscalingsmallH}, results for small values of $H=1.0 \times 10^{-4}$ MeV$^{5/2}$ to $H=1.0 \times 10^{-3}$ MeV$^{5/2}$ are shown. In the left panel, the susceptibility as a function of the reduced temperature $t$ is plotted for different values of $H$. In the right panel, the rescaled susceptibility $H_0 h^{1-1/\delta} \chi$ is plotted as a function of $z=t/h^{1/(\beta \delta)}$ for the same values of $H$. As for the order parameter, we observe that the results collapse onto a single curve after rescaling. No scaling corrections are discernible for these small values of $H$.

However, as we increase the values for $H$, corrections to scaling soon become apparent. 
In Fig.~\ref{fig:suscscalingzlargeH}, we show our results for the range $H=1.0$ MeV$^{5/2}$ to $H=1.0 \times 10^{3}$ MeV$^{5/2}$. Both the susceptibility as a function of $t$ and the rescaled susceptibility as a function of $z$ are shown. For comparison, we also plot the rescaled results for  $H=1.0 \times 10^{-4}$ MeV$^{5/2}$. Deviations from the scaling behavior are already significant for $H=10.0$ MeV$^{5/2}$, in agreement with our observations for the order parameter. 

The corrections to the scaling behavior become more obvious when we again plot the results in Widom-Griffiths scaling form. In Fig.~\ref{fig:suscGriffithslargeH}, the rescaled results for $H=1.0 \times 10^{-4}$ to $H=1.0 \times 10^{3}$ MeV$^{5/2}$ ($H$ covers $7$ orders of magnitude) are shown as a function of $x$ for the range $-1<x<1$. In this plot, the corrections already appear significant for $H>1.0 \times 10^{-2}$ MeV$^{5/2}$.

In conclusion, we were able to obtain the scaling function for the susceptibility in Widom-Griffiths scaling form from the RG results at small values of $H$. This scaling function satisfies the expected relations with the scaling function obtained from the order parameter. In fact, we find that the scaling behavior of the susceptibility is described perfectly by the equation of state obtained from the order parameter. 
The scaling behavior as a function of the scaling variable $z$ can be obtained from the parameterization as well. 
Beyond small values of the field $H$, we find in both scaling forms that corrections to scaling become large and that the deviations from the scaling functions can become significant.

\section{Masses}
\label{sec:masses}

\begin{figure}
\includegraphics[scale=0.3]{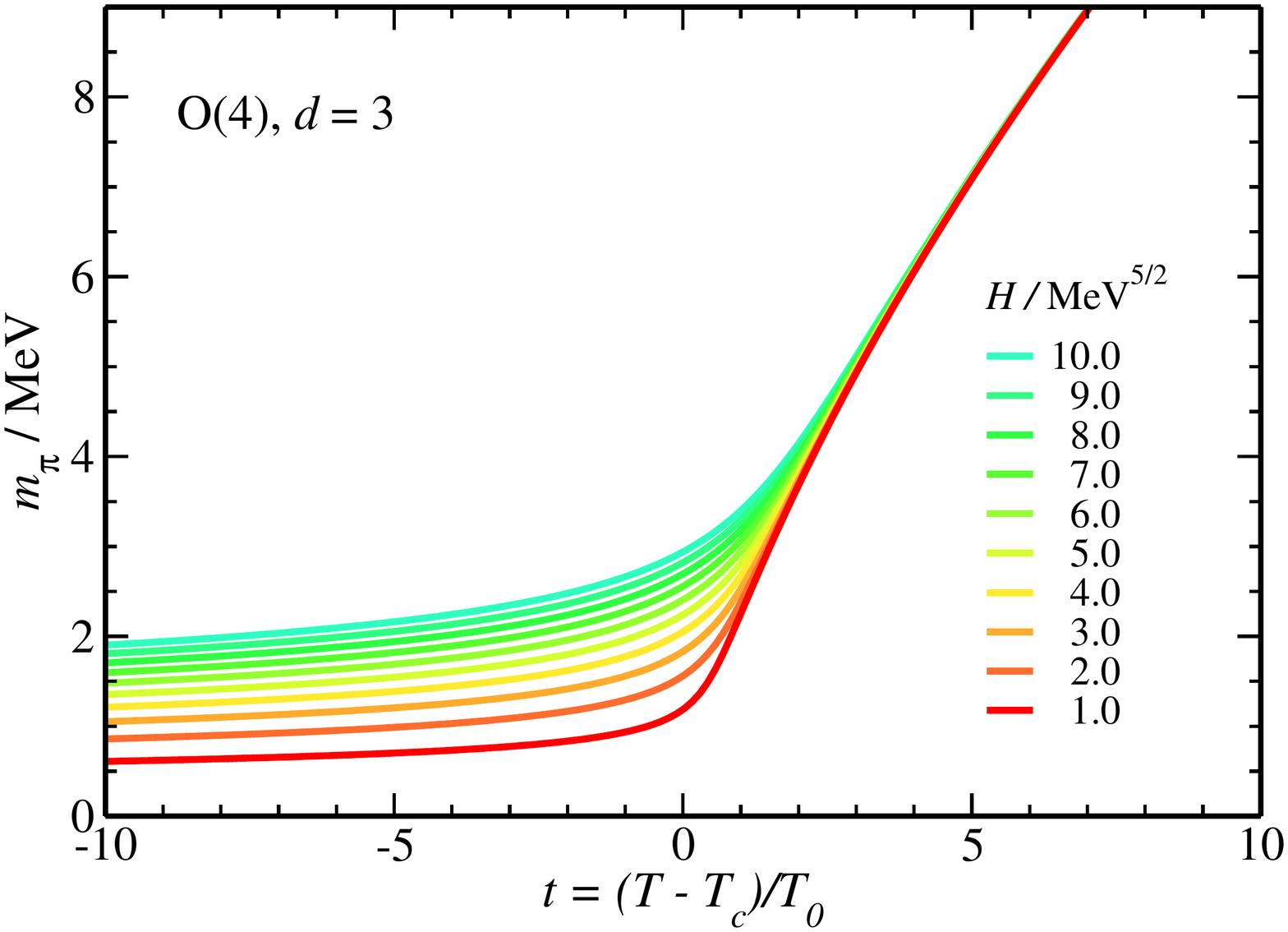}
\includegraphics[scale=0.3]{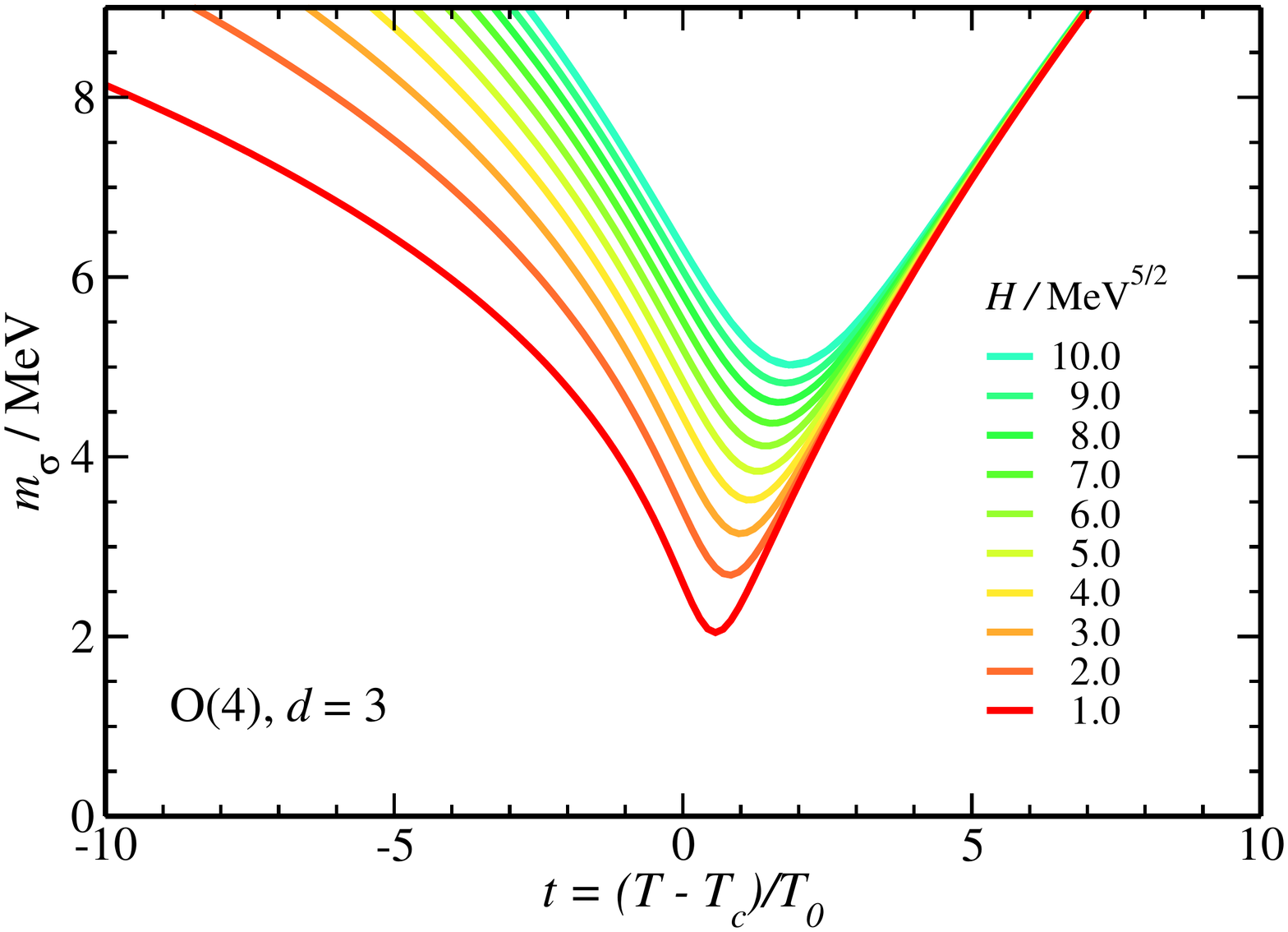}\\
\includegraphics[scale=0.3]{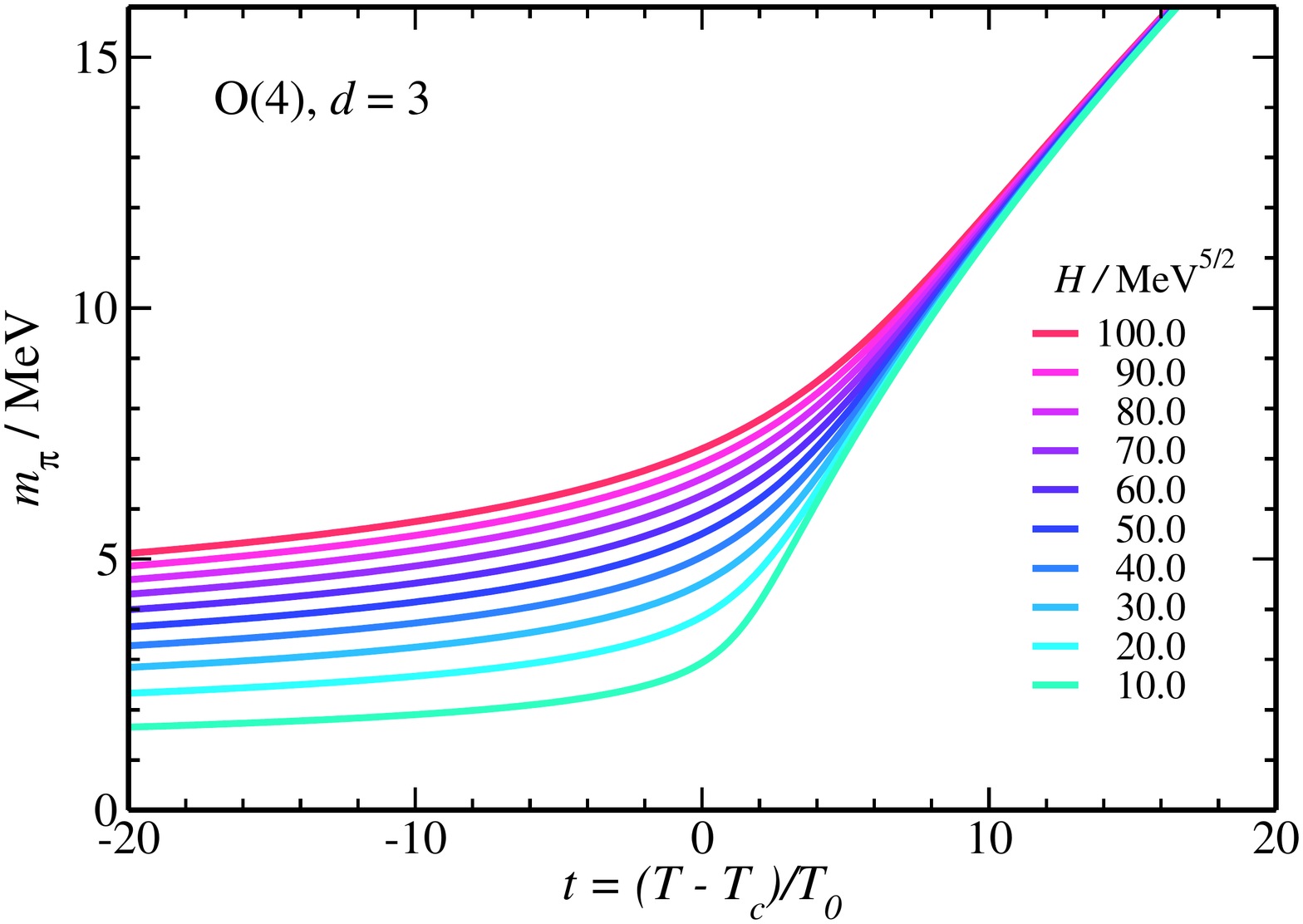}
\includegraphics[scale=0.3]{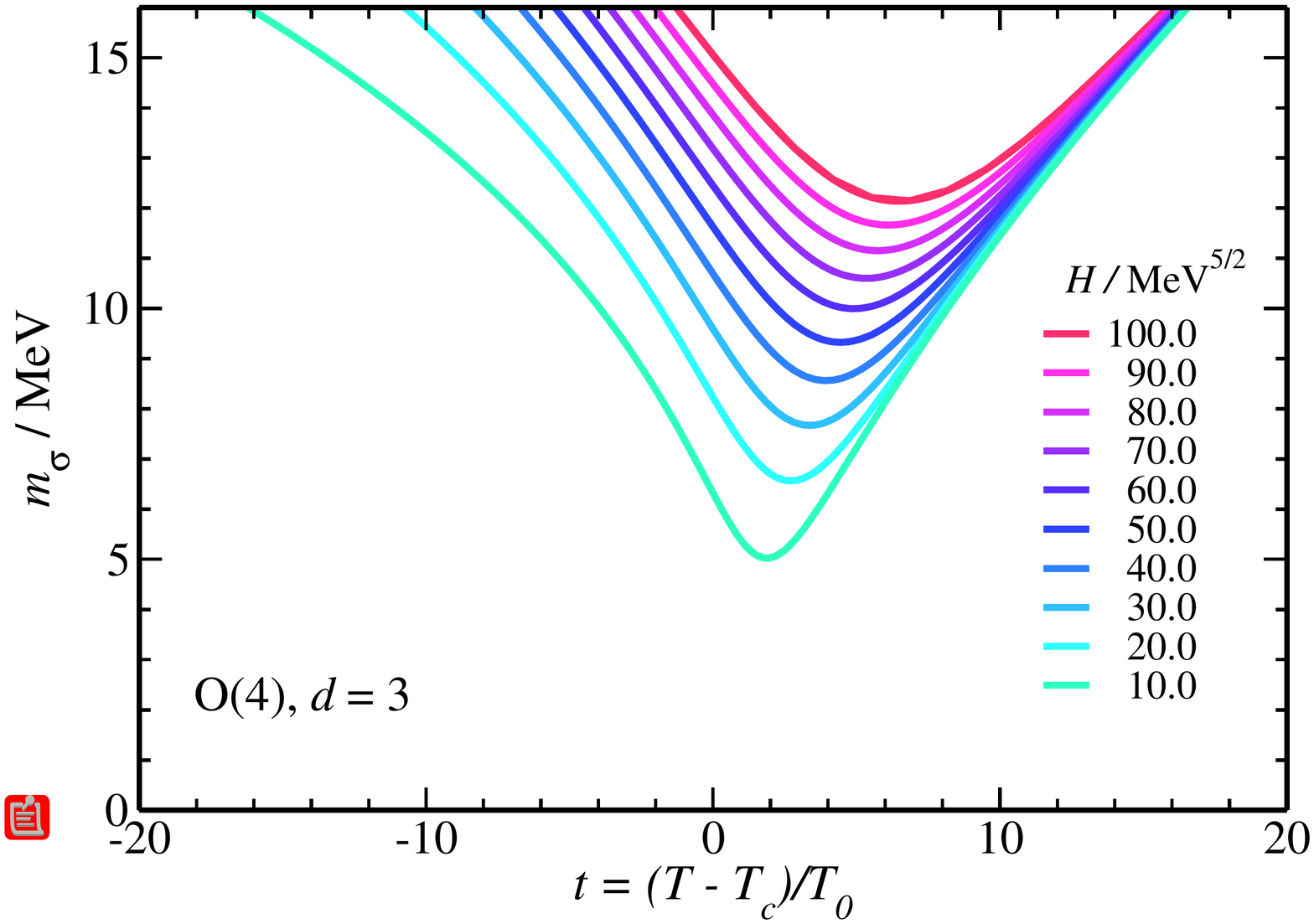}\\
\includegraphics[scale=0.3]{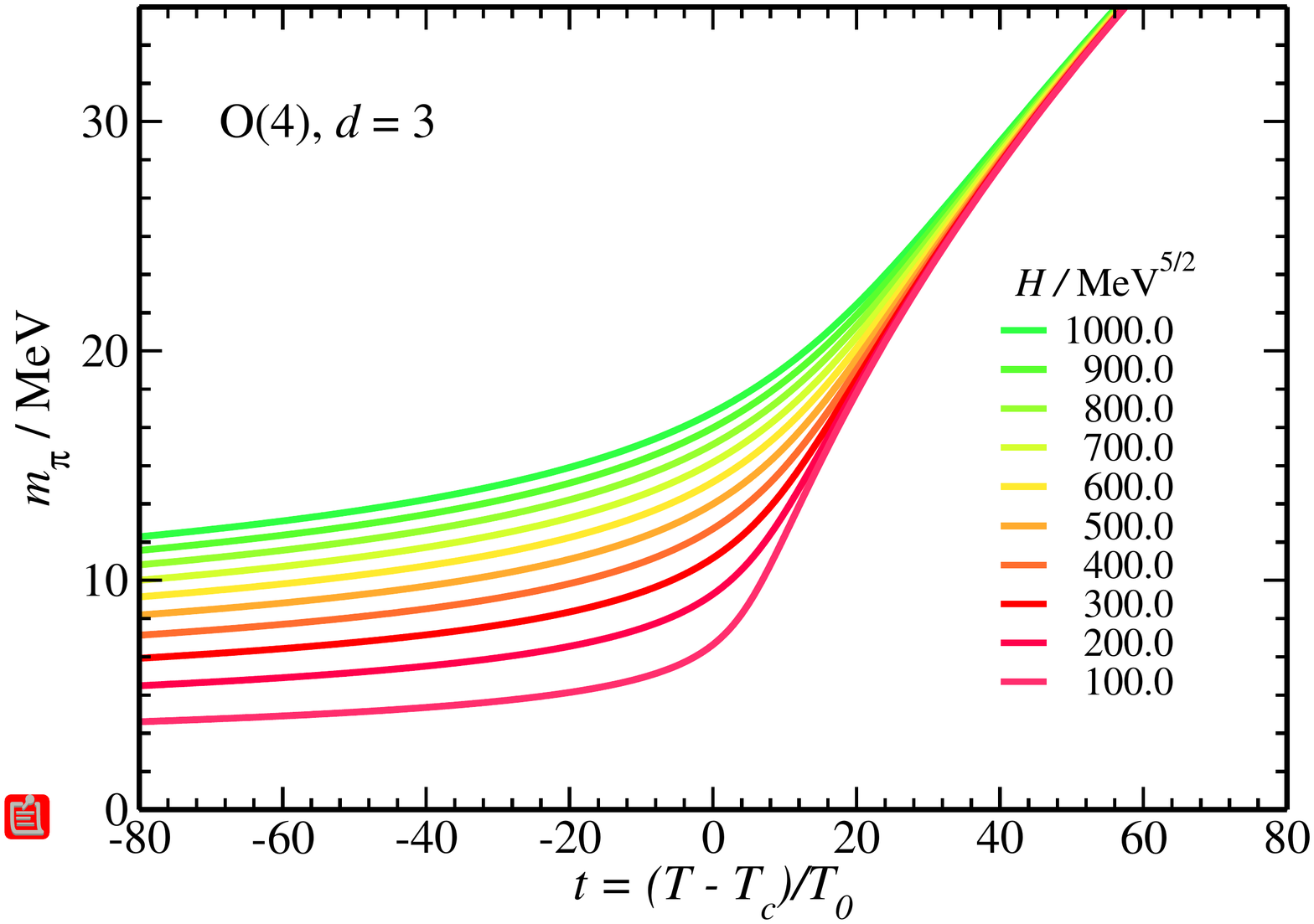}
\includegraphics[scale=0.3]{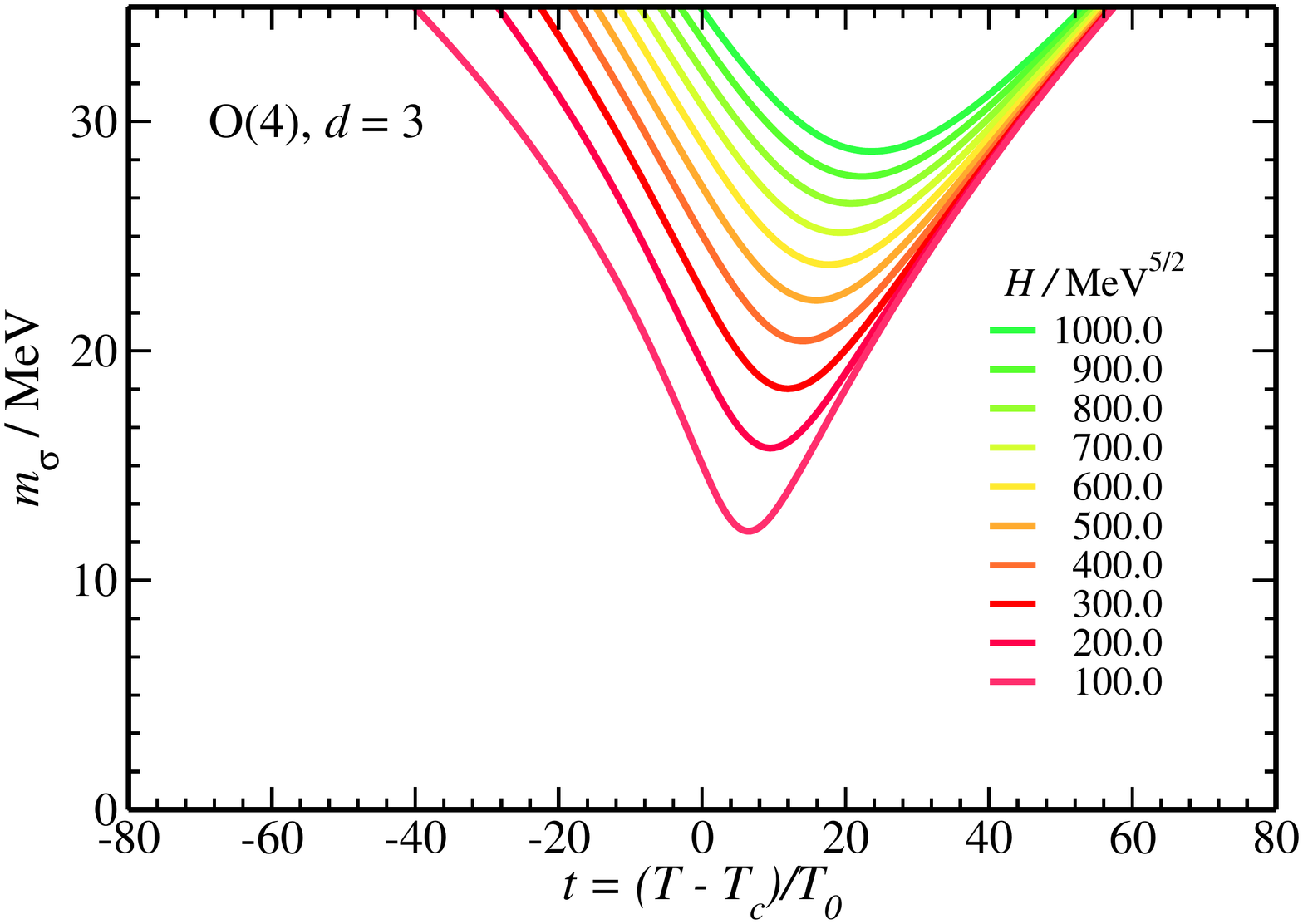}
\caption{\label{fig:masses}Mass of the transversal fluctuations $m_\pi$ (pseudo-Goldstone bosons) vs. reduced temperature $t$ (left column), and mass of the longitudinal fluctuations $m_\sigma$ vs. $t$ (right column) for different values of the symmetry-breaking field $H$. For our O(4)-model with a scale set by the UV cutoff $\Lambda=1.000$ GeV, large scaling violations appear already for values of the external filed $H$ which correspond to $m_\pi$ of less than $10$ MeV.}
\end{figure}

So far we have discussed the universal scaling behavior which is expected to apply to all systems with a critical point governed by O(4) symmetry. But for any scaling analysis, it is an important question how large the scaling region around the critical point actually is. This is not a universal property, but very much dependent on the system in question.
Because the normalization constants $T_0$ and $H_0$ are not universal but depend on the details of the system, a direct extrapolation of the scaling region from one system to a different one is not possible.

Universality only applies to the long-range behavior on small momentum scales, but not to the details of the short-range interactions. For different systems within the same universality class, on short scales in the UV regime the physics can be very different, and thus it is also not possible to fix scales for the scaling region from the UV behavior.

However, in the context of lattice simulations, a very important technique for the scaling analysis are finite-size studies. For a finite-size analysis, the length scale $L$ provides an additional point to compare long-range behavior of different systems, where the systems are comparable. The quantity of interest for this behavior is the ratio of the correlation length to the system size, $\xi/L$. Since the mass of the longitudinal fluctuations is bounded by the mass of the pseudo-Goldstone bosons, $m_\sigma \ge m_\pi$, a useful measure is given by $m_\pi L$. 

We have studied finite-size scaling for the O(4) model with the same choice of parameters as used in this work \cite{Klein:2007gu,Klein:2007qh}. Using the scale set by our choice of $\Lambda$, we find that the fields required to map out the finite-size scaling behavior for volume sizes of a few fm are much larger than the ones employed in this calculation. Consequently, we observe large corrections to scaling in these calculations. 

For the current choice of parameters, the masses are shown in Fig.~\ref{fig:masses}. Compared to 
absolute values on a hadronic scale, the masses of the fluctuations in the scaling region in this calculation are very small, and scaling corrections quickly become large for larger masses. However, it remains possible that we can reach larger, more realistic pion masses while still remaining inside the 
scaling region. A change in the scale $\Lambda$ changes the critical temperature and the values of the normalization constants $T_0$ and $H_0$. This could increase the size of the scaling region, while the absolute value of $m_\pi$ and the dimensionless product $m_\pi L$ could be kept constant.  

The consequences of these observations for the analysis of QCD lattice data are less clear. Since direct comparisons are inadvisable, conclusions must remain somewhat speculative. 
We find in our results for the order parameter for large values of $H$ that the results for the susceptibility as well as the order parameter still appear to scale, i.e. they are close together after rescaling, but show a large deviation from the scaling function.
Similar behavior is observed in some lattice simulation studies, where scaling with the critical exponents seems to take place for the peaks in the chiral susceptibility, but no agreement with the O(4) scaling functions is found \cite{Karsch:1994hm,Aoki:1998wg,Bernard:1999xx}.

\section{Conclusions}
\label{sec:conclusions}
The O(4) scaling function in three dimensions is important for the scaling analysis of systems in this universality class. QCD with two flavors is expected to fall into this class, if the phase transition is second order for two massless quark flavors. In QCD lattice simulations, the symmetry is broken by frequently large quark masses and a scaling analysis is necessary to determine the actual order of the phase transition. In addition, since lattice simulations are performed in a finite volume, finite-size scaling analysis is an important tool. Reliable knowledge about the scaling functions improves the power of this analysis.

We have investigated scaling in the O(4) model in $d=3$ with a non-perturbative Renormalization Group (RG) calculation. In contrast to many earlier investigations of scaling with functional RG methods, we have explicitly included an external symmetry-breaking field $H$. Due to the presence of this field, we chose not to work in a scale-free formulation, but retain the dimensions of all quantities. The scale for the calculation is set by our choice $\Lambda=1.0$ GeV for the initial RG scale. 

We work throughout in a local potential approximation in which the anomalous dimension vanishes, $\eta=0$. 
To ensure consistency of the analysis, we determine the critical exponents $\beta$, $\delta$, and $\nu$, and find good agreement with other RG calculations in this approximation. The values differ systematically from those of lattice Monte-Carlo simulations and from Functional RG calculations with momentum-dependent couplings, which can be explained by the vanishing anomalous dimension. 

We determine the scaling function for the order parameter in Widom-Griffiths scaling form $y(x)$ and as a function $f(z)$ of the scaling variable $z=t/h^{1/(\beta\delta)}$. We find good agreement with the results from the O(4) spin model lattice Monte-Carlo simulations of Engels and Mendes. Differences of the scaling function for the asymptotic behavior are explained by the difference in the values for the critical exponents.
In addition, we explicitly check Griffiths' expansion of the equation of state for asymptotically large values of the scaling variable, and we recover the critical exponent $\gamma$ from the asymptotic behavior. 
The value found in this way is in complete agreement with the one obtained  with the scaling laws.
We obtain a parameterization of the scaling function in Widom-Griffiths form $y(x)$ which provides a very good description of our scaling results for small values of the external symmetry-breaking field $H$.
This parameterization, together with the critical exponents, also provides a parameterization of the scaling function $f(z)$.
This result can be used for comparison in a scaling analysis, although come caution is warranted due to the systematic error in the values of the critical exponents, which is also reflected in the scaling function.

For large values of the symmetry-breaking field, we still observe scaling, but scaling corrections quickly become large. In terms of the scaling variable $z$, the region in which the scaled results fall onto the scaling curve shrinks considerably with increasing $H$. We have covered $7$ orders of magnitude in $H$, from perfect scaling behavior to a region where scaling violations become quite large. We observe that the Widom-Griffiths scaling form is more sensitive to scaling corrections than a rescaling as a function of $z$. 

The scaling function for the longitudinal susceptibility is obtained from a direct calculation of this susceptibility. We confirm an important consequence expected from the critical scaling behavior and  
show that this scaling function is already given by the scaling function of the order parameter. We find very good agreement between the parameters for the equation of state $y=y(x)$ obtained from fits to the order parameter and the susceptibility. We further confirm that the leading large-$x$ behavior of the scaling functions for the order parameter and the susceptibility coincide, which is predicted by the scaling relations and Griffiths' expansion.

Using the scaling form $f(z)$ for the scaling function, we find that the scaling function obtained from the order parameter describes the rescaled susceptibility perfectly, without any additional adjustments of parameters. 
At the same time, this result is a remarkable validation of the functional RG approach to scaling. The critical long-range fluctuations, which are responsible for the critical scaling behavior, are correctly included in the effective potential from which we calculate the observables.

The large scaling corrections we observe for large values of the symmetry-breaking field could have implications for the scaling analysis of lattice QCD results, where scaling behavior is observed, but no agreement with the O(4) scaling function is found.

An obvious improvement of these results can be achieved by including a non-zero anomalous dimension in the calculation, work in this direction is in progress. We have also applied this approach to finite-size scaling in the O(4) model \cite{Klein:2007gu,Klein:2007qh}, a more comprehensive presentation of our results is forthcoming.

Overall, we have obtained a result for the scaling function of the O(4) model in three dimensions which we hope will prove useful, and we have demonstrated very clearly some of the remarkable relations that follow from critical behavior. The non-perturbative RG has proven to be a suitable tool for this application.

\begin{acknowledgments}
The authors would like to thank Holger Gies, Tereza Mendes, and Jan Pawlowski for fruitful discussions.
This work was supported by the Excellence Cluster "Structure and Origin of the Universe" and by the Natural Sciences and Engineering Research Council 
of Canada (NSERC). TRIUMF receives federal funding via a contribution agreement 
through the National Research Council of Canada.
\end{acknowledgments}

\bibliography{O-4-scaling}

\end{document}